\documentclass{aa}  

\usepackage{graphicx}
\usepackage{txfonts}
\usepackage{graphicx}
\usepackage{multirow}
\usepackage{amsmath}
\usepackage{natbib}
\usepackage{xspace}
\usepackage{caption}

\newcommand{\dgr}{\ensuremath{^{\circ}}\xspace}
\newcommand{\dgrs}{\ensuremath{^{\circ}} }
\newcommand{\dgrq}{\ensuremath{^{\circ 2}}\xspace}

\newcommand{\simi}{\ensuremath{\sim}}
\newcommand{\elec}{e\ensuremath{^{-}}}
\newcommand{\eg}{e.g.,}

\newcommand{\sigep}{\ensuremath{\sigma_{e}}\xspace}
\newcommand{\sigepm}{\ensuremath{\sigma_{e,\,m}}\xspace}
\newcommand{\sigept}{\ensuremath{\sigma_{e,\,t}}\xspace}
\newcommand{\sigmat}{\ensuremath{\sigma_{t}}\xspace}
\newcommand{\sigmap}{\ensuremath{\sigma_{p}}\xspace}
\newcommand{\sigmaw}{\ensuremath{\sigma_{w}}\xspace}
\newcommand{\sigmar}{\ensuremath{\sigma_{r}}\xspace}
\newcommand{\domec}{Dome~C\xspace}
\newcommand{\modif}{}

\begin{document}

   \title{Four winters of photometry with ASTEP South at Dome~C, Antarctica}

   \author{N. Crouzet\inst{1,2}, E. Chapellier\inst{3}, T. Guillot\inst{3}, D. M\'ekarnia\inst{3}, A. Agabi\inst{3}, Y. Fante\"\i-Caujolle\inst{3}, L. Abe\inst{3}, J.-P. Rivet\inst{3}, F.-X. Schmider\inst{3}, F. Fressin\inst{4}, E. Bondoux\inst{3}, Z. Challita\inst{5}, C. Pouzenc\inst{6}, F. Valbousquet\inst{7}, D. Bayliss\inst{8}, S. Bonhomme\inst{3}, J.-B. Daban\inst{5}, C. Gouvret\inst{3}, A. Blazit\inst{3}}

   \institute{Instituto de Astrof\'isica de Canarias, C. V\'ia L\'actea s/n, E-38205 La Laguna, Tenerife, Spain \\
   \email{ncrouzet@iac.es}
   \and Universidad de La Laguna, Dept. de Astrof\'isica, E-38206 La Laguna, Tenerife, Spain  
   \and Universit\'e C\^ote d'Azur, Observatoire de la C\^ote d'Azur, CNRS, Laboratoire Lagrange, CS 34229, F-06304 Nice Cedex 4, France
   \and Harvard-Smithsonian Center for Astrophysics, 60 Garden Street, Cambridge, MA 02138, USA
   \and Observatoire Midi-Pyr\'en\'ees, 14 avenue Edouard Belin, 31400 Toulouse, France
   \and Concordia Station, Dome~C, Antarctica
   \and Optique et Vision, 6 bis avenue de l'Esterel, 06160 Juan-les Pins, France
   \and Department of Physics, University of Warwick, Gibbet Hill Road, Coventry CV4 7AL, UK}

   \date{Received ...; accepted...}

 
  \abstract
   {Dome~C in Antarctica is a promising site for photometric observations thanks to the continuous night during the Antarctic winter and favorable weather conditions.}
   {\modif{We developed instruments to assess the quality of this site for photometry in the visible and to detect and characterize variable objects through the Antarctic Search for Transiting ExoPlanets (ASTEP) project}.}
   {\modif{We present the full analysis of four winters of data collected with ASTEP South, a 10~cm refractor pointing continuously toward the celestial south pole. We improved the instrument over the years and developed specific data reduction methods.}}
   {\modif{We achieved nearly continuous observations over the winters. We measure an average sky background of 20 mag arcsec$^{-2}$ in the 579--642~nm bandpass. We built the lightcurves of 6000 stars and developed a model to infer the photometric quality of Dome~C from the lightcurves themselves. The weather is photometric $67.1\pm4.2$~\% of the time and veiled $21.8\pm2.0$~\% of the time. The remaining time corresponds to poor quality data or winter storms. We analyzed the lightcurves of \object{$\sigma$~Oct} and \object{HD~184465} and find that the amplitude of their main frequency varies by a factor of 3.5 and 6.7 over the four years, respectively. We also identify 34 new variable stars and eight new eclipsing binaries with periods ranging from 0.17 to 81 days.}}
   {The phase coverage that we achieved with ASTEP South is exceptional for a ground-based instrument and the data quality enables the detection and study of variable objects. These results demonstrate the high quality of Dome~C for photometry in the visible and for time series observations in general.}
            

   \authorrunning{Crouzet et al.}
   \titlerunning{ASTEP South}
   
   \maketitle
%

\section{Introduction}

Dome~C in Antarctica offers exceptional conditions for time-series observations thanks to the continuous night during the Antarctic winter and very favorable atmospheric conditions. Dome~C is located on a summit of the high Antarctic plateau at an altitude of 3233 meters, 1100 km away from the coast, at coordinates $\rm{75\dgrs 06'\, S, 123\dgrs 21'\, E}$. The site is equipped with the French-Italian Concordia station, which was built between 1999 and 2005 and runs all year long since 2005. The station is designed to host various scientific experiments, and can host about 80 people in summer and 15 people for winter-overs. Site testing for astronomy began in the early 2000s and revealed exceptional conditions: a clear sky most of the time including during the winter, low wind speeds, a dry atmosphere, and low scintillation \citep{Aristidi2003, Aristidi2005a, Lawrence2004a, Ashley2005a, Geissler2006b, Kenyon2006}. Experiments also found a free-atmosphere seeing above a boundary layer of about 30 meters \citep{Agabi2006, Trinquet2008, Aristidi2009}. Night-time astronomical observations started in 2006. They revealed a high duty cycle for spectroscopy and showed that high quality time-series photometry was achievable from this site \citep{Mosser2007a, Strassmeier2008, Briguglio2009}. The continuous coverage during the winter is beneficial to the search for and study of variable stars and transiting exoplanets.

We developed the Antarctic Search for Transiting ExoPlanets (ASTEP) project to search for and characterize transiting exoplanets and to determine the quality of Dome~C for photometry in the visible \citep{Fressin2005a}. The project started in 2005 and two instruments were installed at \domec. The main instrument, ASTEP 400, is a 40~cm Newton telescope designed and built by our team to perform high precision photometry under the extreme conditions of the Antarctic winter \citep{Daban2010}. ASTEP 400 was installed in 2010 and observed nominally over four winters. The results are presented in \citet{Abe2013, Mekarnia2016, Crouzet2016, Chapellier2016} and the lightcurves of more than $300\,000$ stars are publicly available\footnote{\url{https://astep-vo.oca.eu/}}. The 2017 and 2018 winter campaigns were dedicated to the photometric monitoring of \object{$\rm \beta$~Pic} during the transit of the Hill sphere of its planet \object{$\rm \beta$~Pic~b} \citep{Mekarnia2017}. 

A first instrument, ASTEP South, was installed in 2008 to acquire preliminary data for the project, to assess its feasibility, and to identify the challenges of photometric observations from Dome~C. ASTEP South observed over four winters from 2008 to 2011. The results from a preliminary analysis of the first winter are reported in \citet{Crouzet2010}. In this paper, we present the full analysis of the ASTEP South data.

\section{Observations}


ASTEP South consists of a 10~cm refractor, a front-illuminated 4096~$\times$~4096 pixel CCD camera, and a simple mount in a thermalized enclosure. The instrument is completely fixed and points continuously toward the celestial south pole during the Antarctic winter, which minimizes jitter noise and airmass variations during the observations. This setup leads to stars moving circularly on the CCD with a one-sidereal day period and to elongated point spread functions (PSF). The field of view is 3.88~$\times$~3.88\dgrq, corresponding to a pixel size of 3.41 arcsec on the sky. The exposure time is 29~s with 10~s overheads between each exposure. The PSF full width half maximum (FWHM) is nominally 2 pixels, not taking into account the elongation due to rotation. At the border of the field, the rotation rate of the stars on the CCD is 0.15 pixel s$^{-1}$, corresponding to a PSF elongation of 4.33 pixel during one exposure. The observing bandpass is equivalent to a large R band (600$-$900~nm). The enclosure is thermalized to -20{\dgr}C and fans are mounted inside to homogenize the temperature. The enclosure is closed by a double glass window to minimize temperature fluctuations (the outside temperature varies between $-50$ and $-80${\dgr}C during the winter at Dome~C). Software programs were developed by our team to control the camera, run the acquisitions, transfer and save the data. This instrumental setup is presented in more detail in \citet{Crouzet2010}. ASTEP South was installed at the Concordia station in January 2008 and observed during the Antarctic winters 2008, 2009, 2010, and 2011. Spurious reflections presumably due to moonlight were identified on images of the first campaign. We installed a 40~cm long baffle around the enclosure's window in July 2009 to minimize parasitic light reaching the CCD. The data quality is evaluated daily by a software program that performs a basic data reduction, runs a preliminary analysis, and sends the results by email to the team in Nice. With no moving parts, a constant field of view, and dedicated software programs, this first ASTEP instrument observed in automatic mode during almost all the winters and required only limited maintenance. 


The number of science frames collected by ASTEP South over the four observing seasons is given in Table~\ref{tab: image number}. With a size of 32 MB per image, this yields 27.8 terabytes (TB) of science data. The amount of calibrated images is approximately the same because a bias frame was taken between each science frame. This yields a total of about 60~TB of data. The limited internet bandpass at Concordia prevents a full data transfer during the winter. Instead, the data are stored on external hard drives that are shipped from Concordia to Nice at the end of the winter, and a backup copy is kept at Concordia.

\begin{table}
\begin{center}
\caption{Number of science images and amount of data acquired by ASTEP South. Calibration frames are not included.}
\label{tab: image number}
\begin{tabular}{cccc}
\hline\hline
Observing & Number of   & Total size  \\
season & science frames & [TeraBytes]   \\
\hline
2008 & 125,003 & 4.0 \\
2009 & 223,756 & 7.1 \\
2010 & 241,026 & 7.7 \\
2011 & 279,295 & 8.9 \\
\hline
Total & 869,080 & 27.8 \\
 \hline\hline
\end{tabular}
\end{center}
\vspace{-2mm}
\end{table}

\section{Data reduction}

\subsection{Star catalog}

The ASTEP South target star catalog is a combination of UCAC3, UCAC2, TYCHO-2, and APASS. These catalogs were obtained from VizieR at the Centre de Donn\'ees Astronomiques de Strasbourg\footnote{http://vizier.u-strasbg.fr/} and from the American Association of Variable Star Observers Photometric All-Sky Survey website\footnote{http://www.aavso.org/apass}, via requests centered on the celestial south pole (RA $=$ 0\dgr, Dec $=$ $-$90\dgr) with a search radius of 2\dgr. Our main catalog is UCAC3. Most of the bright stars are flagged as unreliable data; we use the TYCHO-2 entries for these stars. TYCHO-2 does not add any stars. We also noticed that a small fraction of stars in our images were missing in UCAC3 but were present in UCAC2; we use the UCAC2 entries for these stars. Finally, APASS is used to gather complementary information on the magnitudes of the targets. We keep all the magnitude and flag information available in these catalogs. For multiple entries, defined as having a separation of less than 7 arcsec (2 pixels in the ASTEP South images), only the brightest star is kept; the other ones may be catalog artifacts or stars that would not be resolved by ASTEP South. We use the UCAC3 fmag magnitude defined in the bandpass 579--642~nm as our reference magnitude, the UCmag defined in the same bandpass for stars present only in UCAC2, and the Sloan i' magnitude for stars present only in APASS.

Due to strong image vignetting (\simi25\% at the edges) and to PSF elongation, the limit magnitude of stars that are visible in the ASTEP South images is lower at the border than at the center of the field. The limit flux $F_{\rm lim}$ is well represented by a linear function after a certain distance from the center:
\begin{equation}
F_{\rm lim} = 
\begin{cases}
F_{\rm lim,\,c} & \text{if} \; r \le r_0 \\
F_{\rm lim,\,c} \; \left( 1 - a \left( r - r_0 \right) \right) & \text{if} \; r > r_0 
\end{cases}
\label{eq: maglim}
\end{equation}
where $F_{\rm lim,\,c}$ is the limit flux at the center of the image, $r$ is the radial distance from the center in pixels, $r_0$ = 300 pixels, and $a=4.284\times10^{-4}$ pixel$^{-1}$. This corresponds to a 1.5 magnitude difference between the center and the edges, which have a limit magnitude fmag$_{\rm lim}$ of 16 and 14.5 respectively. We used this function to select only stars below the limit magnitude after calculating their nominal radial position in the images.

In total, the ASTEP South catalog comprises 5954 target stars including 5830 from UCAC3, 91 from TYCHO-2 (almost all the stars with $\rm fmag < 9.8$), 29 from UCAC2, and four from APASS. A visual inspection of these targets superimposed on a high quality ASTEP South image shows that this catalog is nearly complete for our purpose.

\subsection{Image calibration}


Calibration images were taken during the observations, with a sequence of one bias between every science image, and one dark every ten science images. Although very consuming in terms of disk space, the large number of bias was useful for a sanity check, for example to measure the bias level as a function of the camera's electronics temperature. We removed most of the oscillations shown in Fig.~10 of \citet{Crouzet2010} by improving the thermalization after the first campaign, but non-periodic temperature fluctuations remain. To correct for this, we computed the median value of each bias and interpolated them linearly to obtain the bias level for each science image. A ``masterbias'' and a ``masterdark'' were created for each day by stacking frames together with successive median filters. The dark current itself is negligible and was not detected. The median level of the darks is lower than that of the bias by about 50 analog to digital units (ADU) for an unexplained reason, and the median level of the science images matches that of the darks.

\begin{figure}[htbp]
   \centering
   \includegraphics[width=8cm]{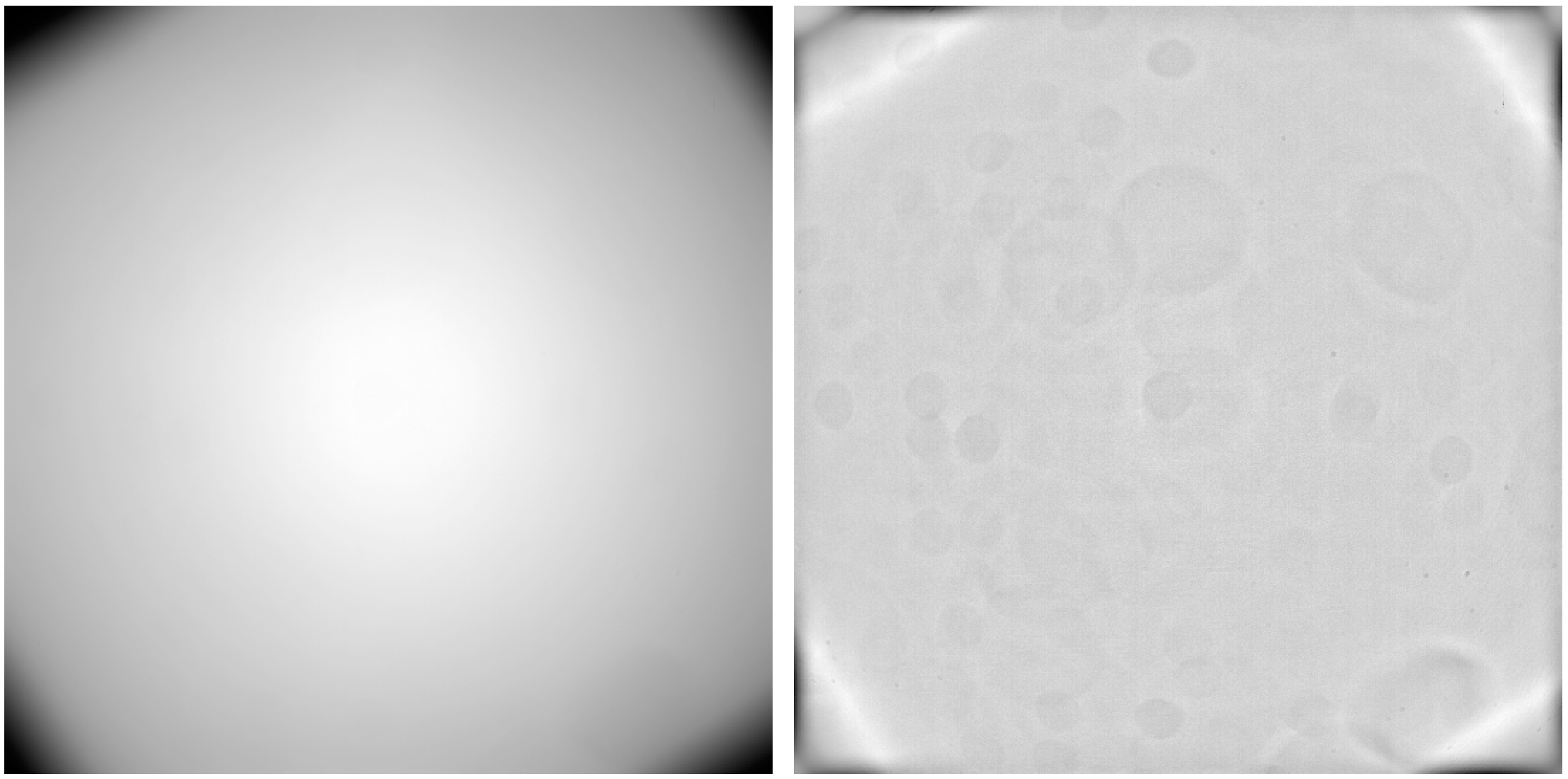}
   \caption{Masterflat (left) and same image after subtracting the low frequency component (right) for 2009. \modif{The vignetting is \simi25\%. The color scale is arbitrary}.}
   \label{fig: master images}
\end{figure}

Around noon local time, the Sun is high enough (below the horizon) to slightly increase the sky background \citep{Crouzet2010}. Although not detectable by naked eye, this illuminates enough the CCD to create ``twilight'' images, which we used to build flat-fields. We used images with median values between 8000 and $30\,000$ ADU (the pixel dynamic range is $65\,535$ ADU). Between a few tens and two hundred such images were recorded daily except in June where the Sun remains too low. Thanks to their motion around the CCD, stars could be removed by stacking images together and applying successive medians. We built one flat-field for each day with the following procedure: bright images were calibrated from bias and dark, corrected from hot pixels, normalized by their median in the $512 \times 512$ pixel central region, and divided into sets of 15 images. We interleaved the different sets instead of grouping consecutive images, so that images in the same set were optimally spaced in time to have stars as far away as possible between images. Images in each set were stacked together with a median filter. We obtained up to seven intermediate flat fields. We stacked them together with an outlier resistant mean to create one flat-field. We found that this method was significantly more efficient at removing stars than applying a unique median filter to all the images together. Finally, these daily flat-fields were stacked together with a median to create ``masterflats'' for each winter. This method yielded high quality masterflats, as shown in Fig.~\ref{fig: master images}.

\begin{figure}[htbp]
   \centering
   \includegraphics[width=9cm]{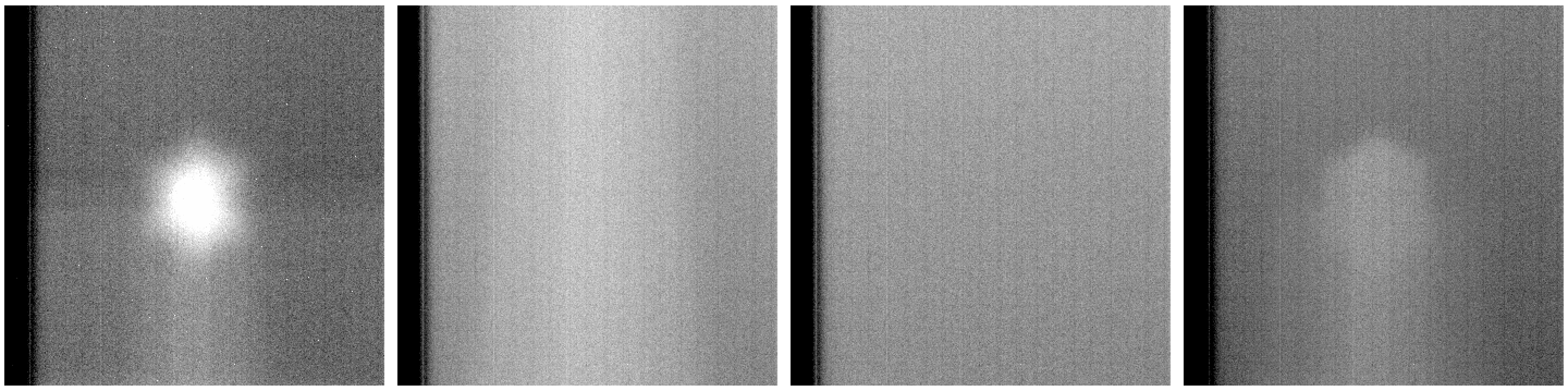}
   \caption{From left to right: masterdark, masterbias, masterbias with shutter effect corrected, and mastermix for July 2nd, 2010. \modif{The color scale is arbitrary}.}
   \label{fig: shutter calibration}
\end{figure}

Random failures of the camera shutter began after two winters of operation at $+5${\dgr}C, corresponding to \simi $393\,000$ open-close cycles. This issue became more severe in 2010 and 2011. We increased the temperature to $+30${\dgr}C around the shutter for the 2011 campaign, which improved its reliability but did not solve the issue completely. Science images, darks, and bias were affected (Fig.~\ref{fig: shutter calibration}). We developed a calibration method to correct for these effects.
The masterbias and masterdark did not provide a satisfactory calibration, so we built a third type of calibration image. First, we cleaned the masterbias from residual light: we removed a low frequency component from each row and replaced it by a linear function between both sides (this linear trend also appeared on bias unaffected by shutter issues).
Then, we adjusted the median value of the masterbias and of the corrected masterbias to match that of the masterdark; the medians were calculated in 200 pixel wide boxes located at the corners. Finally, we stacked these three images together with a median to create a ``mastermix" (Fig.~\ref{fig: shutter calibration}). This median can be seen as a conditional statement: each part of the mastermix corresponds to one of these images. We found that it was a good representation of the shutter effects seen on the science images, and provided a satisfactory calibration.

\begin{figure}[htbp]
   \centering
   \includegraphics[width=8cm]{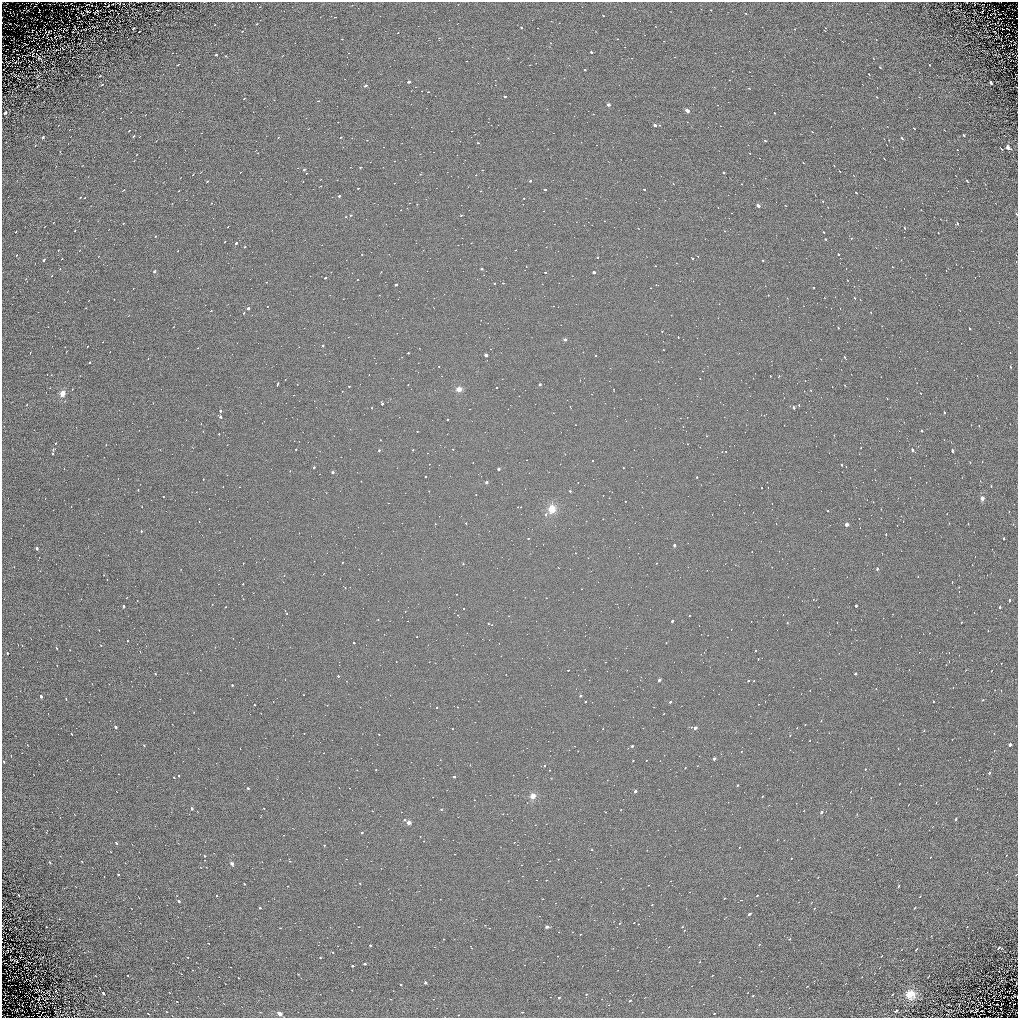}
   \caption{\modif{Example of a calibrated image. The field of view is 3.88~$\times$~3.88\dgrq. The color scale is arbitrary}.}
   \label{fig: calibrated image}
\end{figure}

Our science images were calibrated as:
\begin{equation}
C = \frac{\left( I - M - b \right) \times g}{F}
\label{eq: calib}
\end{equation}
where $C$ is the calibrated image, $I$ is the raw image, $M$ is the masterdark or the mastermix in case of shutter issues, $b$ is the bias variations due to temperature fluctuations (taken around its median value for each day), $g$ is the gain of the CCD, and $F$ is the masterflat. We measured the gain during the characterization of the ASTEP cameras in the lab using the approach of \citet{Leach1980, Leach1987}. We took pairs of images at different illumination levels spanning the full dynamical range of the CCD. Then, the gain is given by $g = 2 \, m/v$, where $m$ is the mean of the image and $v$ is the variance of the differential image (we neglected the read-out noise in this calculation). The $v(m)$ relation was well represented by a linear function below $31\,360$ ADU and by a third order polynomial above. This yielded a constant gain of 2 \elec /ADU below $31\,360$ ADU and an increasing gain from 2 to 2.2 \elec /ADU between $31\,360$ and $60\,000$ ADU. Finally, we corrected for hot pixels. They were present in the masterdarks but not in the masterbias. Pixels in the masterdarks that exceeded their value in the masterbias by more than 150 ADU were flagged for all the images of that day and were replaced by the average of the four adjacent pixels. An example of calibrated image is shown in Fig.~\ref{fig: calibrated image}.

\subsection{Field matching algorithm}

In each science image, we identified point sources using the FIND procedure in IDL, with a threshold intensity of five times the sky standard deviation. These point sources were compared to the ASTEP South catalog using a home-made field matching algorithm. The star coordinates in the catalog (RA, Dec) are converted into image coordinates $(x, y)$. Then, we perform iterations on pairs of stars using the 20 brightest non saturated stars. For each pair in the ASTEP South image and each pair in the catalog, we calculate the geometrical transformation that matches one pair to the other, skipping iterations where both stars do not have approximately the same separation. Then, we apply this transformation to the 400 brightest stars and evaluate the matching by a $\chi^2$, where $\chi$ is the median distance between a star in the transformed image and the closest star in the catalog. The best matching is found very quickly and yields a typical precision of 0.4 pixel. Then, $\chi^2$ is minimized using the simplex method in order to better estimate the transformation parameters. This yields a precision of 0.2 pixel. Finally, we calculate and correct for distortions using a regression in $x$ and $y$ up to the third order, including crossed terms. The final precision on the star positions is 0.1 pixel.

\subsection{Photometric algorithm}

The ASTEP South PSFs are particularly challenging for high-precision photometry. We developed our own photometric algorithm to take into account the elongated PSFs, and their elongation directions and positions that vary in time as the stars rotate on the CCD. The aim was to maximize the measured flux while reducing the sky background, readout noise, hot pixels, cosmic rays, and contamination by surrounding stars compared to circular apertures.

\begin{figure}[htbp]
   \centering
   \includegraphics[width=4cm]{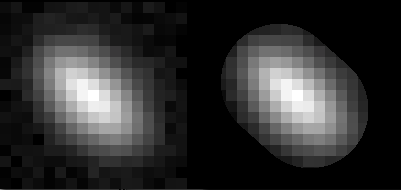}
   \includegraphics[width=3.8cm]{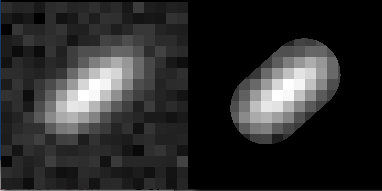}
   \includegraphics[width=3.95cm]{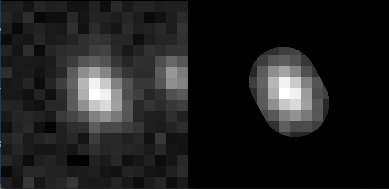}
   \includegraphics[width=3.85cm]{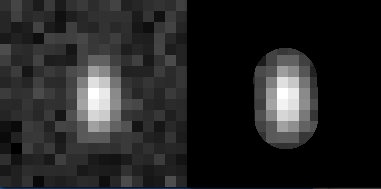}
      \caption{Subimage around the star (left) and photometric aperture computed by the ASTEP South photometric algorithm (right) for stars of magnitude 8.03 (top left), 10.6 (top right), 11.04 (bottom left), and 12.08 (bottom right). \modif{The color scale is arbitrary}.}
   \label{fig: apertures}
\end{figure}

We computed the apertures for each star in each image. The apertures consist of two half-circles joined by a rectangle which length and orientation depend on the position of the star relatively to the celestial south pole (Fig.~\ref{fig: apertures}). The aperture is quasi-circular for stars close to the south pole and is the most elongated at the edges of the field. To measure the stellar flux, we divide each pixels within the aperture into a grid of $11\times11$ subpixels, project the aperture onto this grid, and sum the flux at the subpixel level; this is equivalent to weighting each pixel by the fraction of its area within the aperture. The sky background is calculated in an annulus of 11 pixels starting 5 pixels outside the star aperture, multiplied by the aperture area, and subtracted to the measured flux. 

To calculate the optimal aperture size for each star, defined by the radius of the half-circles, we ran the photometric algorithm on a typical day with radii ranging from 0.5 to 10 pixels by steps of 0.05. The lightcurves were calibrated by a reference lightcurve, calculated from the median of the 30 non saturated brightest stars using an aperture radius of 5 pixels. For each star, we kept the aperture radius that yielded the smallest root mean square (RMS). In the case of neighboring stars, the resulting aperture was generally too large because it contained several stars, and the optimal aperture appeared as a local RMS minimum rather than as a global minimum. This was corrected by forcing the aperture radius to remain within $\pm2$ pixels of the typical aperture for this star's magnitude. In the end, each star had its own optimized aperture size, which remained the same for all the photometric reduction.

After reducing all the days of a winter, we created a single lightcurve file for each star with the flux measurements and information on the aperture position, centroid position, and local sky background. We also built a reference lightcurve using 60 bright stars with low RMS that we identified from one typical day. We normalized these 60 lightcurves by their respective medians, and averaged them using an outlier resistant mean with a 3$\sigma$ cut-off. The resulting reference lightcurve was also normalized by its median.

\subsection{\modif{Correction of systematic effects}}
\label{sec: Systematic effects}

\begin{figure*}[ht!]
   \centering
   \includegraphics[width=5.27cm]{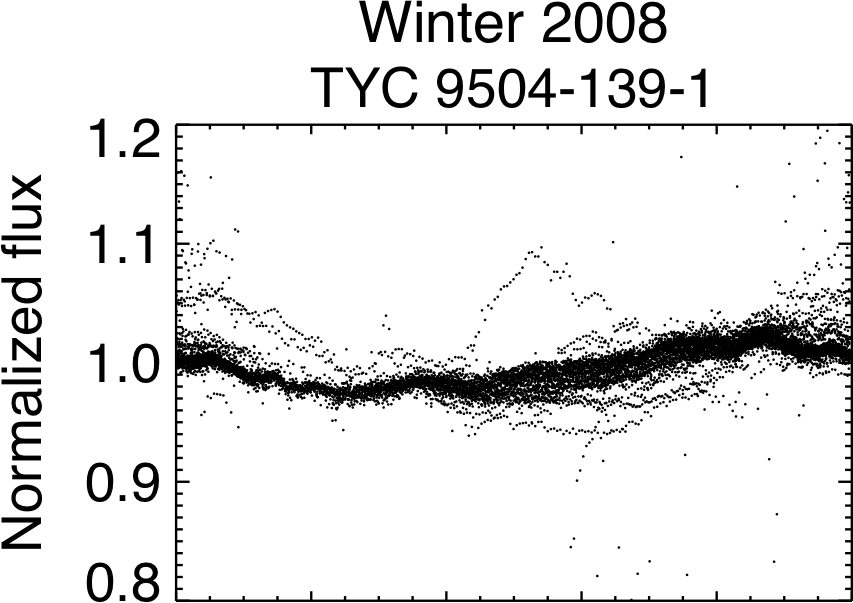}
   \includegraphics[width=4.2cm]{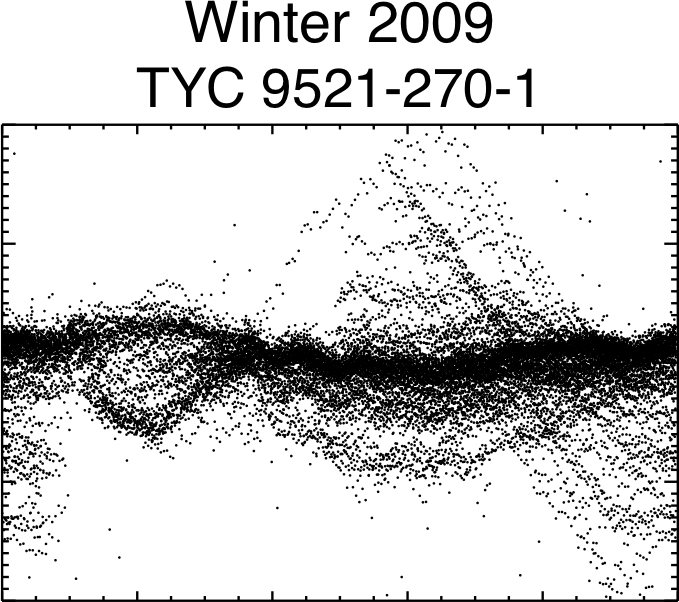}
   \includegraphics[width=4.2cm]{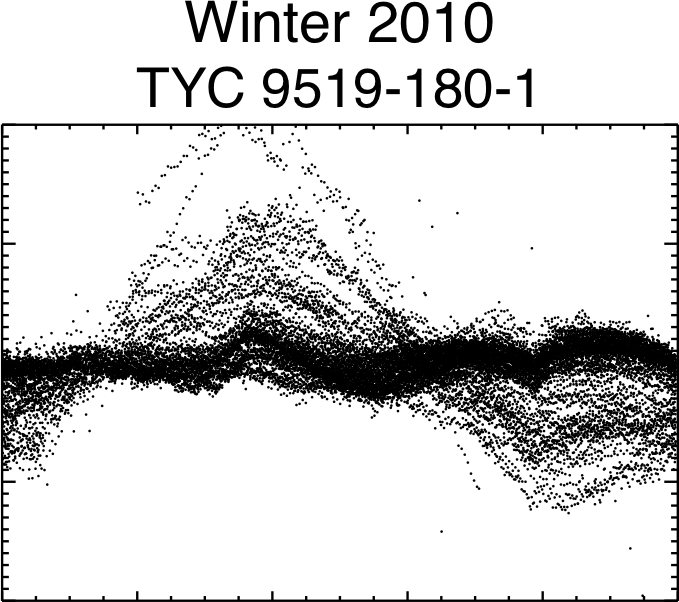}
   \includegraphics[width=4.2cm]{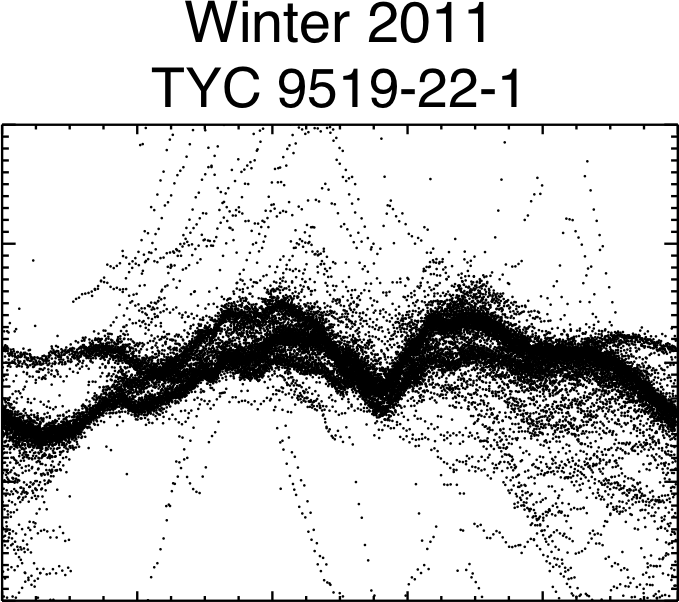}
   \includegraphics[width=5.27cm]{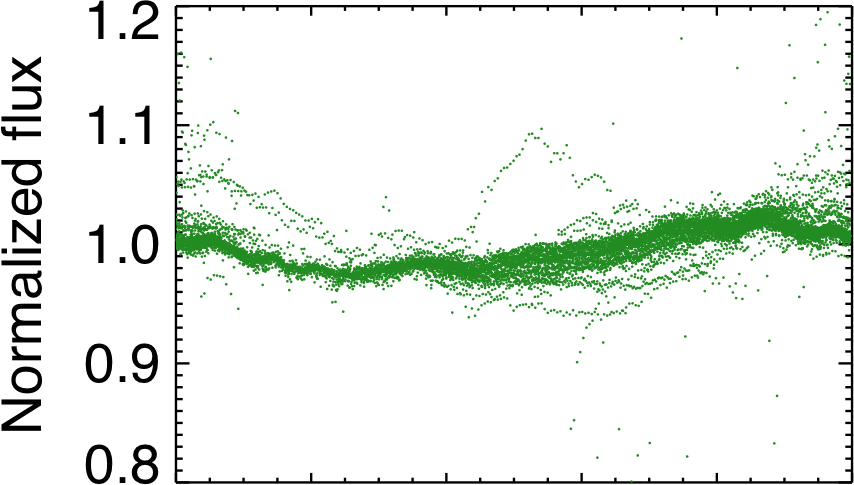}
   \includegraphics[width=4.2cm]{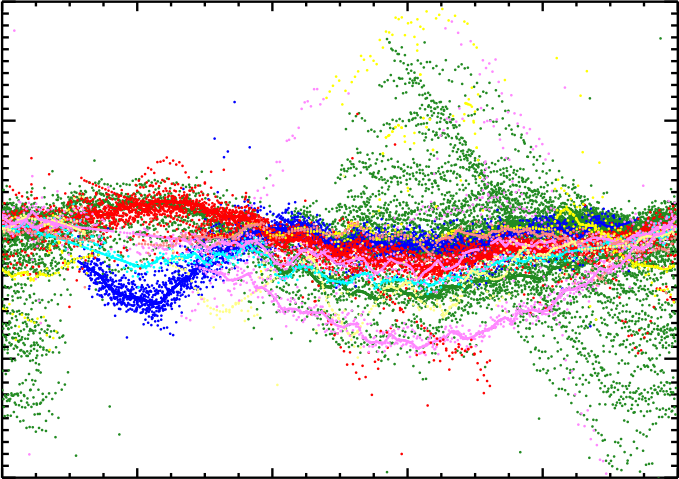}
   \includegraphics[width=4.2cm]{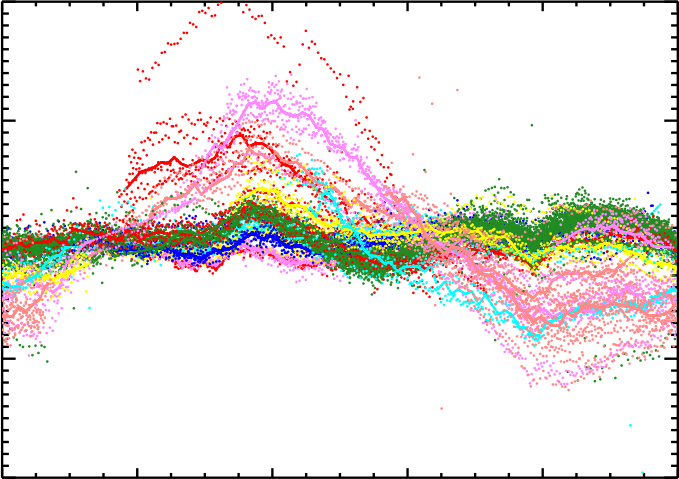}
   \includegraphics[width=4.2cm]{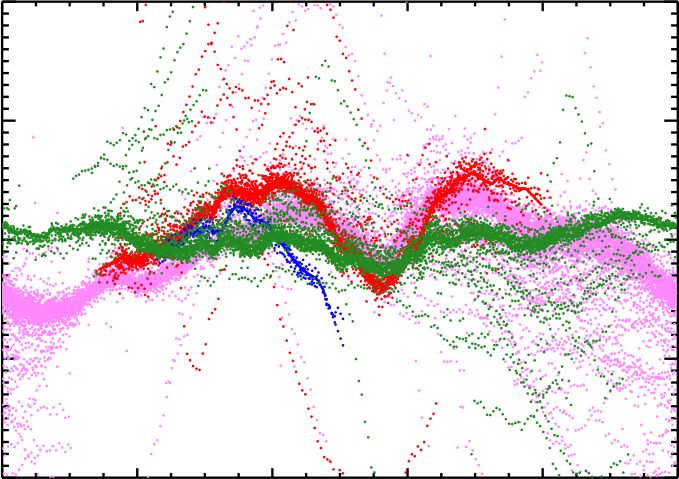}
   \includegraphics[width=5.27cm]{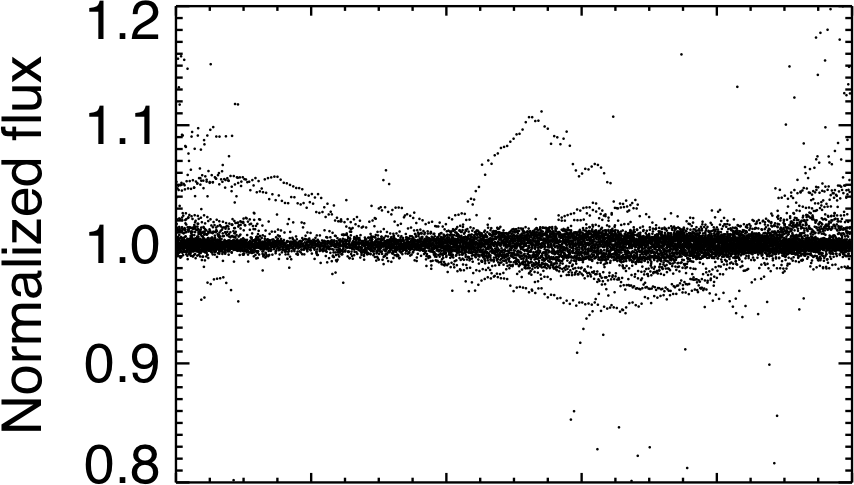}
   \includegraphics[width=4.2cm]{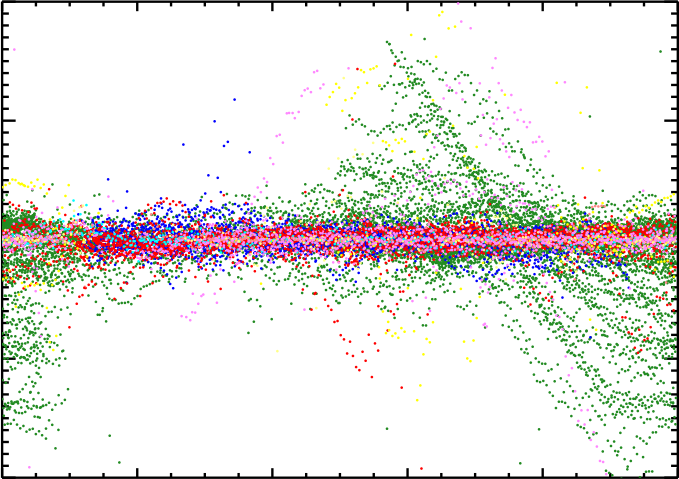}
   \includegraphics[width=4.2cm]{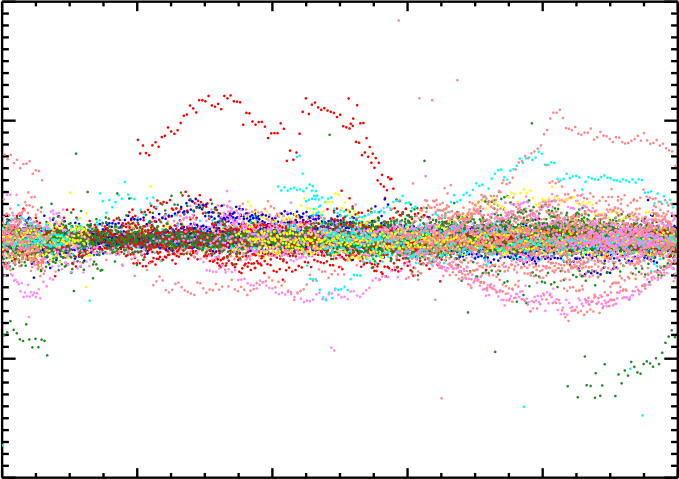}
   \includegraphics[width=4.2cm]{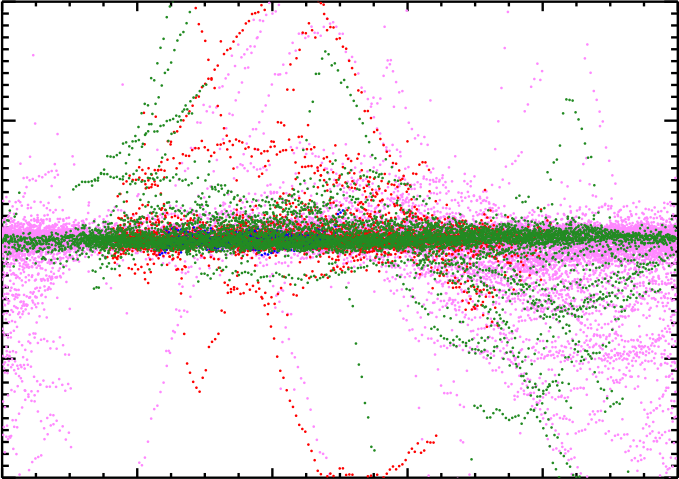}
   \includegraphics[width=5.27cm]{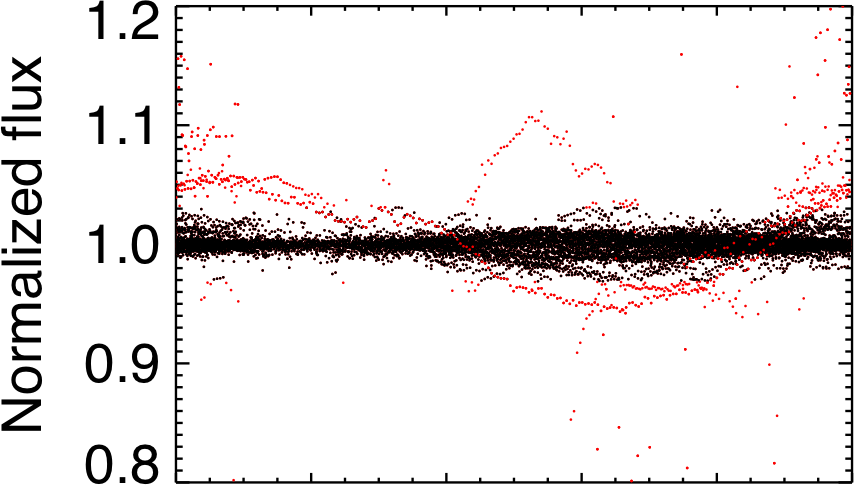}
   \includegraphics[width=4.2cm]{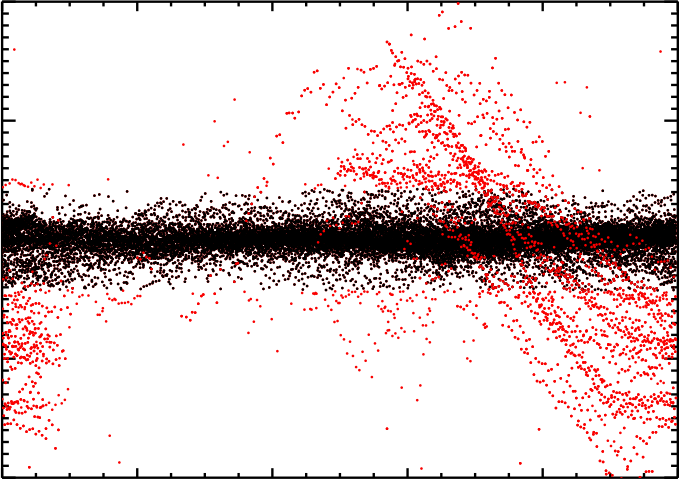}
   \includegraphics[width=4.2cm]{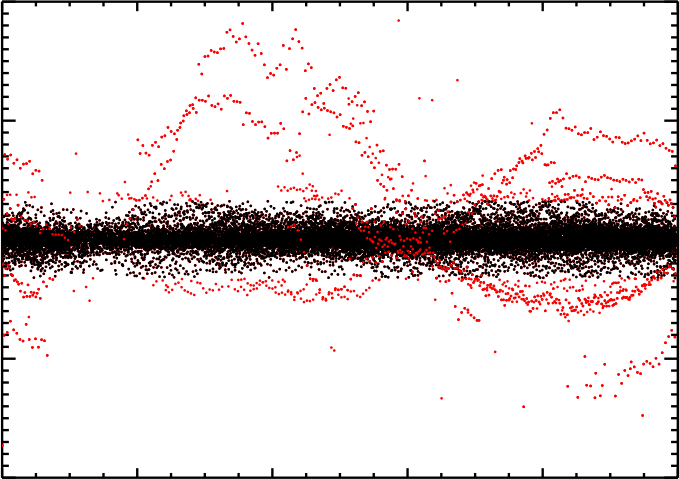}
   \includegraphics[width=4.2cm]{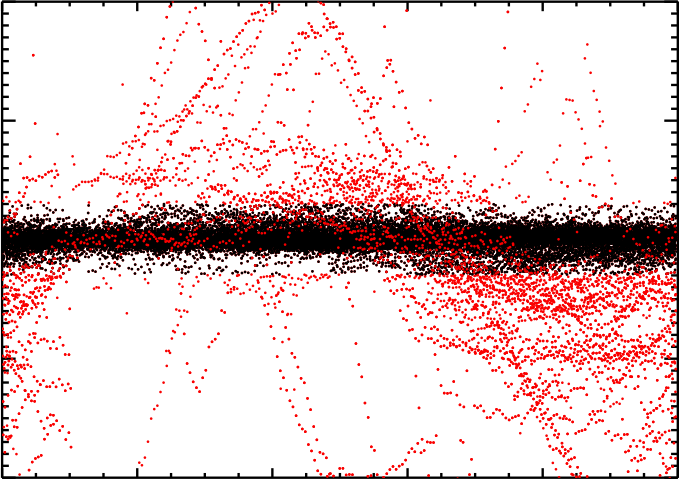}
   \includegraphics[width=5.27cm]{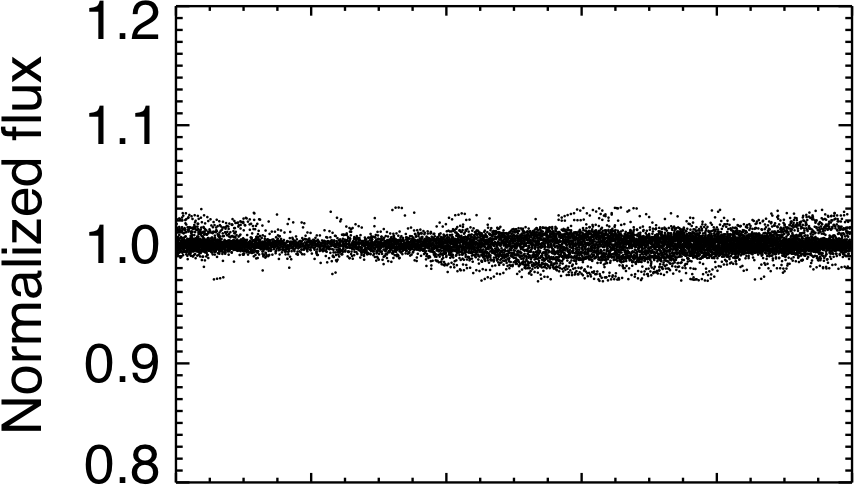}
   \includegraphics[width=4.2cm]{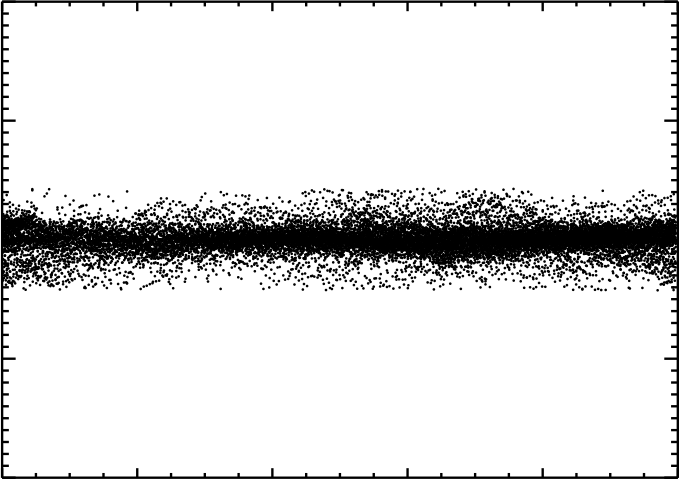}
   \includegraphics[width=4.2cm]{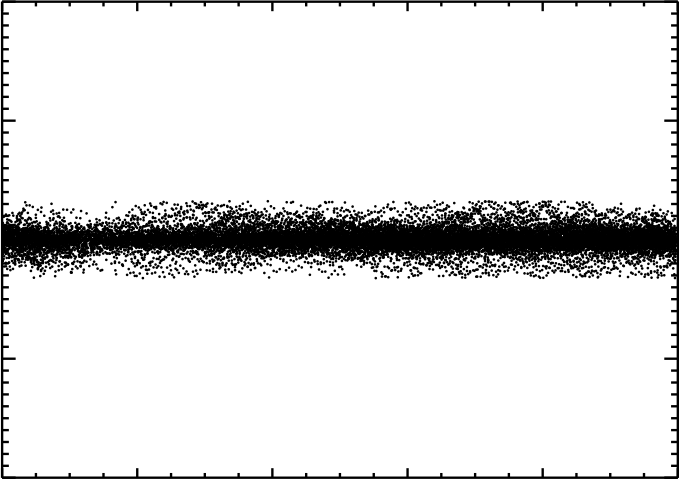}
   \includegraphics[width=4.2cm]{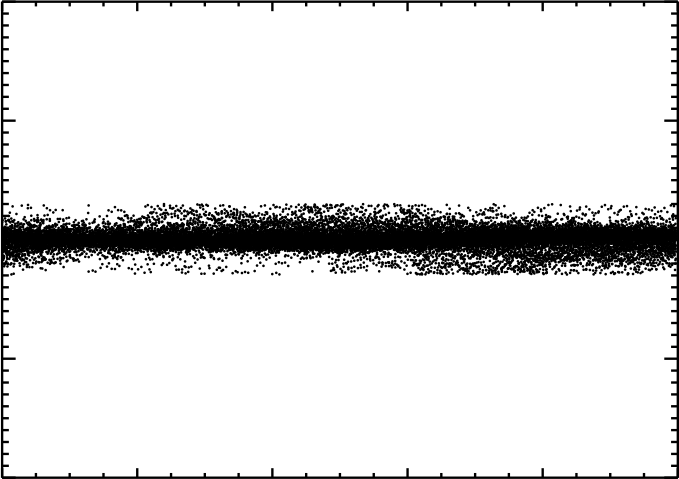}
   \includegraphics[width=5.27cm]{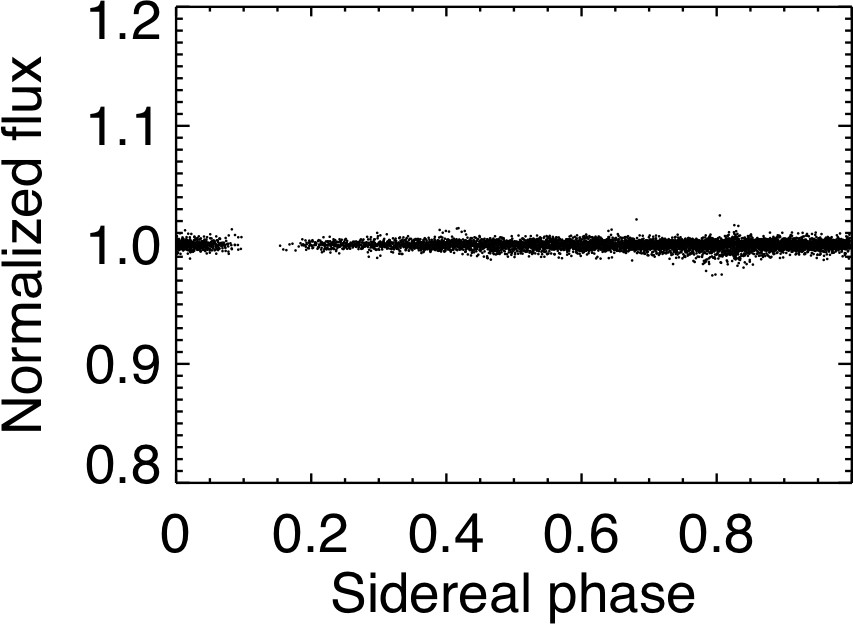}
   \includegraphics[width=4.2cm]{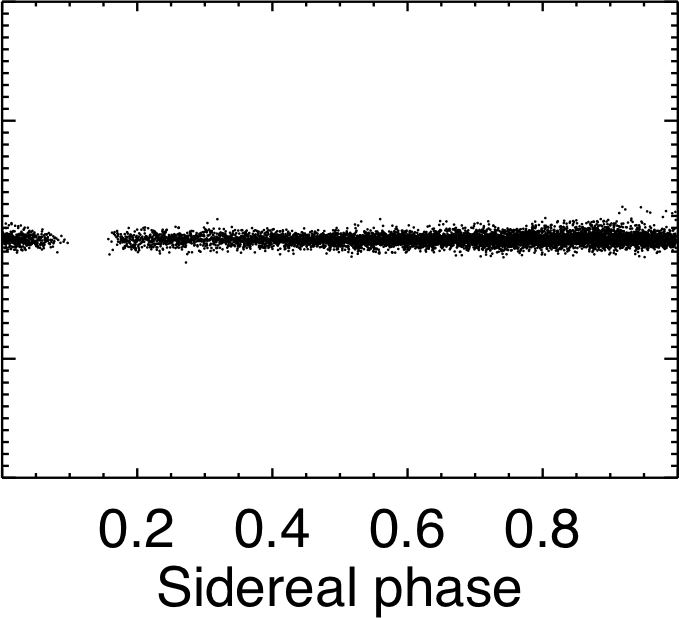}
   \includegraphics[width=4.2cm]{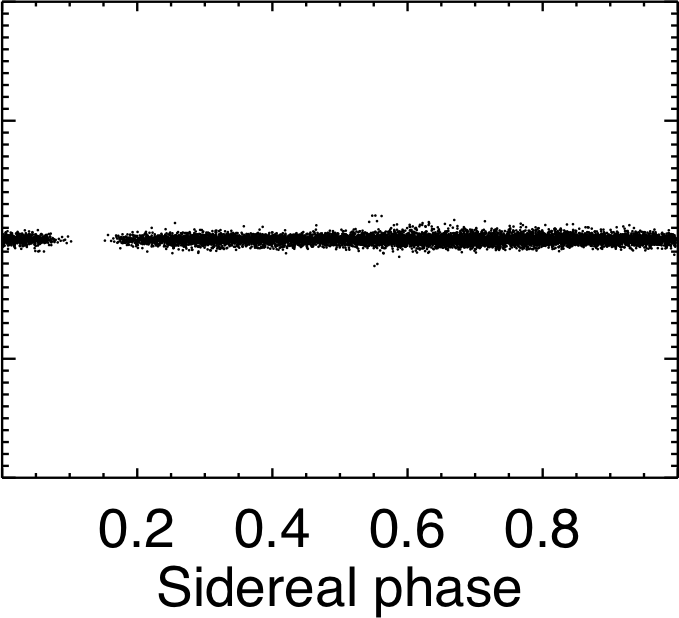}
   \includegraphics[width=4.22cm]{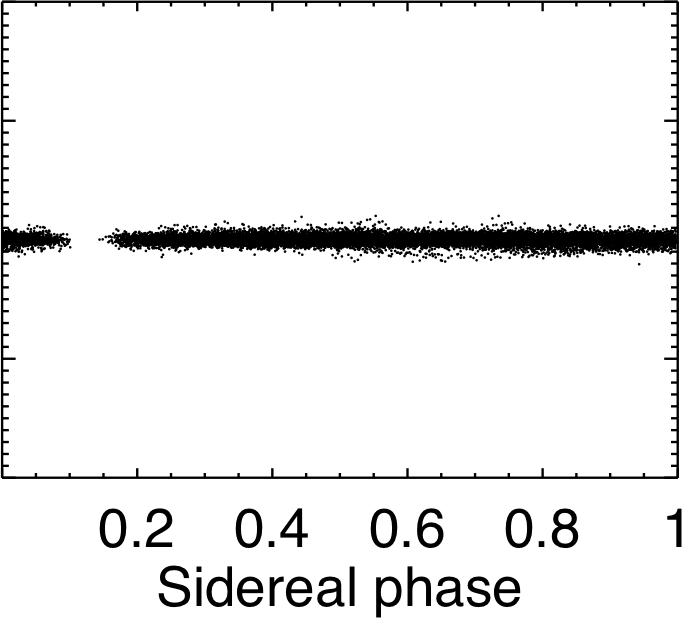}
   \caption{Correction of the one-sidereal day variations. From left to right: winters 2008 to 2011. One star is given as an example for each winter; the stars' IDs are indicated. The lightcurves are phase-folded at a one sidereal day period. First row: initial lightcurve. Second row: split into different data sets. Each set is represented by a color and the recovered pattern is shown as a line of the same color. Third row: lightcurve after division by the patterns. Fourth row: identification of outliers, indicated in red. Fifth row: lightcurve with outliers removed. Sixth row: final lightcurve after applying all the photometric correction and data selection steps.}
   \label{fig: corday}
\end{figure*}

We divided the target stars' lightcurves by the reference lightcurve and binned them per 400 second intervals (6.7 minutes, approximately ten points). Then, we identified systematic variations and applied several corrections. The largest variations were due to the rotation of stars on the CCD and were on the order of a few percent, sometimes up to 10\%. They formed a pattern with a one-sidereal day period, which was different for each star and was usually repeatable. However, several complications arose. First, we sometimes modified the pointing during the winter, for example after improving the thermal control or after other interventions on the instrument; this changed the one-day patterns. Second, each pattern was split into two or more patterns that are shifted in flux, corresponding to different states of the FWHM (see Sect.~\ref{sec: fwhm}). Third, we noticed flat-field variations over the winter possibly due to mechanical deformations and temperature variations inside the box, and we built a flat-field for each image calibration method (with or without shutter correction). In the end, we used between one and four flat-fields over the winter. These pointing, FWHM, and flat-field effects were mixed together and yielded variable one-sidereal day patterns. To recover them, we split the data into sets that were as consistent as possible according to these parameters; in practice, each set corresponds to a specific time interval during the winter.
Then, we recovered the patterns for each star and each set individually: we fold the star lightcurve over a given time interval at a one-sidereal day period, split it into 100 bins, calculate the median flux for each bin, interpolate them linearly, and divide the lightcurve by this function (Fig.~\ref{fig: corday}). We applied this procedure twice. In some cases, because of poor weather, large FWHM variations, or small data sets, the patterns could not be recovered. The 2008 and 2011 data were divided into one and four sets respectively and were well corrected. The 2009 and 2010 data were much less stable and were divided into 12 and 16 sets respectively; some of them were completely discarded, whereas several large sets comprised most of the good data and were well corrected.

We converted fluxes into magnitudes and normalized them by their mean magnitude. From this point, we worked with the magnitude residuals. Then, we calibrated each star by a set of reference stars chosen as follows. We divided each lightcurve by all others and sorted the resulting lightcurves by their mean absolute deviation (MAD). The stars yielding the lowest MADs were used as reference stars and their lightcurves were averaged to build a reference lightcurve. We repeated this procedure for each star (so in the end each star had its own set of reference stars). We used three reference stars for the 600 brightest stars ($\rm fmag < 11.9$) and ten reference stars for fainter stars.

We corrected for remaining systematic variations. We gathered the lightcurves in a star, epoch array with stars sorted by ascending magnitudes. Systematic trends appeared along the epochs and were common to stars of similar magnitudes. We corrected for them on a row by row (epoch by epoch) basis by computing a median smoothing over 200 stars and subtracting it to the initial row. This method yielded a better correction than Systematic Removal \citep[SYSREM,][]{Tamuz2005} for these data at this stage of the pipeline, because it is insensitive to variable stars, which are often interpreted as systematic effects by SYSREM. Our method provided more uniform lightcurves overall. After these corrections, we computed the average standard deviation for each epoch $\sigma_{e}$ using stars of magnitudes between 12 and 13.5.

\subsection{\modif{Data selection and final lightcurves}}
\label{sec: Data selection}

In the process, we flagged outliers departing by more than $4\,\sigma$ from each lightcurve, where $\sigma$ is the standard deviation of the lightcurve. We flagged entirely the epochs with more than 50\% of outliers, as well as the epochs with less than 500 detected stars. Considering individual stars, we flagged entirely the days with more than 50\% of outliers, because the remaining points were generally off the regular one-sidereal day pattern. We flagged the epochs with a median sky background greater than 500 ADU. We flagged the epochs with $\sigma_{e} > 2\%$ because they correspond to the tail of the $\sigma_{e}$ distribution (see Sect.~\ref{sec: photometric quality}). For 2009, a limit at 3\% would be more appropriate considering all the data; however the distribution is bi-modal, one mode corresponding to a nominal FHWM (\simi 2 pixels) and the other to a larger FWHM ($>$ 3 pixels). To keep only nominal data and for consistency with the other years, we kept the limit $\sigma_{e} < 2\%$. We also flagged the epochs at which more than 30\% of the 1000 brightest stars had bad data (these epochs were often of poor quality for all the stars), and those at which more than 10\% of the 100 brightest stars had bad data (we found that this was a good proxy for data taken around full Moon). Finally, we flagged lonely epochs (epochs with less than 20\% of good epochs within their 40 closest neighbours), which were often missed by other flagging constraints. \modif{The flagging information was used in the analysis of the Dome~C photometric quality (Sect.~\ref{sec: photometric quality}) but the flagged epochs were discarded in the final lightcurves and in the search for variable objects (Sect.~\ref{sec: study of variable stars and eclipsing binaries})}. \modif{The final lightcurves of the 5954 target stars are publicly available on the ASTEP lightcurve database website}\footnote{\url{https://astep-vo.oca.eu/}} and at the CDS. Examples of lightcurves of variable objects are shown in Fig.~\ref{fig: lightcurve examples}.

\subsection{Lightcurve precision}  
\label{sec: rms diagram}

\begin{figure}[htbp]
   \centering
   \includegraphics[width=8cm]{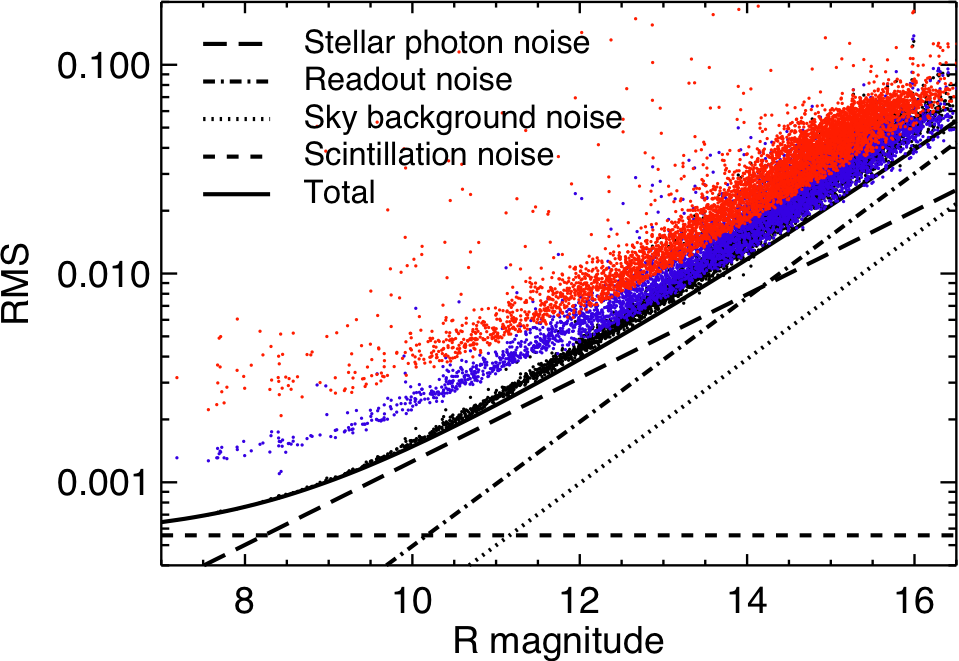}
   \caption{RMS of the four winter lightcurves as a function of stellar $R$ magnitude. The theoretical stellar photon noise, readout noise, sky background noise, scintillation noise, and their quadratic sum are shown as long-dashed, dot-dashed, dotted, dashed, and plain lines respectively, in the case of circular PSFs and apertures. The theoretical noise for each star taking into account the PSF and aperture elongation is shown as black dots. The true and point-to-point RMS are shown as red and blue dots, respectively.}
   \label{fig: rms diagram}
\end{figure}

We calculated the precision of the four-winter lightcurves over 6.7 minute bins. We computed the true lightcurve RMS ($\sigma$) and the point-to-point RMS ($\sigma_p$) which does not take into account correlated noise. We also computed theoretical estimates of the stellar photon noise, readout noise, sky background noise, scintillation noise, and their quadratic sum, first considering circular PSFs and circular apertures, then taking into account the PSF elongation and specific aperture of each star ($\sigma_t$). The scintillation noise is obtained from Eq.~1 of \citet{Young1967} \citep[see also][]{Dravins1998, Ryan1998}. By analyzing turbulence profiles, \citet{Kenyon2006} suggest a potential gain of 3.6 at Dome~C compared to Chilean sites. We did not apply this factor here: the scintillation noise remains well below other noise sources for ASTEP South, and using larger telescopes or studying a few bright stars would be necessary to confirm this gain. The lightcurve RMS as a function of stellar magnitude is shown in Fig.~\ref{fig: rms diagram} and the average values for various magnitudes are reported in Table~\ref{tab: lightcurve RMS}.

At the faint end, the point-to-point RMS is well explained by the theoretical RMS for elongated PSFs. At magnitudes 10 and 12, the point-to-point RMS is larger than the theoretical RMS by a factor 1.5 and 1.2 respectively, indicating that additional sources of uncorrelated noise is present and affect mostly bright stars. The true RMS is larger than the theoretical RMS by a factor 2.6, 1.8, 1.6, and 1.3 at magnitudes 10, 12, 14, and 15.5 respectively, indicating that correlated noise also remains in the lightcurves. For most stars, the true RMS of the four-winter lightcurves is within a factor of two of the theoretical estimates. This could be explained by the particular issues related to the challenging observing mode of ASTEP South. Overall, this shows that good photometry over very long periods is achievable at Dome~C.

\begin{table}
\begin{center}
\caption{RMS of the four-winter lightcurves over 6.7 minute bins for different magnitudes, in percent. $\sigma_t$ is the theoretical RMS, $\sigma_p$ is the point-to-point RMS, and $\sigma$ is the true RMS.}
\label{tab: lightcurve RMS}
\begin{tabular}{lcccc}
\hline\hline
$R$ mag               &   10     &    12     &    14     &    15.5  \\
\hline
\hspace{3mm} $\sigma_t$   &     0.16   &  0.45    &  1.54    &  4.08 \\
\hspace{3mm} $\sigma_p$  &     0.23   &  0.55   &   1.79   &   4.14 \\
\hspace{3mm} $\sigma$      &     0.41   & 0.81    &  2.41    &  5.47 \\
\hline\hline
\end{tabular}
\end{center}
\end{table}

\section{Instrumental parameters and sky background}

Besides stellar fluxes, we extracted several parameters in order to evaluate the functioning of the instrument and the quality of Dome~C for photometry. We present the results in this section, which complements the analysis of \citet{Crouzet2010} for the first winter.

\subsection{Pointing variations} 

\begin{figure}[htbp]
   \centering
   \includegraphics[width=4.68cm]{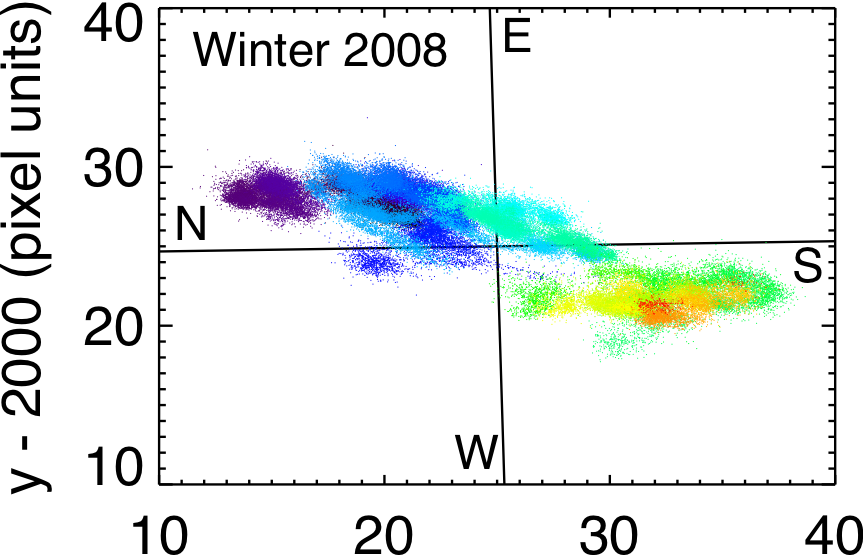}\hfill
   \includegraphics[width=4.1cm]{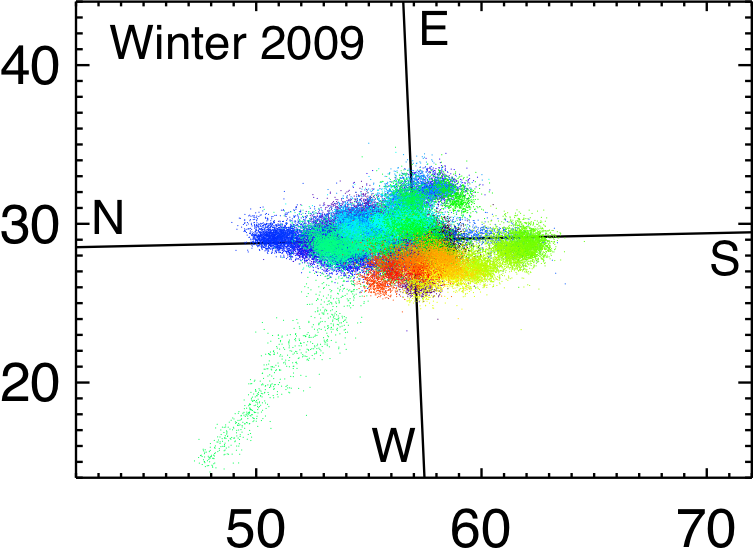}\\
   \vspace{1mm}
   \includegraphics[width=4.745cm]{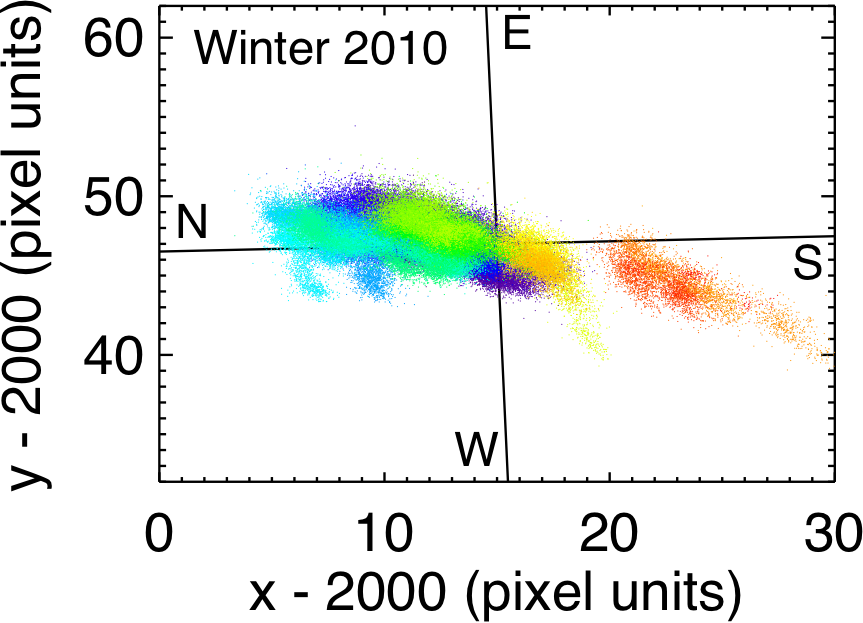}\hfill
   \includegraphics[width=4.15cm]{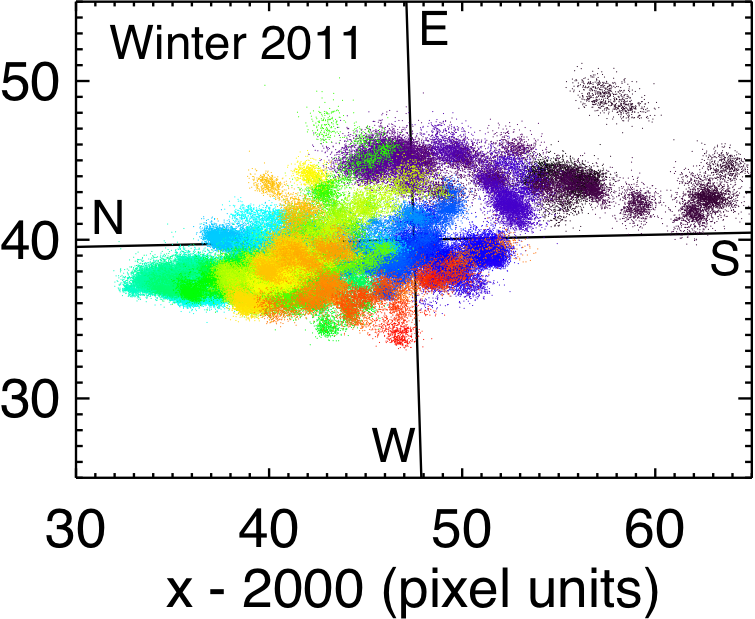}
      \caption{Position of the celestial south pole on the CCD during the winters 2008, 2009, 2010, and 2011. For each winter, only the longest period without repointing the instrument is shown. Increasing time is represented by black, purple, blue, green, yellow, orange, and red. The orientation of the CCD (north, south, east, west) is indicated.}
   \label{fig: pointing variations}
\end{figure}

We investigated the mechanical stability of ASTEP South by measuring the motion of the celestial south pole on the CCD (Fig.~\ref{fig: pointing variations}). Within one day, the south pole is stable within one pixel: the standard deviation from its average position is typically 0.6 and 0.5 pixel in $x$ and $y$, respectively. The long term motion over the winters is given in Table~\ref{tab: south pole motion} and averages to 0.18 and 0.09 pixel per day in $x$ and $y$, respectively. This long term motion is oriented close to the north-south axis which is the vertical axis of the instrument; this is particularly evident in 2008 and 2010. This motion may correspond to a bending of the instrument under its own weight over time, modulated by thermal fluctuations inside the box affecting its deformation coefficients. A motion of the ice below the instrument is also possible. Overall, this motion is very small and would be completely negligible for any instrument equipped with a guiding system operating at Dome~C.

\begin{table}
\begin{center}
\caption{\modif{Motion of the south pole on the CCD in the $x$ and $y$ directions for each winter ($\Delta x$ and $\Delta y$) calculated over the longest time interval without repointing the instrument ($\Delta t$). The pixel scale is 3.41 arcsec per pixel.}}
\label{tab: south pole motion}
\begin{tabular}{cccc}
\hline\hline
               &   $\Delta x$     &    $\Delta y$     &    $\Delta t$    \\
               &   [pixel unit]     &    [pixel unit]      &    [day]     \\
\hline
2008     &   25   &  10   &  128    \\
2009     &   13   &   8    &    86    \\
2010     &   25   &   11  &  115     \\
2011     &   33   &  16   &  211    \\
\hline\hline
\end{tabular}
\end{center}
\end{table}

\subsection{Point spread function variations} 
\label{sec: fwhm}

\begin{figure}[htbp]
   \centering
   \includegraphics[width=4.58cm]{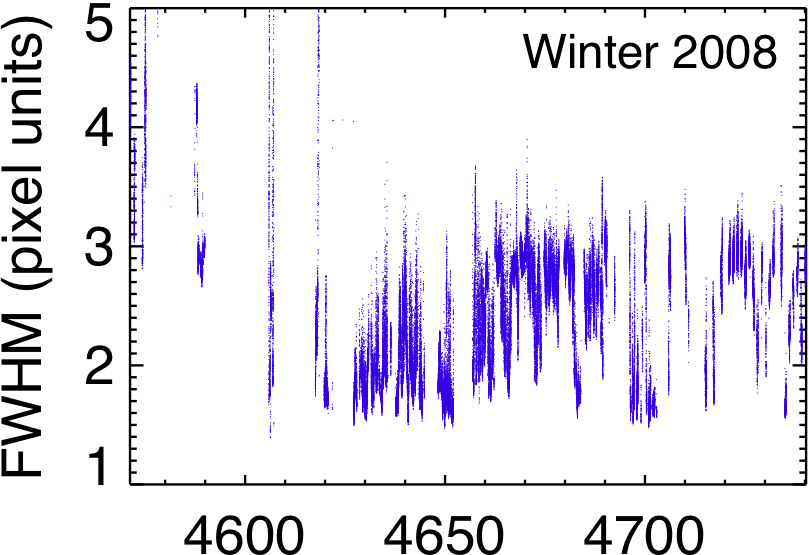}\hfill
   \includegraphics[width=4.1cm]{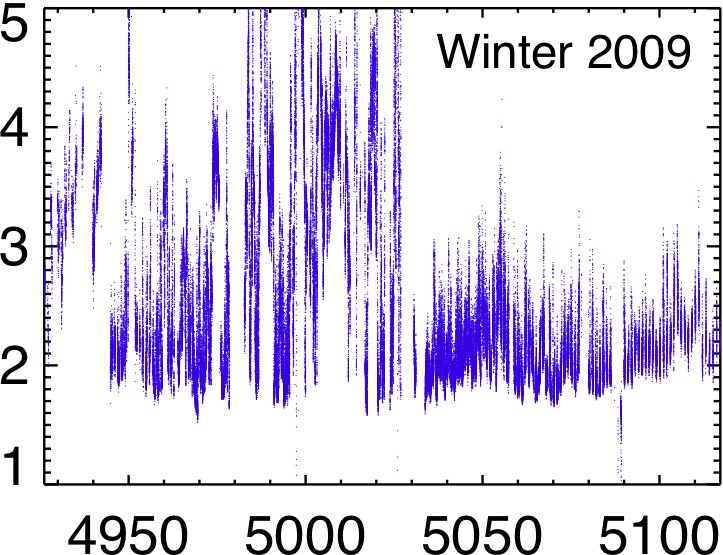}\\
   \vspace{1mm}
   \includegraphics[width=4.58cm]{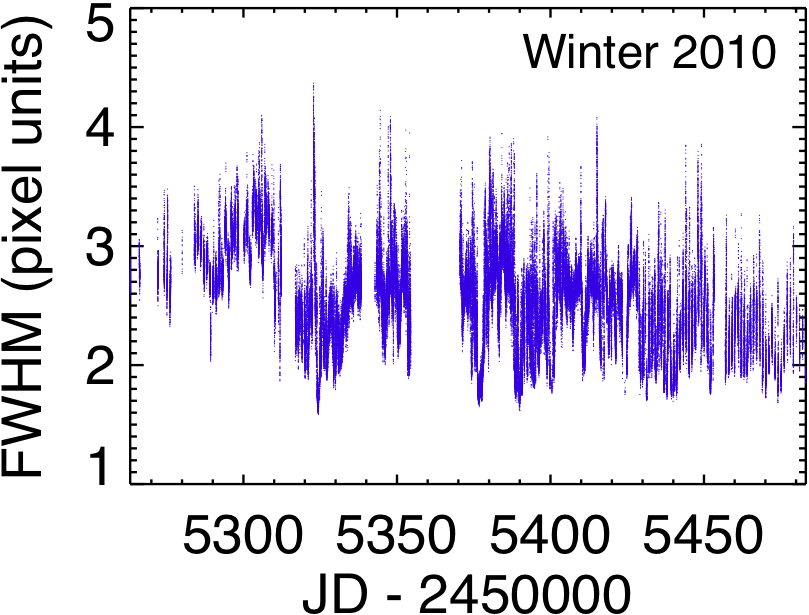}\hfill
   \includegraphics[width=4.1cm]{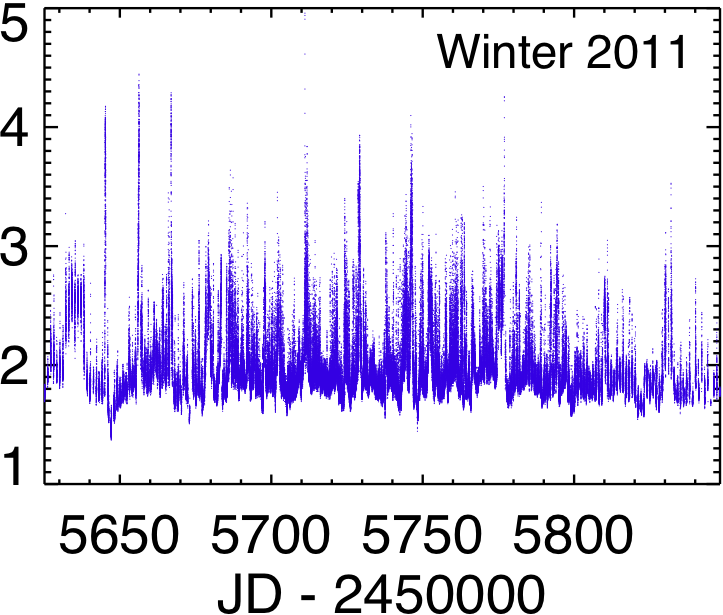}
      \caption{PSF FWHM during the winters 2008, 2009, 2010, and 2011. Short-term variations (timescale < 1 day) are due to seeing variations, and long-term variations (timescale > 1 day) are due to poor thermal stability causing mechanical deformations of the instrument.}
   \label{fig: fwhm}
\end{figure}

We measured the PSF variations using the 20 brightest stars in the $1000\times1000$ pixel central region of the image, where the elongation is negligible. We fitted the PSFs with a Gaussian function to derive the FWHM. This calculation was done by the day-to-day analysis software program and was used as an instrument control parameter. The nominal FWHM for ASTEP South is 2 pixels, and its variations over the four winters are shown in Fig.~\ref{fig: fwhm}. 
Thermal stability is critical to a stable FWHM. For example, during the first part of the 2009 winter, \modif{the FWHM suffered extreme variations because of a fan failure. The recorded temperature inside the enclosure (around the probe) varied between $-22$ and $-18${\dgr}C but the actual variations and gradients were probably much stronger, inducing thermal deformations of the instrument. After restarting the fans, the recorded temperature remained stable within 0.2{\dgr}C of the setting point and the FWHM remained stable between 2 and 2.5 pixels}. Long-term variations (timescale > 1 day) in 2008 and 2010 are also correlated with thermal variations. Because any intervention on the instrument during the Antarctic winter is extremely delicate, the thermal control was not modified during these winters. Under stable thermal conditions, the FWHM recorded by ASTEP South is directly correlated with the atmospheric seeing, which yields short-term variations (timescale < 1 day). This has been shown for the 2008 winter by \citet{Crouzet2010} and is also evident during the second half of the 2009 winter and during the 2011 winter, where increases of the FWHM correspond to poor seeing conditions.

As for any photometric instrument, periods affected by large FWHM variations result in poor photometry. In particular, FWHM variations are correlated with the one-sidereal day patterns, with different patterns for FWHM below and above a limit of either 2.5 or 3 pixels. Above this value, the pattern can be highly variable from day to day and was not always recovered, in particular in 2009 and 2010. These data were generally eliminated from the lightcurves.

\subsection{Sky background} 
\label{sec: Sky background}

\begin{figure*}[htbp]
   \centering
   \includegraphics[width=4.2cm]{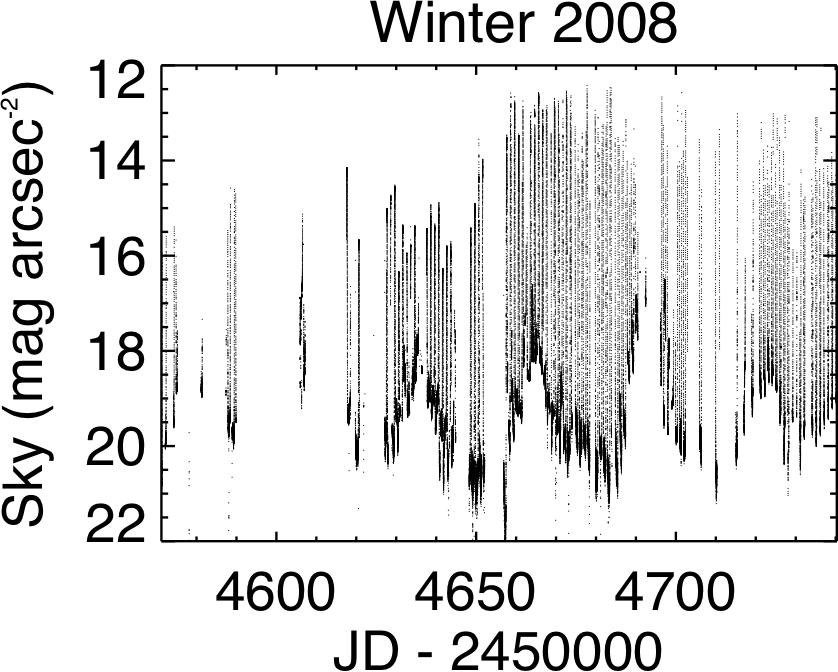}
   \vspace{1mm}
   \includegraphics[width=3.4cm]{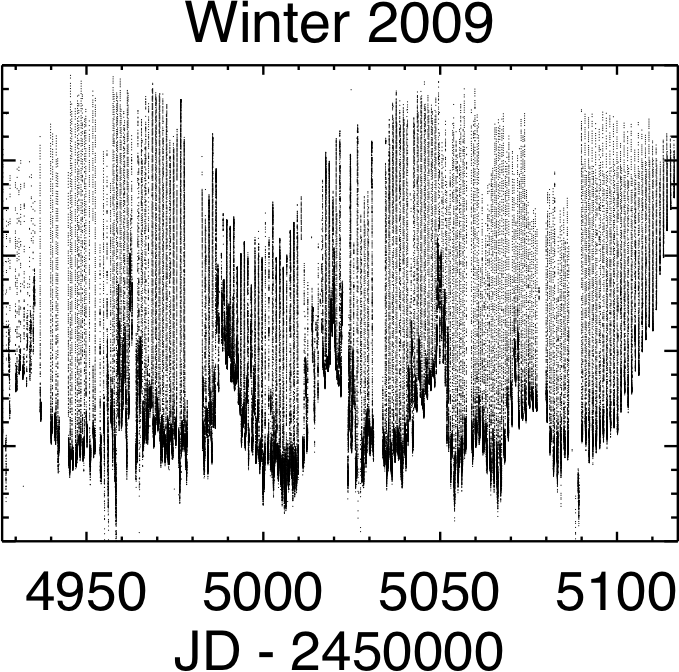}
   \includegraphics[width=3.4cm]{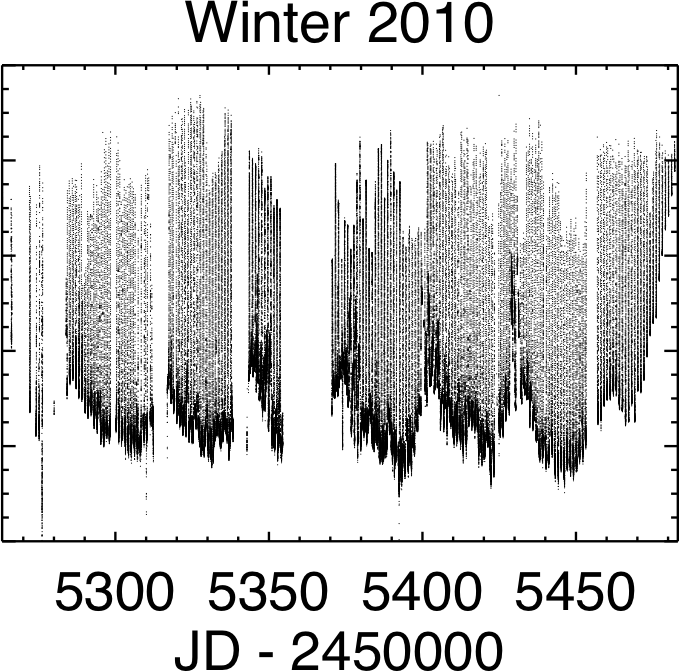}
   \includegraphics[width=3.4cm]{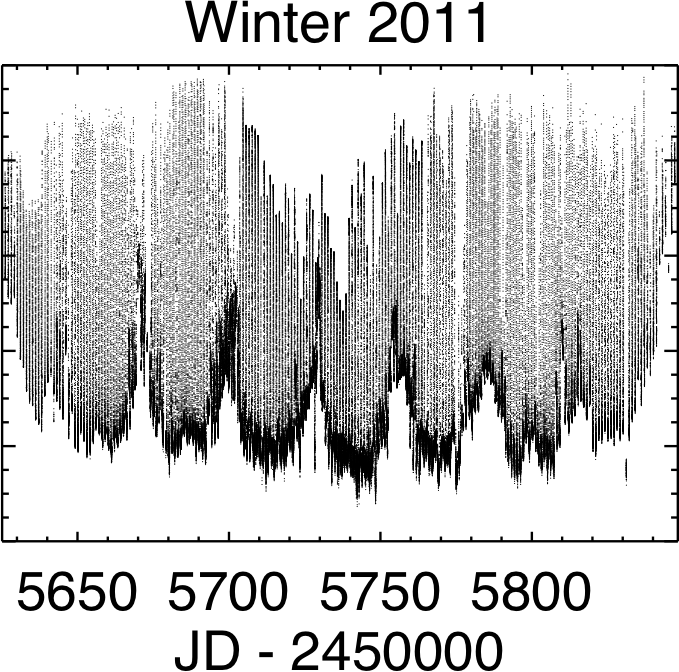}
   \includegraphics[width=3.4cm]{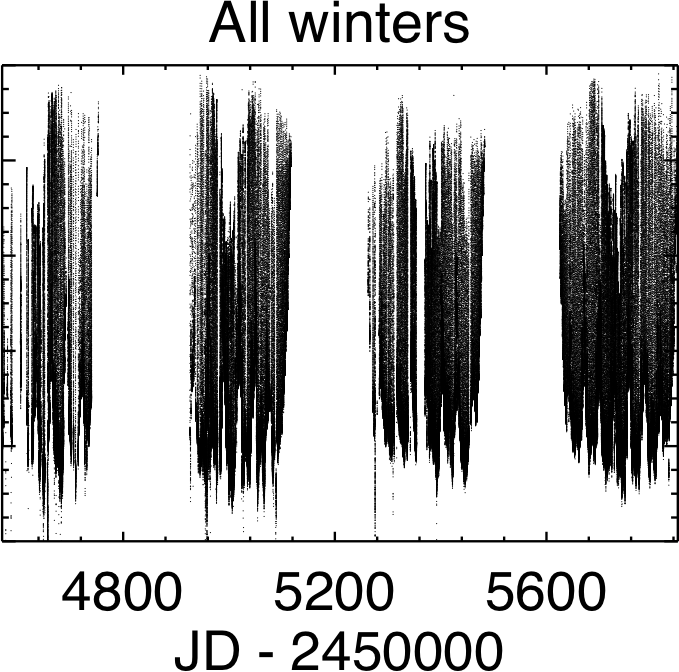}
   \includegraphics[width=4.25cm]{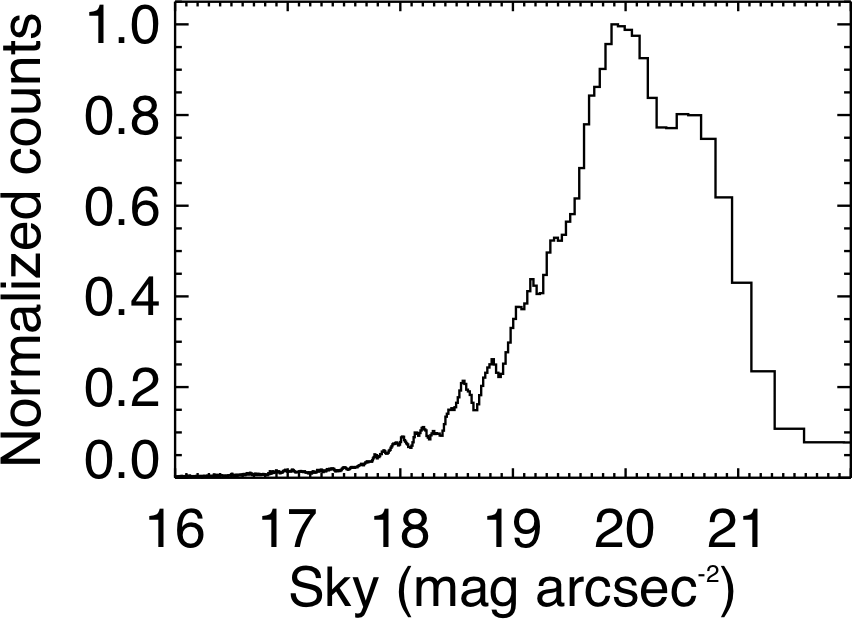}
   \vspace{1mm}
   \includegraphics[width=3.4cm]{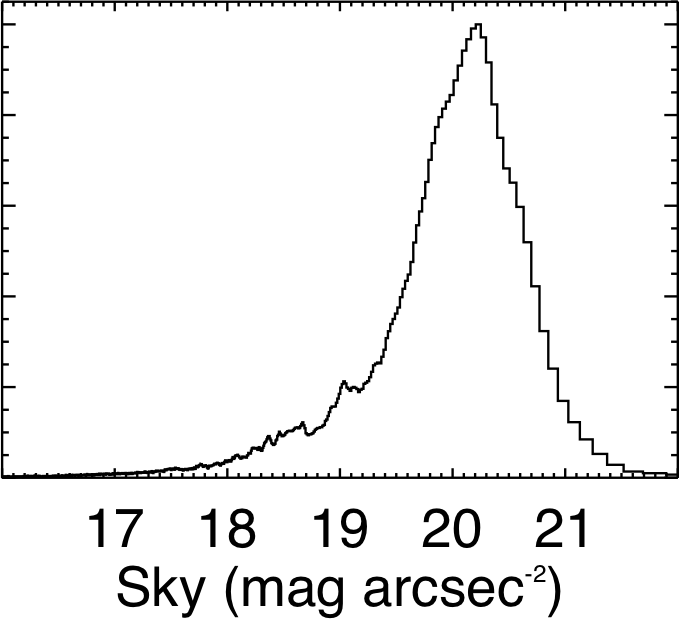}
   \includegraphics[width=3.4cm]{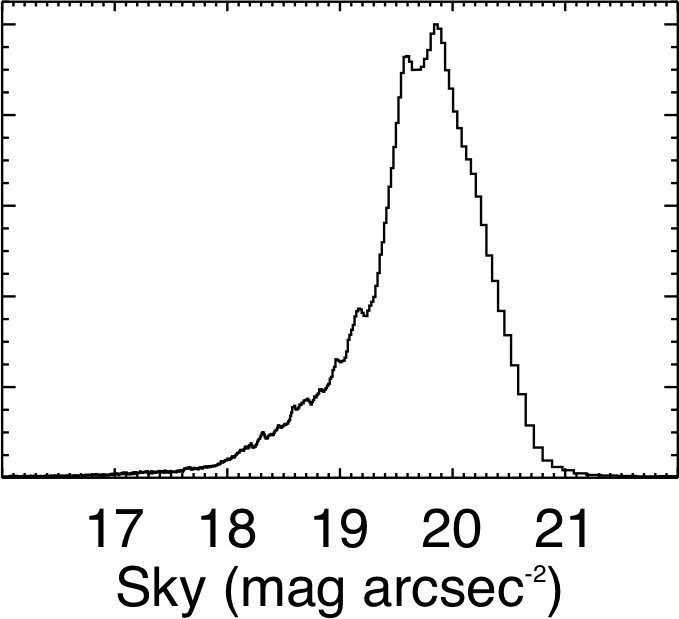}
   \includegraphics[width=3.4cm]{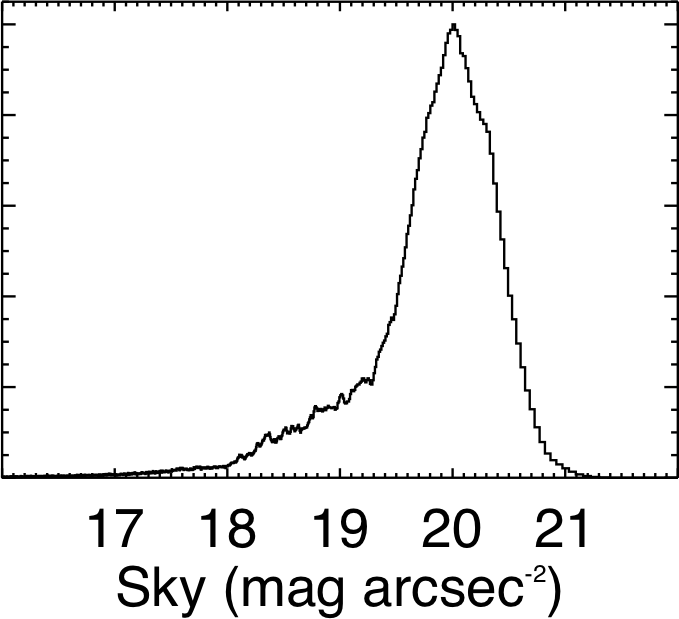}
   \includegraphics[width=3.55cm]{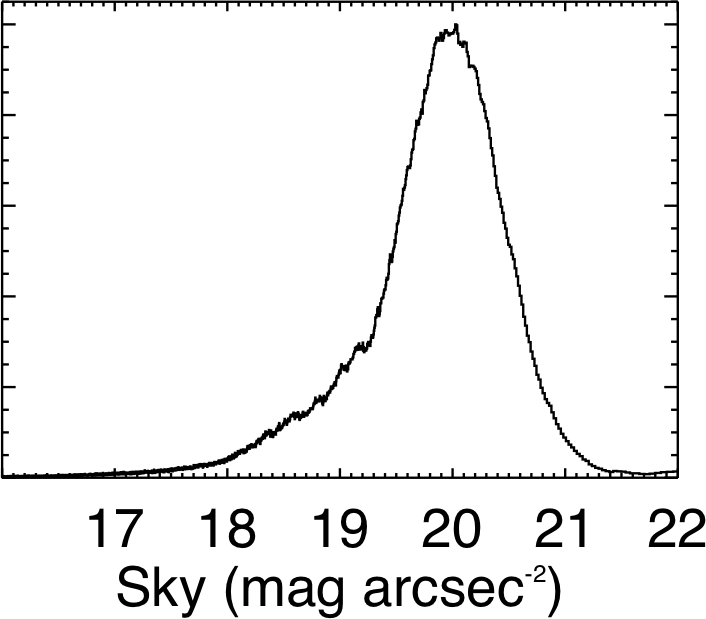}
   \includegraphics[width=4.2cm]{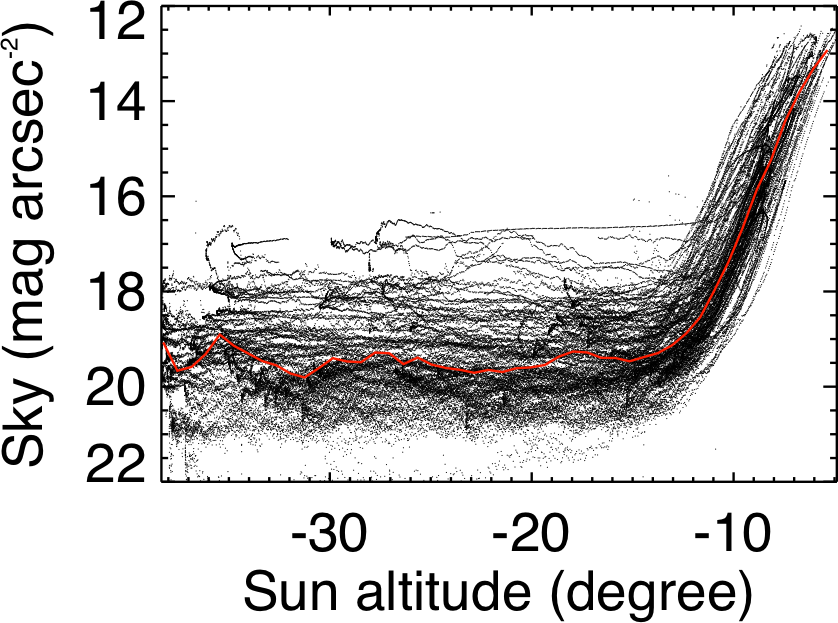}
   \vspace{1mm}
   \includegraphics[width=3.4cm]{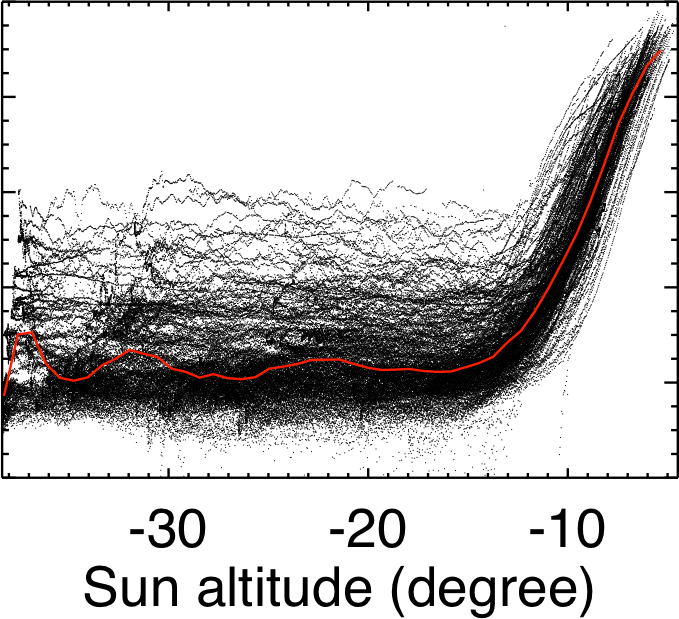}
   \includegraphics[width=3.4cm]{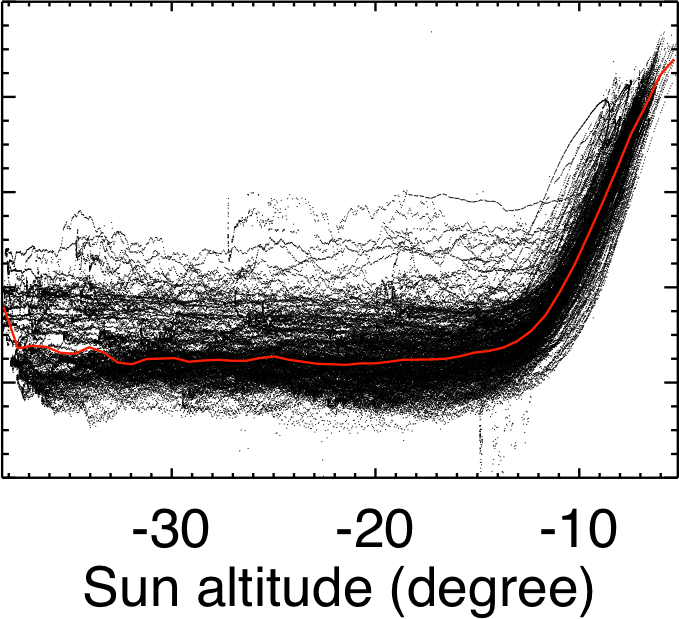}
   \includegraphics[width=3.4cm]{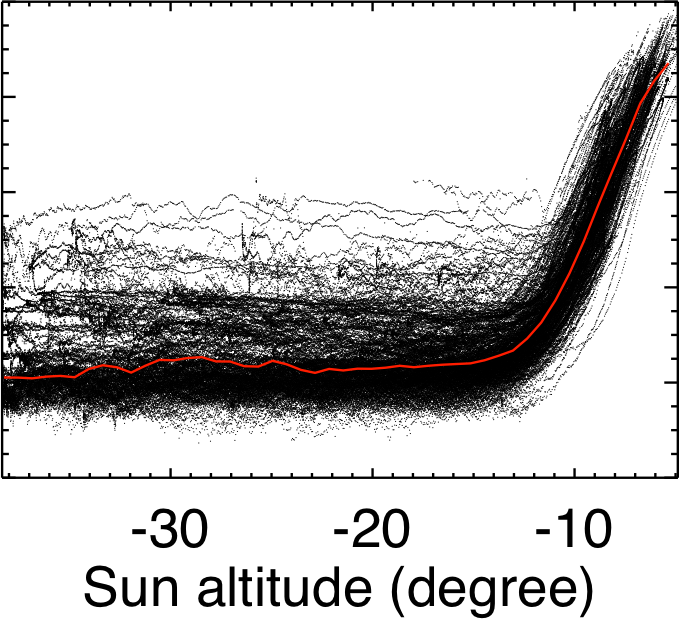}
   \includegraphics[width=3.4cm]{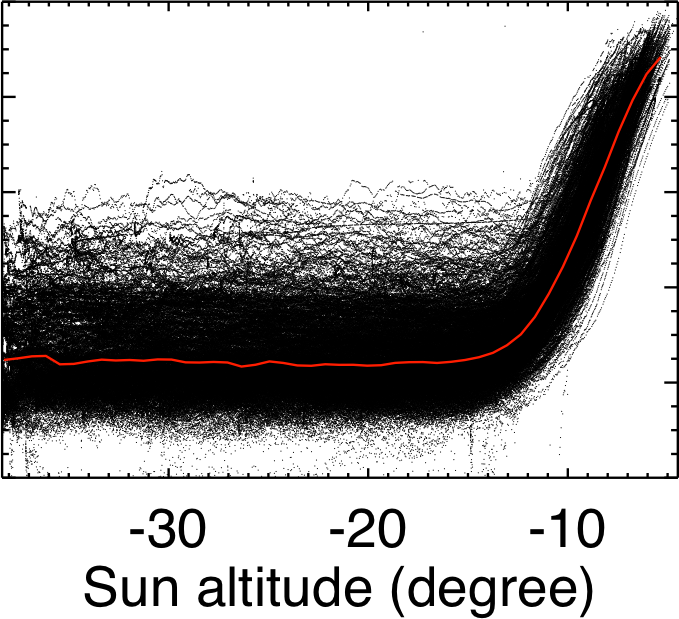}
   \includegraphics[width=4.335cm]{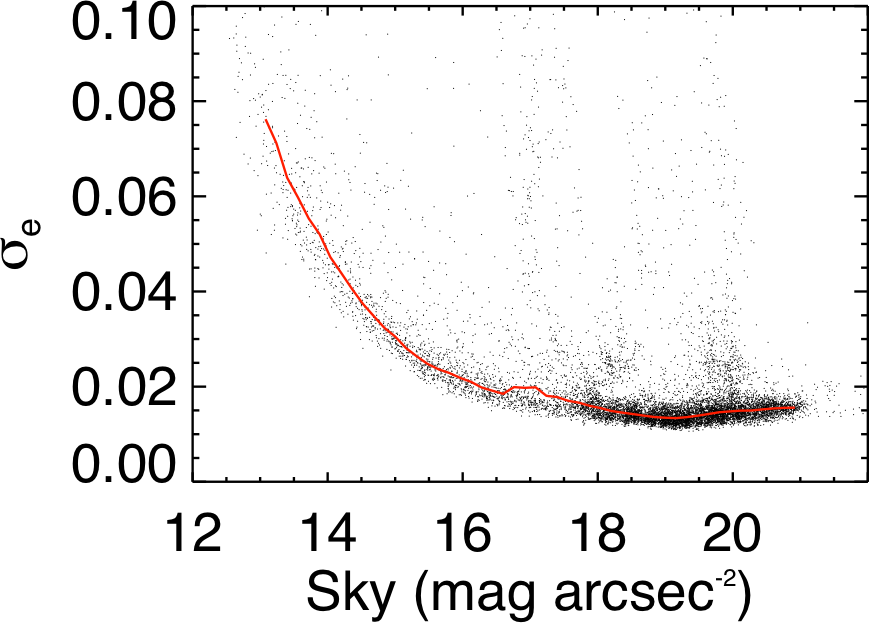}
   \vspace{1mm}
   \includegraphics[width=3.4cm]{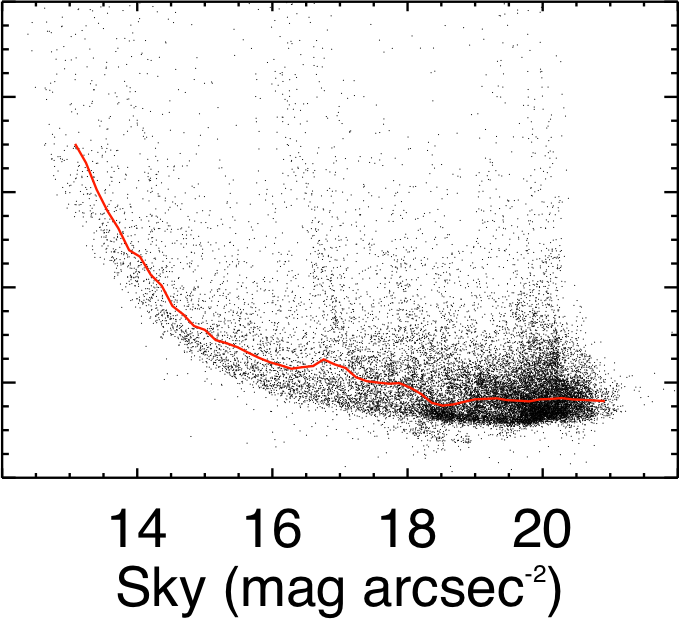}
   \includegraphics[width=3.4cm]{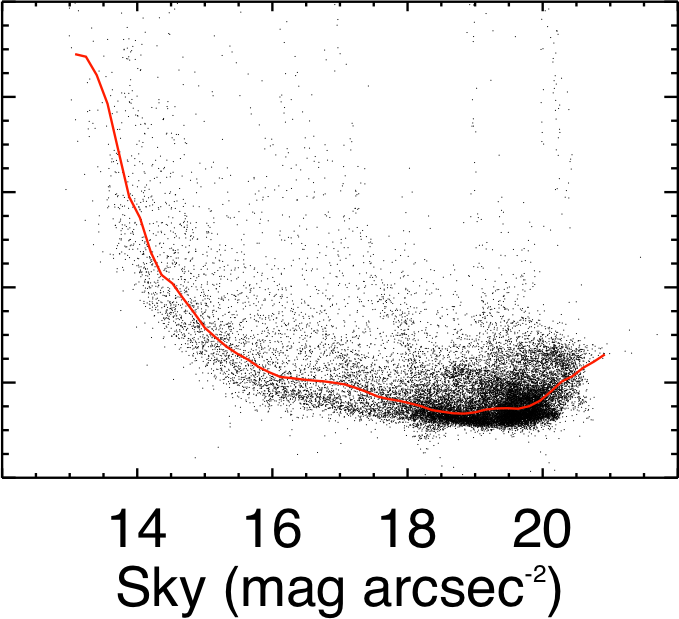}
   \includegraphics[width=3.4cm]{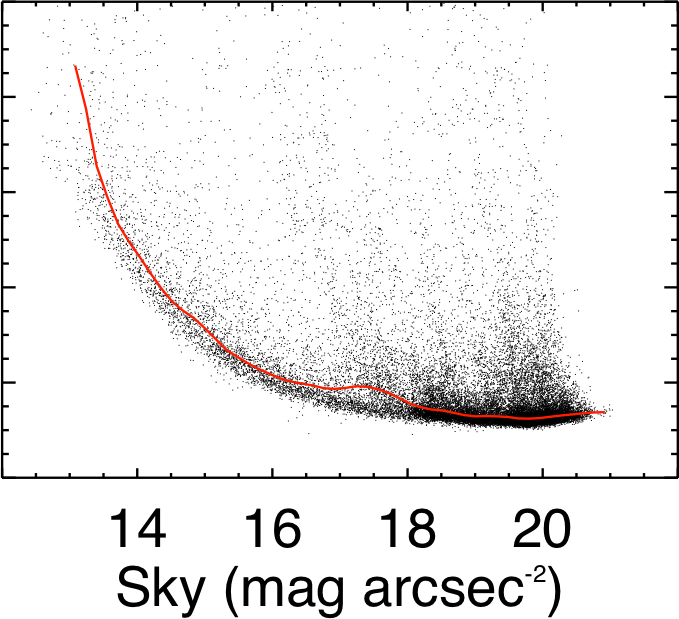}
   \includegraphics[width=3.535cm]{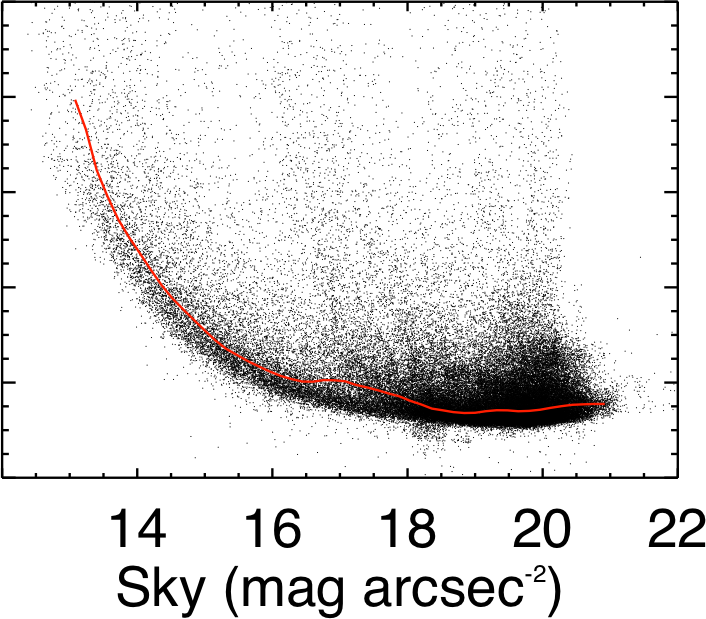}
   \includegraphics[width=4.35cm]{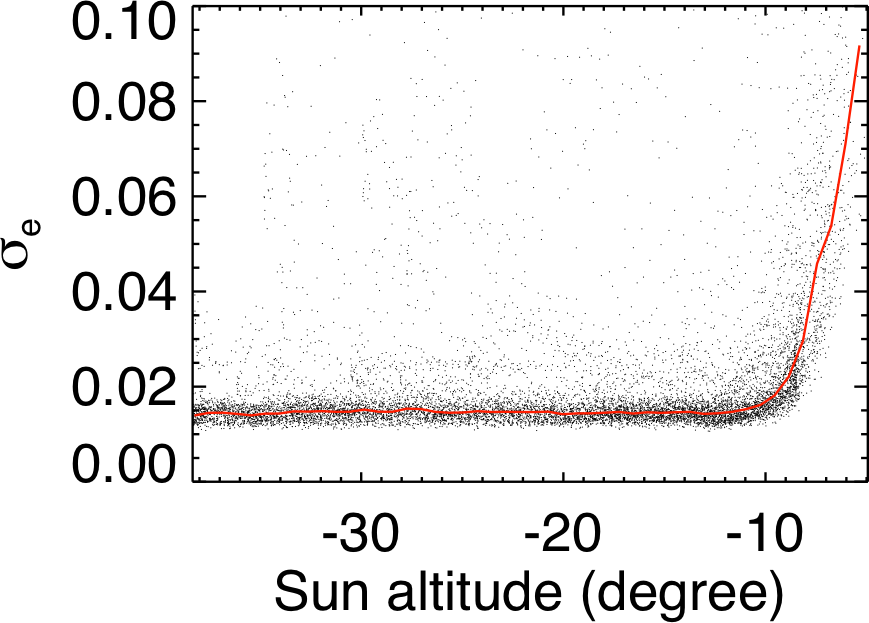}
   \includegraphics[width=3.4cm]{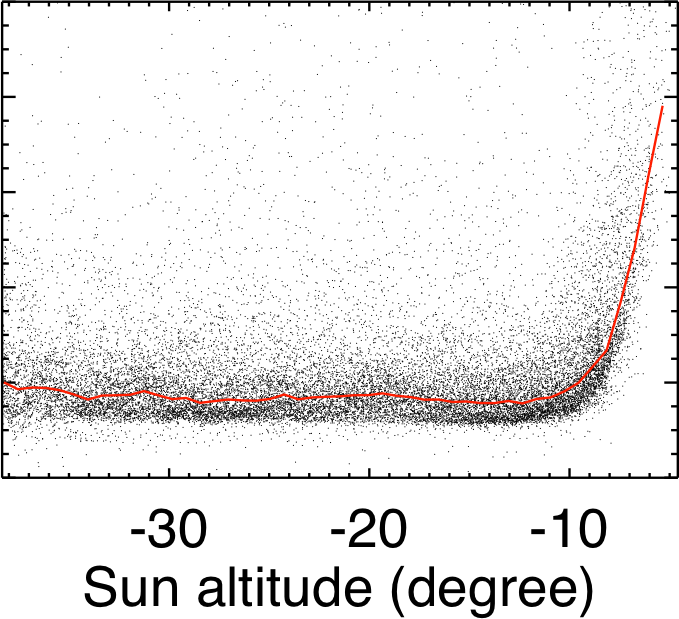}
   \includegraphics[width=3.4cm]{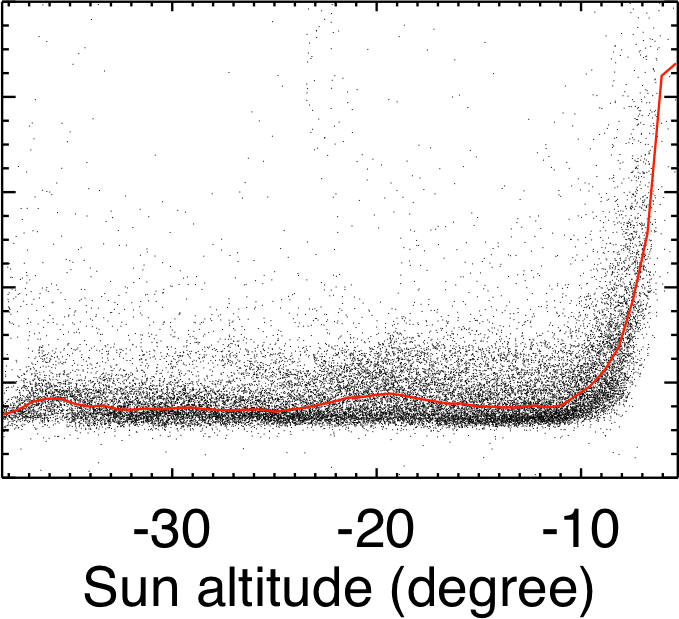}
   \includegraphics[width=3.4cm]{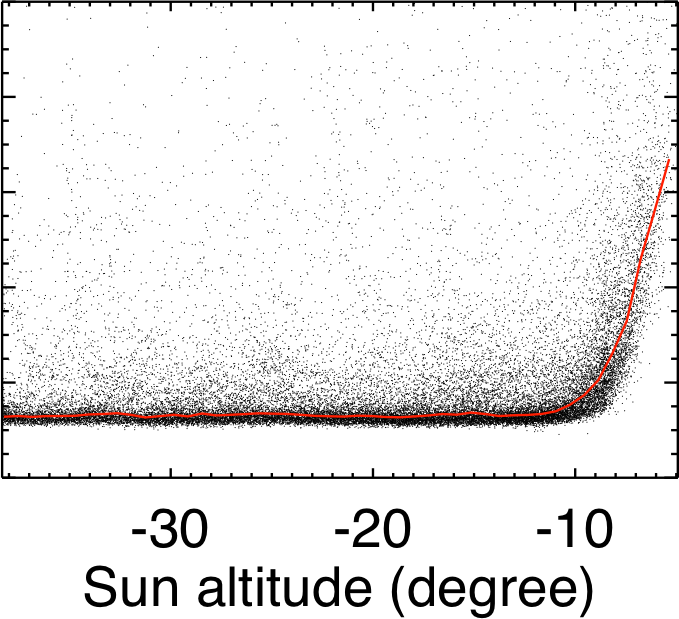}
   \includegraphics[width=3.4cm]{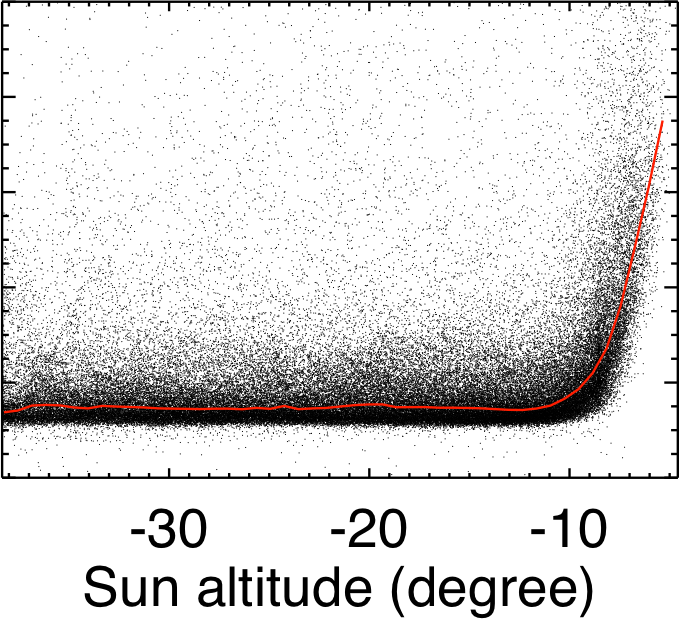}
   \caption{Sky background and related parameters for the winters 2008 to 2011 and for all the winters combined, from left to right. From top to bottom: sky background as a function of time, sky background distribution, sky background as a function of Sun altitude, epoch RMS $\sigma_e$ (see text) as a function of sky background, and $\sigma_e$ as a function of Sun altitude. The sky background histograms were calculated in flux units and are displayed in magnitude units. For the last three rows, we computed an outlier resistant mean using 50 bins across the abscissa range (red lines).}
   \label{fig: sky background}
\end{figure*}

The sky background was computed in each image using the IDL SKY procedure, corrected for the zenith angle, converted from photo-electrons to instrumental magnitude, and calibrated to a fmag magnitude (579--642~nm bandpass). We performed this latter calibration by fitting a linear function between the stars' instrumental magnitudes and their UCAC3 fmag magnitudes, independently for each winter. The sky background and its correlations with other parameters are shown in Fig.~\ref{fig: sky background}. 
The first row shows the sky background as a function of time. The Sun increases the sky background every day around noon and the Moon cycle is clearly visible, as already reported in \cite{Crouzet2010}.
The second row displays the histograms of sky values. They are composed of three regions: a broad distribution centered around the nominal dark sky value (20 mag arcsec$^{-2}$), a region between 19.5 and 18 mag arcsec$^{-2}$ that corresponds to the illumination around full Moon, and a long tail for values brighter than 18 mag arcsec$^{-2}$ that corresponds to the illumination by the Sun around noon.
To measure the average dark sky background at Dome~C, we combined the distributions of the four winters and fitted the main peak with a Gaussian function. We did this calculation in flux units and converted the mean and 1$\sigma$ values back to magnitudes. We find a dark sky background at Dome~C of $20.0^{+0.4}_{-0.3}$ mag arcsec$^{-2}$ in the 579--642~nm bandpass. The uncertainty is partly due to real variations of the dark sky background: in 2011, the sky is darker around mid-winter (\simi20.5 mag arcsec$^{-2}$) and brighter closer to the equinoxes (\simi19.5 mag arcsec$^{-2}$). This trend is less clear for the other winters due to less consistent and less complete coverage. For comparison, the dark sky background in mag arcsec$^{-2}$ in the R band is 20.4 at Mauna Kea\footnote{\url{http://www.gemini.edu/sciops/telescopes-and-sites/observing-condition-constraints/optical-sky-background}}, 20.9 at Paranal \citep{Patat2008}, 21.0 at La Palma\footnote{\url{http://www.ing.iac.es/Astronomy/observing/conditions/skybr/skybr.html\#vsol}}, and 21.2 at Cerro Tololo and La Silla \citep{Krisciunas2007}. Thus, the sky background at Dome~C is comparable to that of the best observing sites. 

The sky background as a function of the Sun altitude $h_{Sun}$ is shown in the third row of Fig.~\ref{fig: sky background}. A limit $h_{Sun} < -15^{\circ}$ can be used to define dark sky conditions, whereas the sky background increases quickly for $h_{Sun} > -13^{\circ}$. 
Finally, we calculated the lightcurve dispersion at each epoch $\sigma_e$ using the stars in the magnitude interval 12--13.5, which represents the photometric quality at that epoch (see Sect.~\ref{sec: photometric quality}). We show $\sigma_e$ as a function of the sky background and Sun altitude in the fourth and fifth rows, respectively. We find that $\sigma_e$ starts increasing noticeably when the sky background becomes brighter than $18 \; \rm mag\,arcsec^{-2}$. We find that $\sigma_e$ remains at its average level while $h_{Sun} < -12^{\circ}$ and increases steeply for $h_{Sun} > -10^{\circ}$. These numbers give the conditions under which good quality photometry can be achieved at Dome~C as seen by our 10~cm diameter instrument.

\section{Photometric quality of Dome~C}
\label{sec: photometric quality}

The long, continuous, and consistent database provided by ASTEP South is well suited to measure the weather quality during the winter at Dome~C. The clear sky fraction for the winter 2008 was measured by \cite{Crouzet2010} using the number of detected stars in the field of view. Here, we use the quality of the lightcurves themselves to derive the photometric quality. In the following calculations, we used only the subset of stars with R magnitudes between 12 and 13.5 (1090 control stars).
The final lightcurves are in the form of magnitude residuals. We calculated the standard deviation of the residuals at each epoch $\sigma_{e}$ using the control stars. This parameter provides a direct measurement of the data quality at each epoch: large dispersions in the residuals correspond to low quality epochs. As shown in Fig.~\ref{fig: sigep}, the $\sigma_{e}$ distribution is composed of a bulk and a tail containing the high and low quality data, respectively.
To differentiate the bulk and the tail, we built a model for $\sigma_{e}$ and compared it to the measured distribution. In the following, \sigepm and \sigept refer to the measured and theoretical \sigep respectively. We calculated \sigept in several steps as described next.

\subsection{Theoretical noise}
\label{sec: Theoretical noise}

First, we calculated the theoretical noise that was expected for each star at each epoch taking into account the Poisson noise, read-out noise, and sky background noise. We did not consider scintillation noise because it is negligible in the magnitude range $12-13.5$ (see Fig.~\ref{fig: rms diagram}). Each star had its own photometric aperture which remained fixed in size for the four years of data. The fraction of the PSF included in the aperture depended on its FWHM, which varied in time. We calculated this fraction for each star at each epoch by integrating numerically the PSF over the photometric aperture. The PSF was taken as a two-dimensional Gaussian function elongated in one direction, its FWHM was that measured by ASTEP South, and the elongation was calculated as a function of the distance of the star from the celestial south pole. The photometric aperture was the same as used in the reduction pipeline.

We calculated the nominal flux for each star as the median of the flux measured under excellent conditions, which we defined as non-flagged epochs falling within the 20\% largest number of stars and the 20\% smallest FWHM. Then, we calculated the theoretical flux for each star at each epoch by scaling the nominal flux to the fraction of PSF within the aperture, and used it to derive theoretical estimates of the stellar Poisson noise, read-out noise, and sky background noise, relatively to the stellar signal.
In these calculations and in the following, we included a 10\% uncertainty on the measured FWHM and sky background values to account for measurement errors.

As shown in Sect.~\ref{sec: rms diagram}, these theoretical noises do not explain fully the measured lightcurves' RMS. Thus, we added unidentified noise sources and split their contributions into two components \sigmaw and \sigmar, defined as:
\begin{equation}
\label{eq: sigma wr}
\begin{aligned}
\sigmaw^2 = \sigmap^2-\sigmat^2 \;\;\;\;{\rm and} \;\;\;\;
\sigmar^2 = \sigma^2-\sigmap^2
\end{aligned}
\end{equation}
where \sigmat is the theoretical RMS, \sigmap is the point-to-point RMS, and $\sigma$ is the true RMS. \sigmaw corresponds to a white (uncorrelated) noise component and \sigmar as a red (correlated) noise component. We obtained \sigmaw and \sigmar for each star directly from their lightcurve (see Fig.~\ref{fig: rms diagram}); these lightcurves were cleaned from bad data points including a cut-off in \sigep so they correspond to good weather conditions. We expressed the noise $\sigma_{s,\,e}$ for star $s$ at epoch $e$ as:
\begin{equation}
\label{eq: sigma_s_ep} 
\sigma_{s,\,e}^2 = \sigma_{phot,\,s,\,e}^2+\sigma_{read,\,s,\,e}^2+\sigma_{sky,\,s,\,e}^2+\sigma_{w,\,s}^2+\sigma_{r,\,s}^2
\end{equation}
where $\sigma_{phot,\,s,\,e}$, $\sigma_{read,\,s,\,e}$, and $\sigma_{sky,\,s,\,e}$ are the stellar photon noise, read-out noise, and sky background noise, and $\sigma_{w,\,s}$ and $\sigma_{r,\,s}$ are the additional white and red noise contributions.

\begin{figure}[htbp]
   \centering
   \includegraphics[width=8cm]{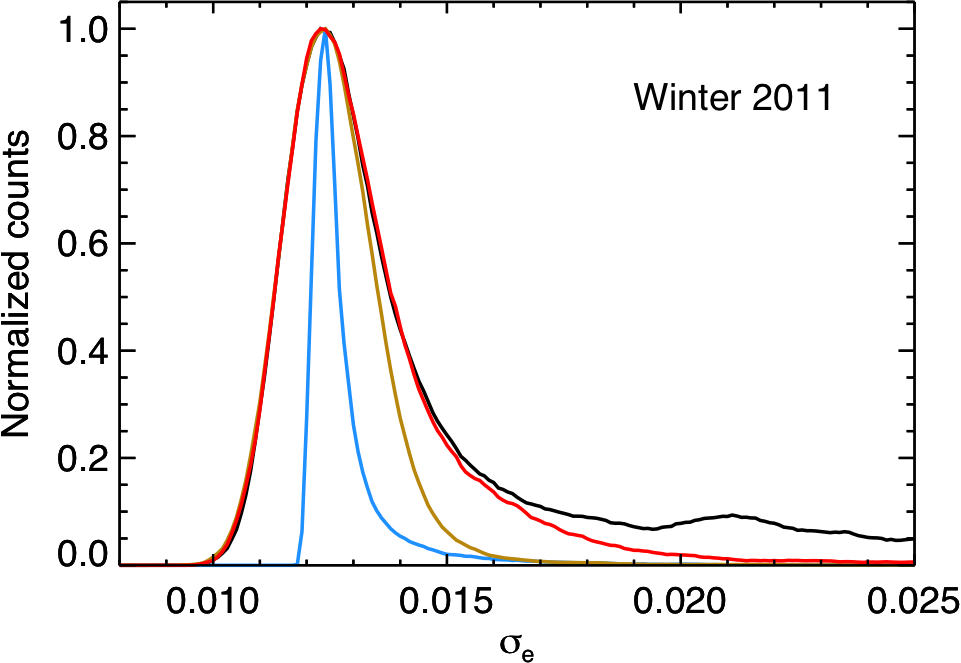}
   \caption{Construction of a theoretical model for the distribution of $\sigma_{e}$. The blue line shows a model taking into account the theoretical noises plus the white and red noise values, as defined in Eq.~\ref{eq: sigma_s_ep}. The brown line shows a model after introducing a correlative component between stars, as defined in Eq.~\ref{eq: sigma_s_ep 2}. The red line shows a model after introducing correlations with the FWHM and the sky background, as defined in Eq.~\ref{eq: sigma_s_ep 3}. The black line is the measured distribution $\sigma_{e,\,m}$. Only the 2011 winter is shown as an example.}
   \label{fig: sigep steps}
\end{figure}

Finally, we computed the theoretical noise for epoch $e$, \sigept, as the median of $\sigma_{s,\,e}$ over the control stars. As shown in Fig.~\ref{fig: sigep steps}, the resulting \sigept distribution is well centered on the \sigepm distribution but is very narrow. This is because the correlation of $\sigma_{s,\,e}$ between stars (the correlative nature of $\sigma_{r,\,s}$) was not accounted for: $\sigma_{r,\,s}$ is only an amplitude. Increasing the random errors on the FHWM and sky background measurements broadened only slightly the \sigept distribution. In fact, more complex correlated noise was present and was not captured yet.

\subsection{Addition of a blind correlative component}

A complete model of the variations affecting the ASTEP South lightcurves was beyond the scope of this work. Instead, we used an empirical approach and included a correlative component in the \sigmar parameter. We modeled the correlations in the following way. We divided the lightcurves into segments of about 30 minute duration. We associated each segment to a correlated noise value randomly drawn from a normal distribution of mean zero and standard deviation $\beta$, noted $\mathcal{N}(\beta)$. These values were the same for all stars, meaning that the lightcurve residuals varied in the same way for all stars at a given epoch. For star $s$, we defined the red noise distribution $\mathcal{R}$ along the segments by:
\begin{equation}
\label{eq: sigma R} 
\mathcal{R} = \sigma_{r,\,s} \; \Big(1 + \mathcal{N}(\beta)\Big)
\end{equation}
The red noise value $\sigma_{\mathcal{R},s}$ for star $s$ was taken as the realization of $\mathcal{R}$ on the segments and remained constant on each segment.
The theoretical noise $\sigma_{s,\,e}$ for star $s$ at epoch $e$ was then defined as:
\begin{equation}
\label{eq: sigma_s_ep 2} 
\sigma_{s,\,e}^2 = \sigma_{phot,\,s,\,e}^2+\sigma_{read,\,s,\,e}^2+\sigma_{sky,\,s,\,e}^2+\sigma_{w,\,s}^2+\sigma_{\mathcal{R},\,s,\,e}^2
\end{equation}
where $\sigma_{\mathcal{R},\,s,\,e}$ was then correlated between stars. The theoretical noise for epoch $e$, \sigept, was taken as the median of $\sigma_{s,\,e}$ over the control stars.

The amount of correlation encoded in the $\beta$ parameter was unknown; the larger it was, the broader was the \sigept distribution. We adjusted $\beta$ by matching the left side of the \sigept histogram to that of \sigepm. Indeed, the broadening on the left side of the \sigepm histogram was due to noise that was present even in good weather conditions, whereas the broadening on the right side was also due to bad weather, high sky background, etc. The resulting \sigept distribution reproduced well the bulk of the \sigepm distribution, as shown in Fig.~\ref{fig: sigep steps}. 

In this model, the sizes of the segments did not matter as long as the stars were affected in the same way. The $\beta$ parameter represented the amount of correlated noise at a given time in the star domain, whereas the correlated noise is generally measured on a lightcurve, that is for a given star in the time domain \citep[\eg][]{Gillon2006}. The value of $\beta$ depended on the number of correlated stars: if less stars were correlated, $\beta$ had to be larger to reproduce the \sigepm distribution; different groups of stars may also have had different correlations. Algorithms such as SYSREM \citep{Tamuz2005}, trend filtering algorithm \citep[TFA,][]{Kovacs2005} or others using a principal component analysis aim at identifying and removing such correlations (although the use of SYSREM on the ASTEP South data did not improve the lightcurves compared to our own procedures). Our approach was equivalent to convoluting the \sigept distribution from Eq.~\ref{eq: sigma_s_ep} with a Gaussian function, but related this broadening to an amount of correlated noise that could be included in the noise budget. A more general approach based on the covariance matrix could also have been used \citep{Pont2006}.

\subsection{Correlations with the FWHM and sky background}

The tail of the \sigepm histogram contained the bad weather data, but also included other noise sources which effects were not symmetrical with respect to the mode of the histogram and which were not included in our model. We investigated potential correlations of \sigepm with the various recorded parameters for data taken under good weather conditions, and found clear correlations with the PSF FWHM and the sky background.

\begin{figure}[htbp]
   \centering
   \includegraphics[width=4.45cm]{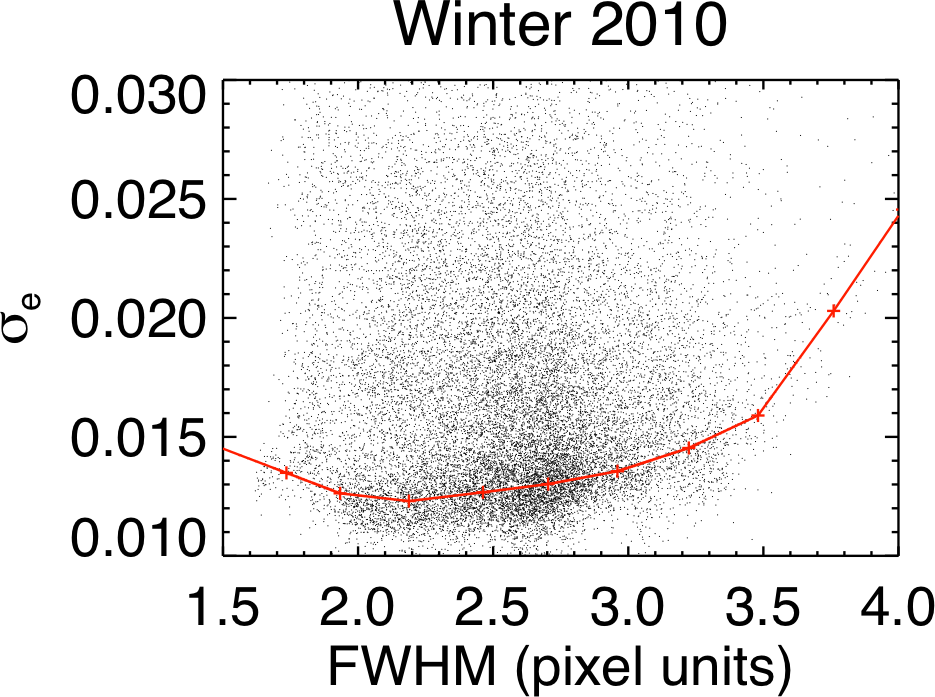}\hfill\vspace{1mm}
   \includegraphics[width=4.45cm]{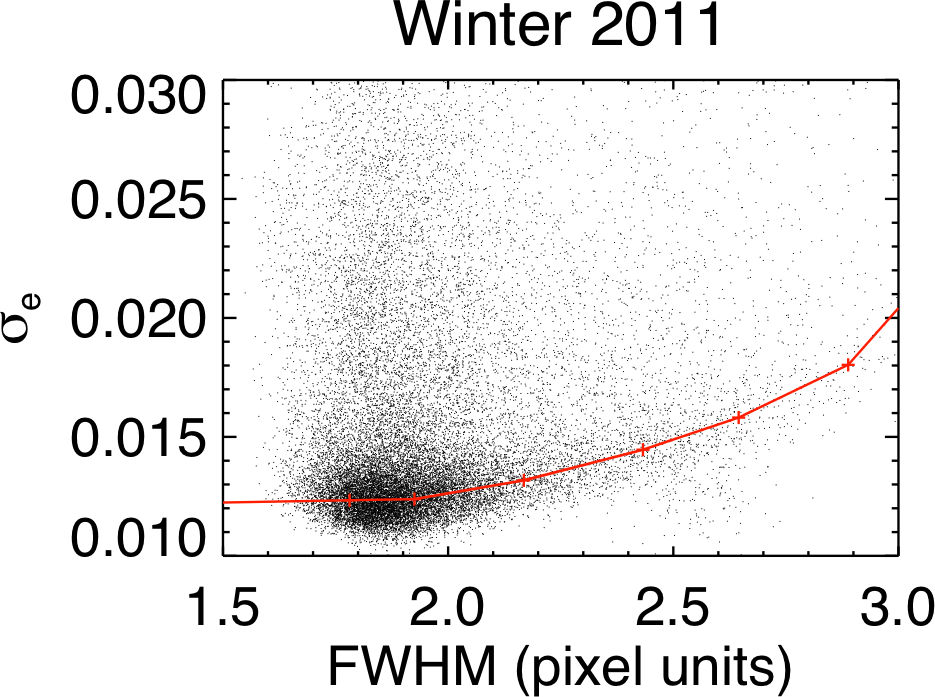}
   \includegraphics[width=4.25cm]{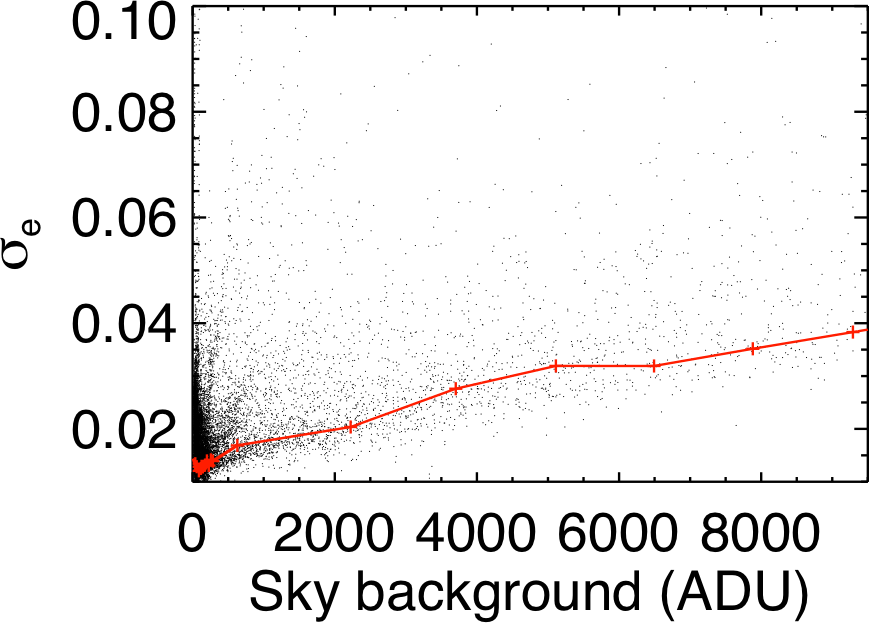}\hspace{2mm}\vspace{1mm}
   \includegraphics[width=4.25cm]{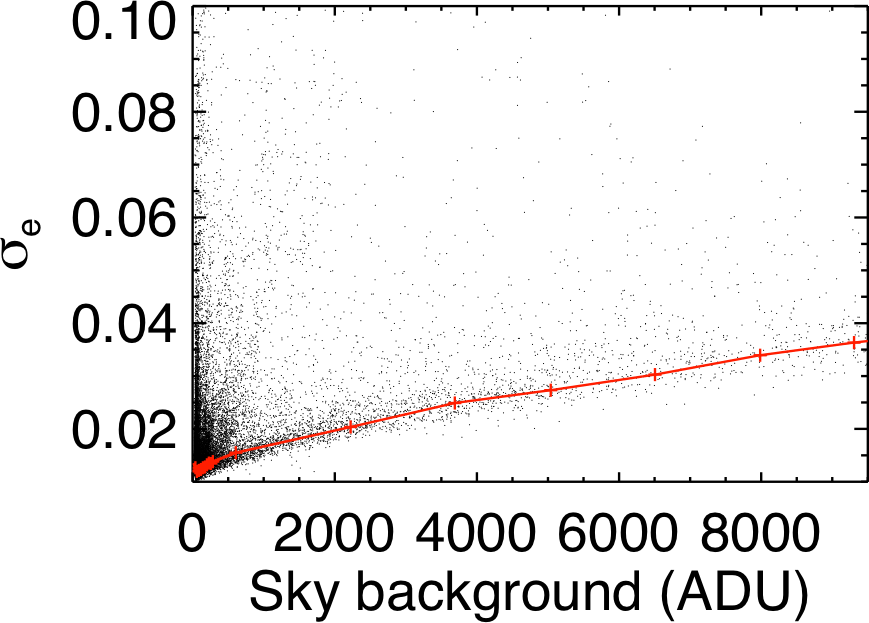}
   \includegraphics[width=4.5cm]{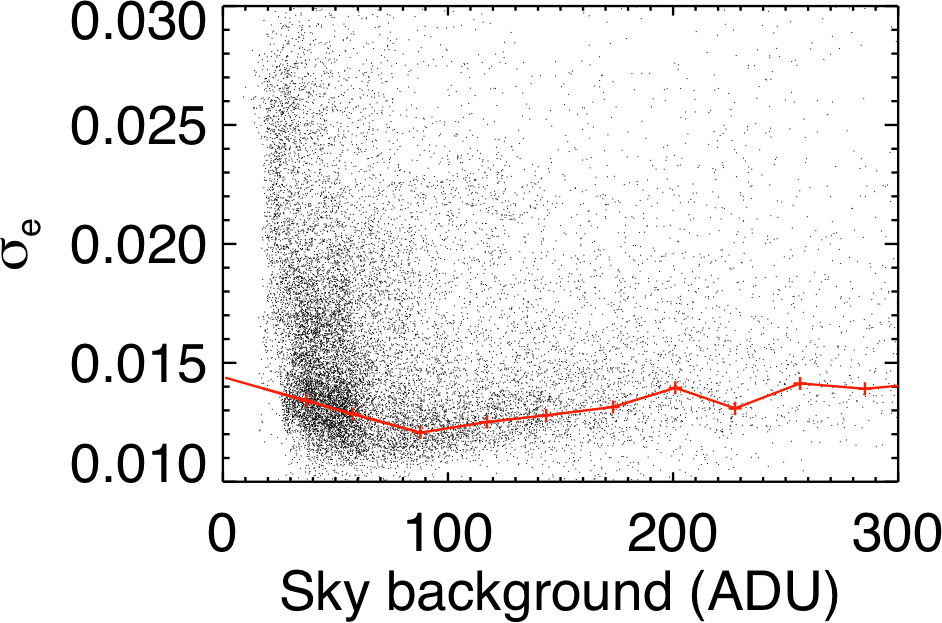}\hfill
   \includegraphics[width=4.5cm]{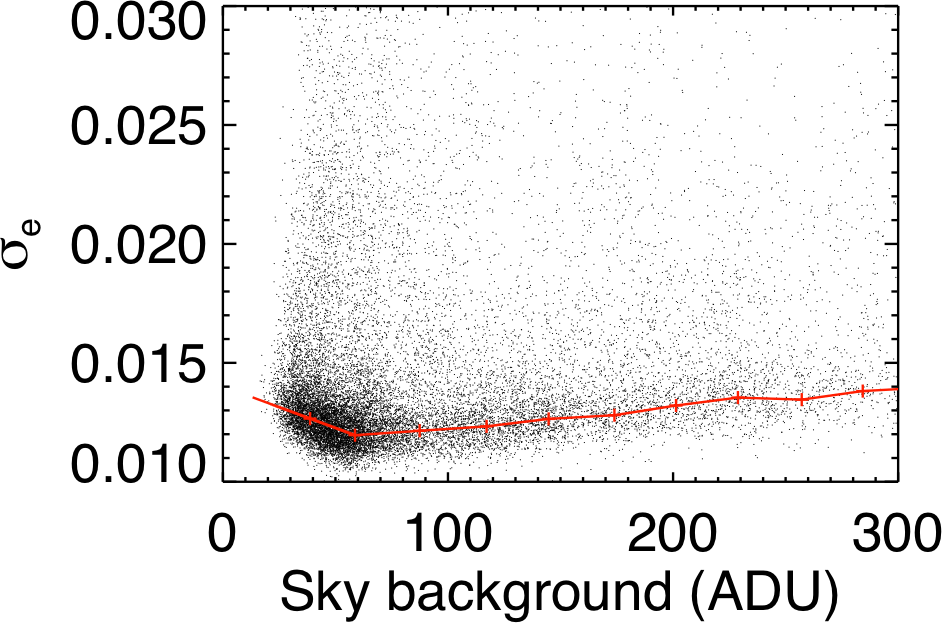}
   \caption{Correlations of $\sigma_{e}$ with the PSF FWHM (top) and the sky background (middle and bottom) for the winters 2010 (left) and 2011 (right). Measurements are indicated as black dots. The correlations extracted from good weather data are shown as red lines. The sky background correlation plots are shown in two different scales.}
   \label{fig: fwhm sky}
\end{figure}

In spite of the correction by reference stars, FWHM variations affected the photometry more than predicted in Sect.~\ref{sec: Theoretical noise}. To derive the correlation between \sigepm and the FWHM, we divided the FWHM measurements into ten intervals, and calculated the median FWHM and the median \sigepm in each interval. In this process, we removed the \sigepm outliers using a ten iteration sigma-clipping procedure with a threshold of 1.5 times the standard deviation, to ensure that the correlation was extracted from good weather data only. The extracted correlation function was denoted $f$ and is shown for the winters 2010 and 2011 in Fig.~\ref{fig: fwhm sky}. We interpolated $f$ linearly to each measured FWHM (after including a 10\% uncertainty on the measurements) and calculated the associated noise $\sigma_{F,\,e}$ by:
\begin{equation}
\sigma_{F,\,e}^2 = f^2 - f_{0}^2
\end{equation}
where $f_{0}$ was the minimum of $f$. This ensured that $\sigma_{F,\,e}$ only accounted for additional noise caused by FWHM variations. We used the same $\sigma_{F,\,e}$ value for all stars at a given epoch.

We used the same method for the sky background noise. We calculated the correlation function $g$ using 22 intervals, with smaller sizes at lower sky values. $g$ is shown for the winters 2010 and 2011 in Fig.~\ref{fig: fwhm sky}. We interpolated $g$ at each measured sky value (after including a 10\% uncertainty on the measurements) and calculated the associated noise $\sigma_{S,\,e}$ by:
\begin{equation}
\sigma_{S,\,e}^2 = g^2 - g_{0}^2
\end{equation}
where $g_{0}$ was the minimum of $g$. This empirical model was close to the theoretical estimate from Poisson noise but with some differences at low sky values, possibly related to image calibration.

We also had to modify the theoretical relative photon noise, read-out noise, and sky noise terms in Equations \ref{eq: sigma_s_ep} and \ref{eq: sigma_s_ep 2} because part of their contributions was then contained in $\sigma_{F,\,e}$ and $\sigma_{S,\,e}$. To account for this, we calculated the stellar fluxes for a unique value of FWHM (that of $f_{0}$), take a unique sky value (that of $g_{0}$), and recomputed the theoretical noise estimates. These calculations were still different for each star because each had its own photometric aperture, but they were equal for all epochs. The FWHM and sky background variations were also responsible for a certain amount of red noise, so the contribution of $\sigma_{\mathcal{R},\,s,\,e}$ had to be reduced. Some amount may have been considered as white noise (larger Poisson noise for larger FWHM or sky background), but we reduced only $\sigma_{\mathcal{R},\,s,\,e}$ and not $\sigma_{w}$ (the final result was equivalent). We defined a new red noise distribution $\mathcal{R}'$ as:
\begin{equation}
\mathcal{R}' = \alpha \; \mathcal{R}
\end{equation}
with $\alpha < 1$ and $\mathcal{R}$ defined in Eq.~\ref{eq: sigma R}. The corresponding red noise value was noted $\sigma_{\mathcal{R}',\,s,\,e}$. Our final model was:
\begin{equation}
\label{eq: sigma_s_ep 3} 
\sigma_{s,\,e}^2 = \sigma_{phot,\,s}^2+\sigma_{read,\,s}^2+\sigma_{sky,\,s}^2+\sigma_{w,\,s}^2+\sigma_{F,\,e}^2+\sigma_{S,\,e}^2+\sigma_{\mathcal{R}',\,s,\,e}^2
\end{equation}
As before, the theoretical noise for epoch $e$, \sigept, was taken as the median of $\sigma_{s,\,e}$ over the control stars.
We derived $\alpha$ by matching the left sides of the \sigept and \sigepm histograms. The final \sigept distributions for each winter are shown in Fig.~\ref{fig: sigep}. We also defined the ratio $r$ of the joint contribution of $\sigma_{F,\,e}$ and $\sigma_{S,\,e}$ in the red noise budget compared to the unknown sources $\sigma_{\mathcal{R}'}$ as:
\begin{equation}
r = \sqrt{1/\alpha^2 - 1}
\end{equation}
The two parameters of our model, $\beta$ and $\alpha$, and the ratio $r$ are reported for each winter in Table~\ref{tab: histogram parameters}. \modif{For the four winters, $\sigma_{F,\,e}$ and $\sigma_{S,\,e}$ have a larger contribution than $\sigma_{\mathcal{R}',\,s,\,e}$ ($r > 1$), so they are the main sources of correlated noise. For 2011, $r$ is lower than for the other winters, which is consistent with the stable FWHM during that winter.}
Following this approach, we could in principle have split entirely the red noise into its various components. However, no obvious correlations were found between \sigepm and other recorded parameters that were independent of the FWHM and the sky background, so we kept $\sigma_{\mathcal{R}',\,s,\,e}$ as a global representation of the remaining components.

\begin{table}
\begin{center}
\caption{Model parameters $\beta$ and $\alpha$ for the epoch noise distribution \sigept and ratio of the red noise contributions $r$. \modif{The values and uncertainties were obtained after repeating the procedure 100 times.}}
\label{tab: histogram parameters}
\begin{tabular}{c|cc|c}
\hline
\hline
Winter & $\beta$ & $\alpha$ & $r$  \\
\hline
2008 &  0.22 $\pm$ 0.09  &  0.50 $\pm$ 0.01  &  1.73  \\
2009 &  0.17 $\pm$ 0.07  &  0.38 $\pm$ 0.01  &  2.44 \\
2010 &  0.39 $\pm$ 0.09  &  0.37 $\pm$ 0.01  &  2.49  \\
2011 &  0.21 $\pm$ 0.02  &  0.65 $\pm$ 0.01  &  1.18 \\
\hline
\hline
\end{tabular}
\end{center}
\end{table}

\subsection{Weather statistics}

\begin{figure}[htbp]
   \centering
   \includegraphics[width=4.45cm]{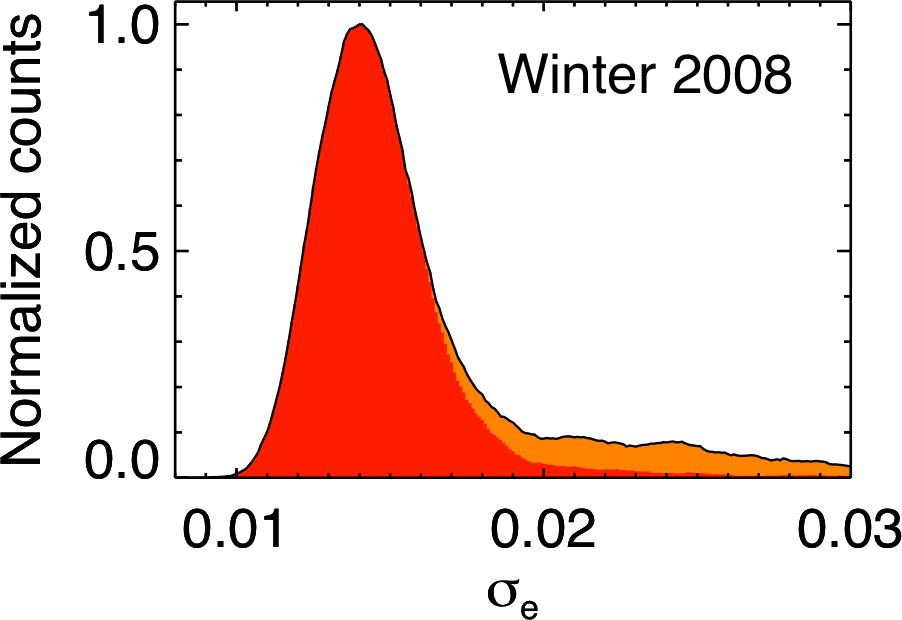}\vspace{1mm}
   \includegraphics[width=4.45cm]{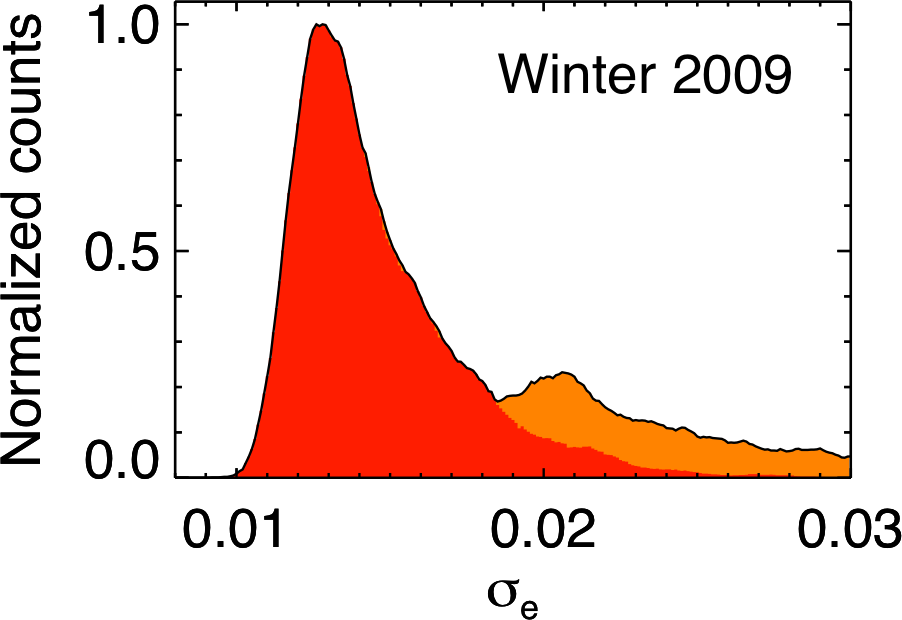}
   \includegraphics[width=4.45cm]{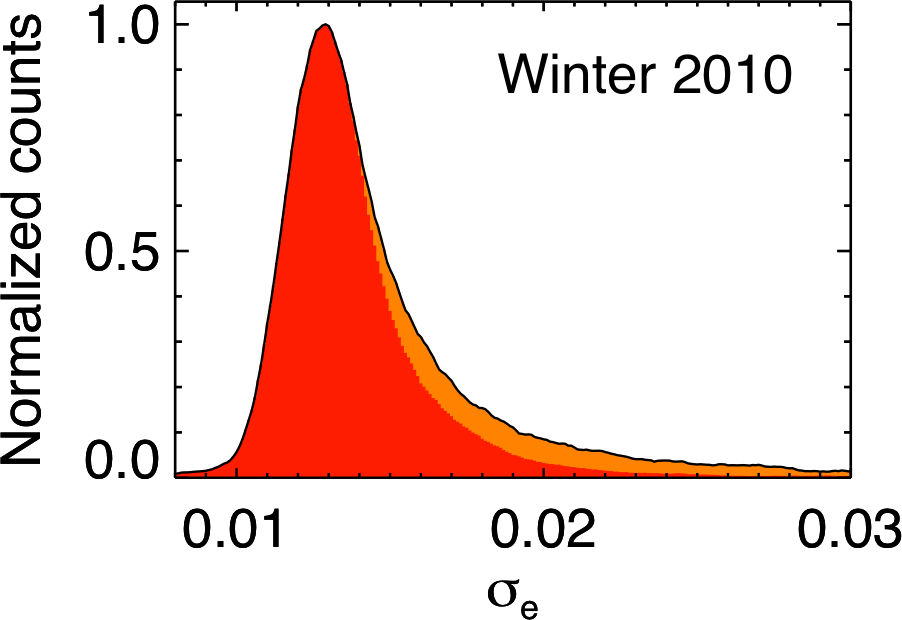}
   \includegraphics[width=4.45cm]{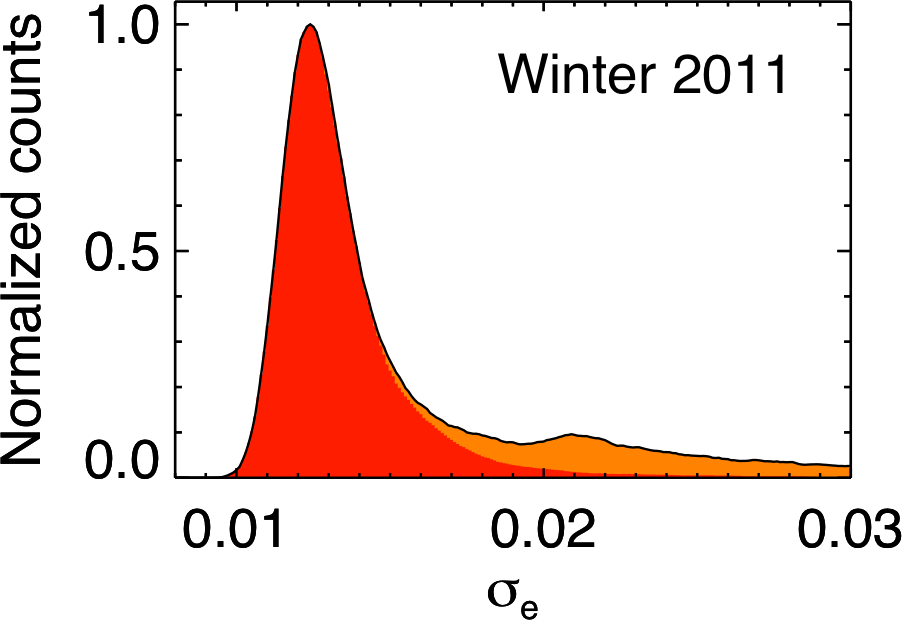}
   \caption{Normalized distribution of $\sigma_{e}$ for the winters 2008, 2009, 2010, and 2011 (black lines). Periods of photometric and veiled weather as defined by our model are indicated in red and orange respectively.}
   \label{fig: sigep}
\end{figure}

We attributed to veiled weather the values of \sigepm in excess of our model. We defined the ``photometric'' and ``veiled'' weather fractions as the areas of \sigepm below and above \sigept respectively, as illustrated in Fig.~\ref{fig: sigep}. We repeated this procedure 100 times with different random distributions for the definition of the correlation segments, for $\mathcal{R}$, and for the uncertainties on the FWHM and sky background measurements. We took the mean and standard deviation of the results to obtain the photometric and veiled weather statistics and their uncertainties.  
In addition, small fractions of the lightcurves were of reasonable quality but were not well corrected by the initial one-sidereal day period variation removal. They were kept but flagged early in the pipeline, and did not have a $\sigma_e$ value. Whether these data were affected by instrumental effects or poor weather conditions could not be inferred by our method. We classified these periods manually by interpreting the variations using the recorded parameters, the daily reports of the winter-over crew, and knowledge of the instrument behavior. We adopted a relative uncertainty of 10\% for the corresponding fractions. These periods tend to increase the veiled weather fraction, and some of them correspond to bad weather.
Furthermore, bad data were also eliminated during the photometric calibration process, and the instrument was stopped during ``white-outs'' (winter storms lasting one to a few days). We included these periods and adopted a relative uncertainty of 10\% for the corresponding fractions. \modif{Finally, in the first part of 2009 (JD $<$ 2455032), the instrument and lightcurves suffered extreme variations and our model did not reproduce the measured distribution accurately, so we derived the weather statistics only from the second part of the winter.}

\begin{figure}[htbp]
   \centering
   \includegraphics[width=4.4cm]{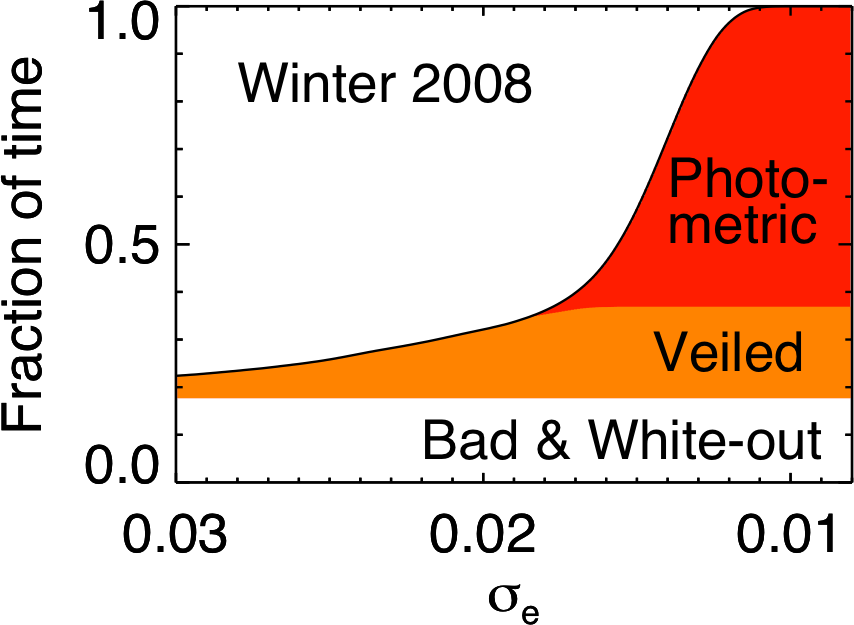}\vspace{1mm}
   \includegraphics[width=4.4cm]{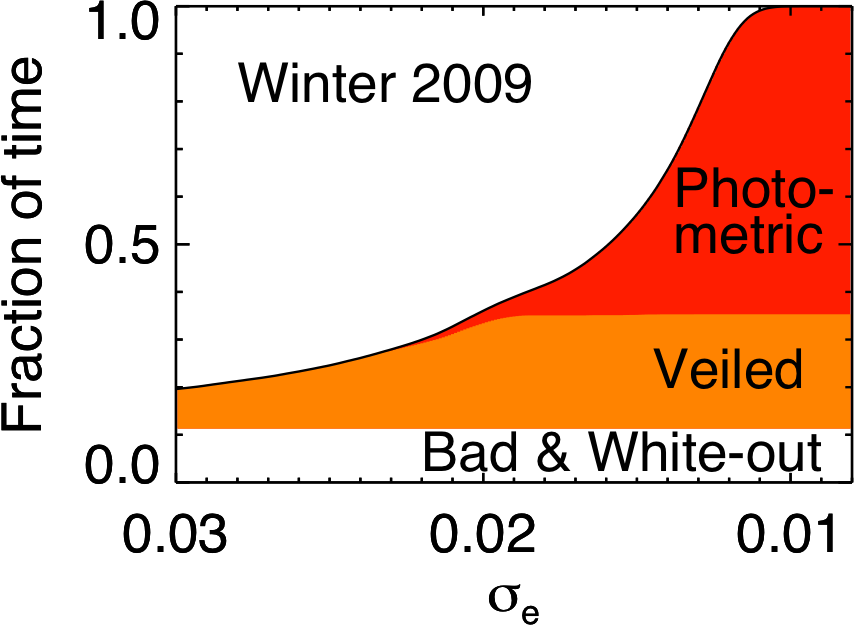}
   \includegraphics[width=4.4cm]{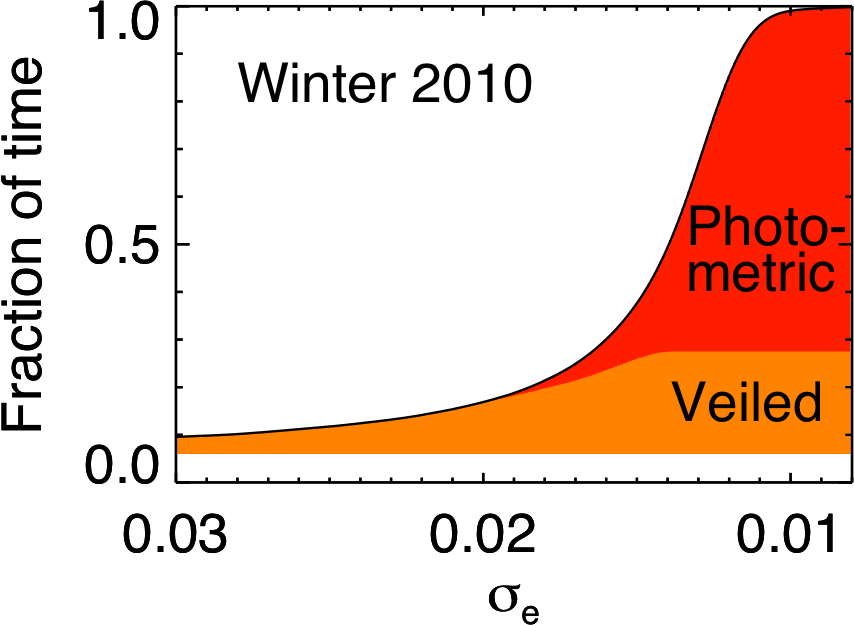}
   \includegraphics[width=4.4cm]{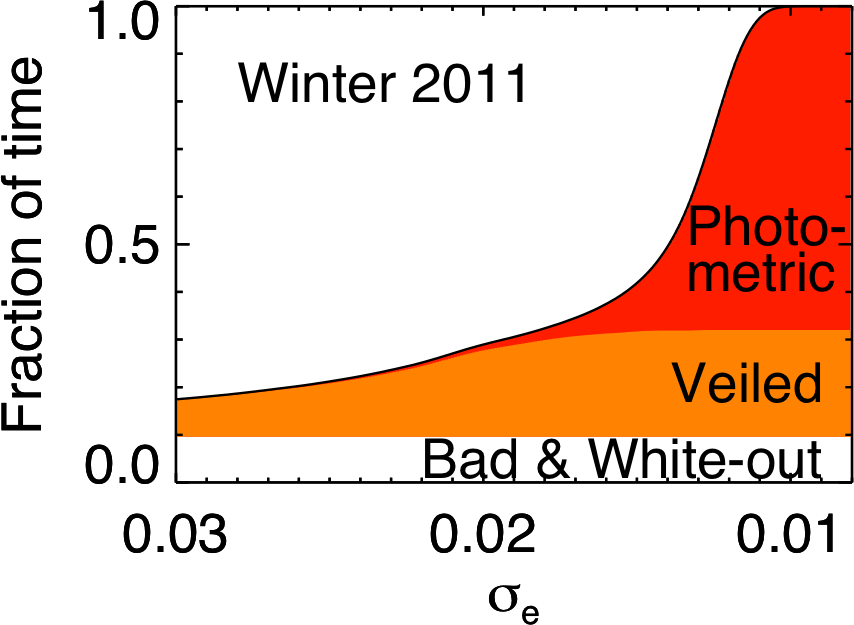}
   \caption{Cumulative normalized distribution of $\sigma_{e}$ for the winters 2008, 2009, 2010, and 2011 (black lines). Periods of photometric and veiled weather are indicated in red and orange respectively, and include adjustments for periods without a $\sigma_{e}$ value (see text). Periods of bad weather and white-out are indicated in white.}
   \label{fig: sigep cumul}
\end{figure}

After these adjustments, we computed the final weather statistics for the winters at Dome~C, which are reported in Table~\ref{tab: weather statistics}. The cumulative distributions of photometric weather, veiled weather, and bad weather plus white-out fractions are shown in Fig.~\ref{fig: sigep cumul}. We also calculated the average weather statistics for the winter at Dome~C by taking the mean and standard deviation over the four winters.
On average, we find a photometric weather fraction of 67.1 $\pm$ 4.2~\%, a veiled weather fraction of $21.8\pm2.0$~\%, a bad weather fraction of $3.3\pm2.0$~\%, and a white-out fraction of $7.9 \pm 6.0$~\%. The photometric weather fraction is comparable to that of the best astronomical sites at temperate latitudes \citep[see][for a more detailed comparison]{Crouzet2010}. These statistics are consistent with those obtained from a different method based on the number of detected stars using the 2008 winter \citep{Crouzet2010}. In particular, both methods are in excellent agreement for the 2008 winter.

\begin{table*}
\begin{center}
\caption{Weather statistics for the winter at Dome~C as observed with ASTEP South. All numbers are in percent.}
\label{tab: weather statistics}
\begin{tabular}{cccccc}
\hline
\hline
      &  2008  &  2009  &  2010  &  2011  &  Average  \\
\hline
Excluding white-outs: & & & & & \\
Photometric  &  75.6 $\pm$  2.2  &  69.6 $\pm$  1.2  &  75.3 $\pm$  2.7  &  70.9 $\pm$  1.2  &  72.9 $\pm$  3.1  \\
Veiled  &  23.1 $\pm$  2.2  &  25.8 $\pm$  1.1  &  22.3 $\pm$  1.7  &  23.3 $\pm$  1.1  &  23.7 $\pm$  1.5  \\
Bad  &   1.3 $\pm$  0.1  &   4.5 $\pm$  0.5  &   2.3 $\pm$  0.2  &   5.8 $\pm$  0.6  &   3.5 $\pm$  2.0  \\
\hline
Including white-outs: & & & & & \\
Photometric  &  63.0 $\pm$  2.2  &  64.7 $\pm$  1.2  &  72.5 $\pm$  2.6  &  68.0 $\pm$  1.2  &  67.1 $\pm$  4.2  \\
Veiled  &  19.2 $\pm$  1.8  &  24.0 $\pm$  1.1  &  21.5 $\pm$  1.6  &  22.4 $\pm$  1.1  &  21.8 $\pm$  2.0  \\
Bad  &   1.1 $\pm$  0.1  &   4.2 $\pm$  0.4  &   2.2 $\pm$  0.2  &   5.5 $\pm$  0.6  &   3.3 $\pm$  2.0  \\
White-out  &  16.6 $\pm$  1.7  &   7.1 $\pm$  0.7  &   3.8 $\pm$  0.4  &   4.1 $\pm$  0.4  &   7.9 $\pm$  6.0  \\
\hline
\hline
\end{tabular}
\end{center}
\end{table*}

\subsection{Validation}

\begin{figure}[htbp]
   \centering
   \includegraphics[width=4.5cm]{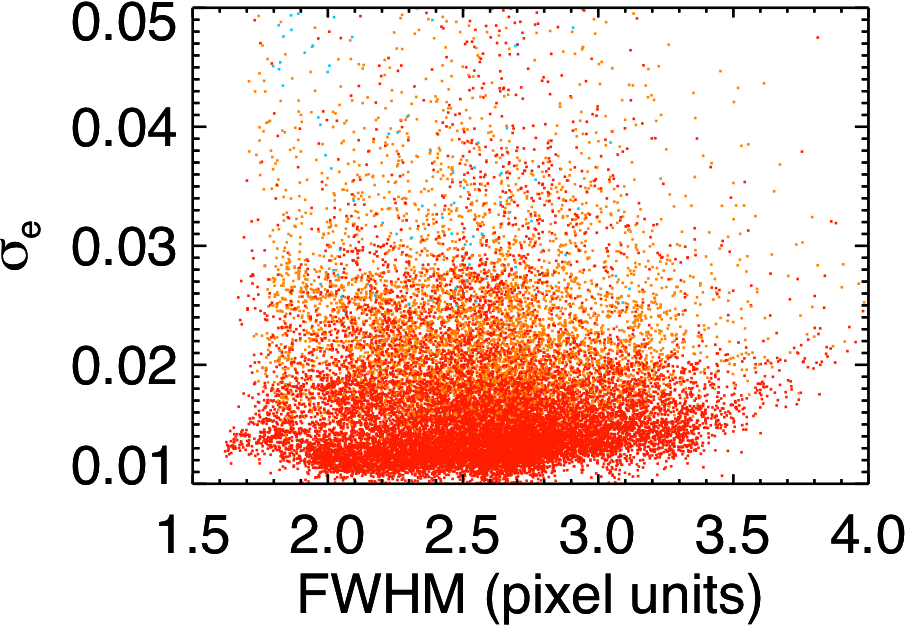}\vspace{1mm}\hfill
   \includegraphics[width=4.5cm]{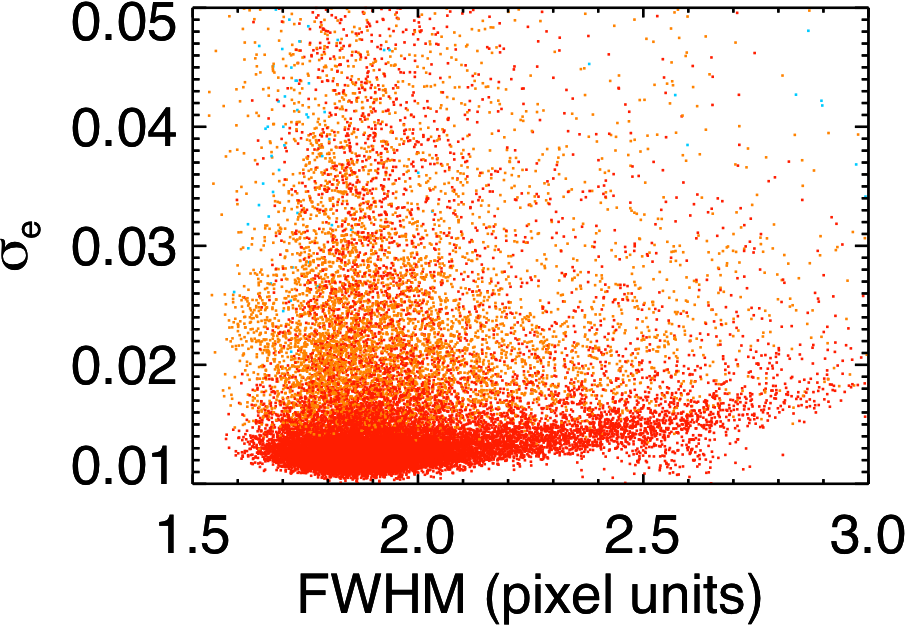}
   \includegraphics[width=4.5cm]{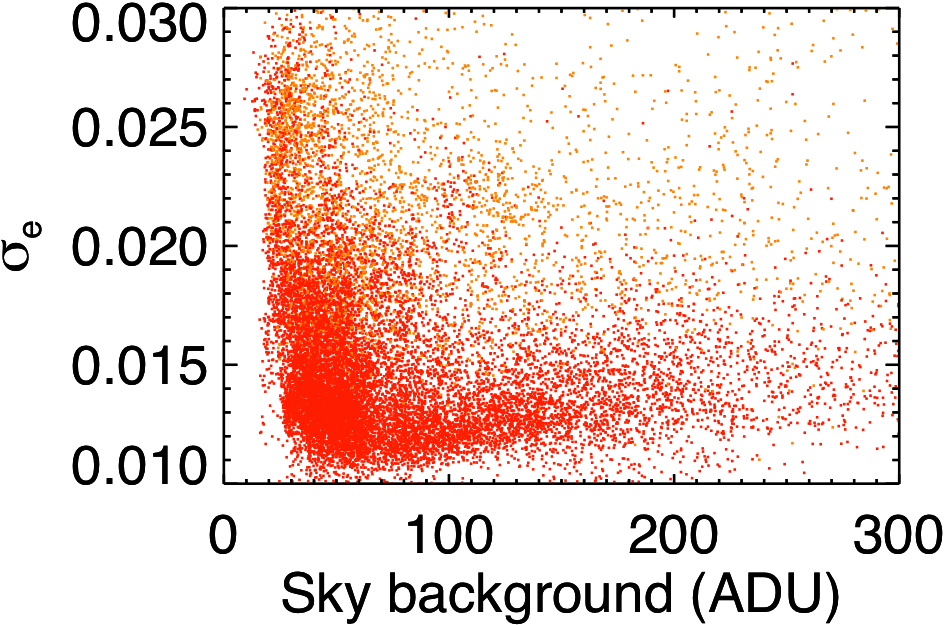}\vspace{1mm}\hfill
   \includegraphics[width=4.5cm]{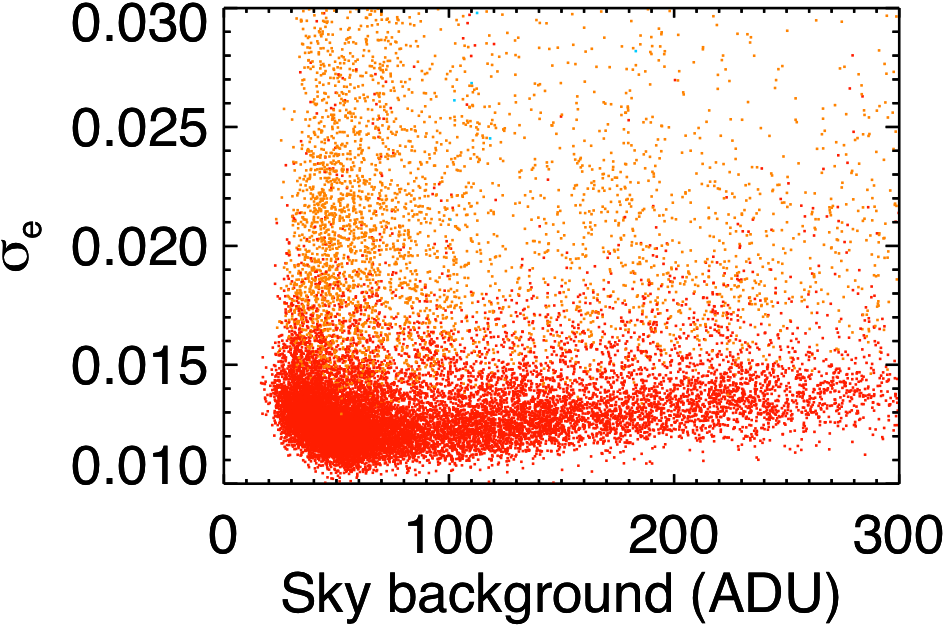}
   \includegraphics[width=4.4cm]{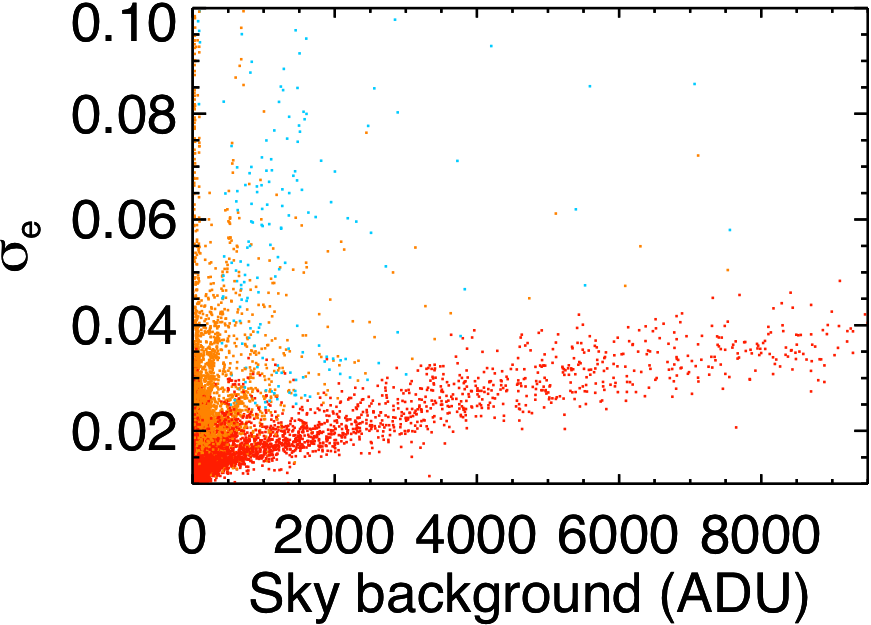}\hfill
   \includegraphics[width=4.4cm]{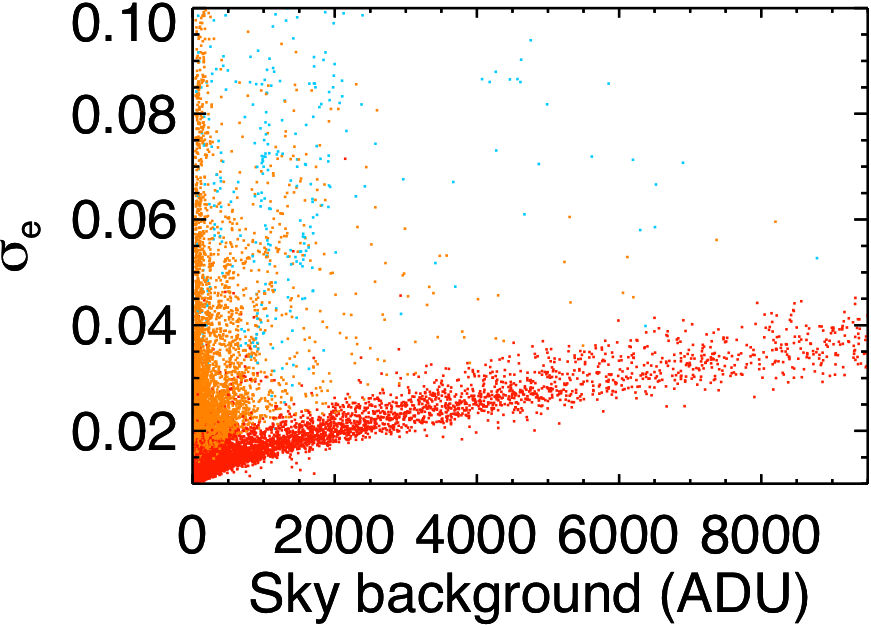}
   \caption{Correlations of $\sigma_{e}$ with the PSF FWHM (top) and the sky background (middle and bottom) for the winters 2010 (left) and 2011 (right). Photometric and veiled epochs as inferred from our model are indicated in red and orange, respectively. Bad epochs are indicated in blue. The sky background plots are shown in two different scales.}
   \label{fig: fwhm sky validation}
\end{figure}

\begin{figure}[htbp]
   \centering
   \includegraphics[width=4.5cm]{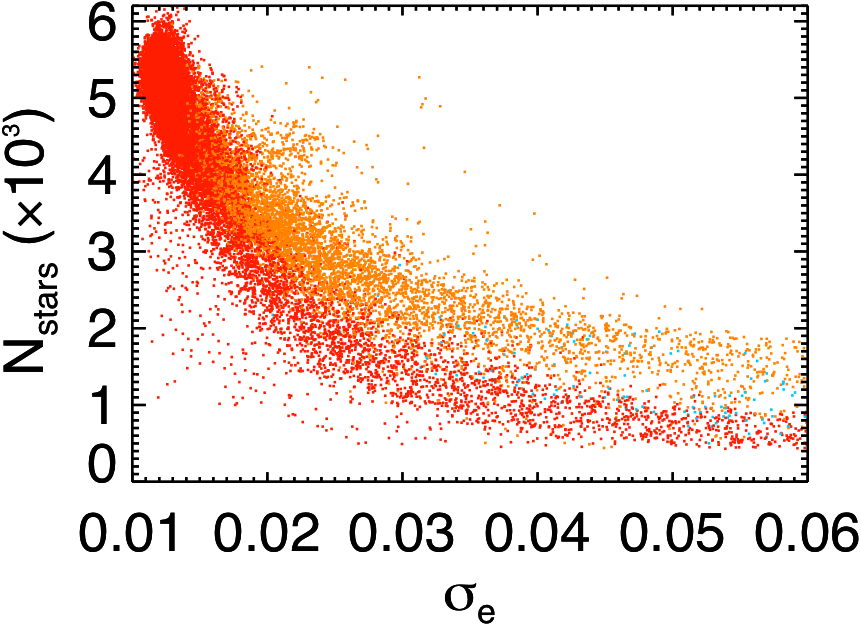}\vspace{1mm}\hfill
   \includegraphics[width=4.5cm]{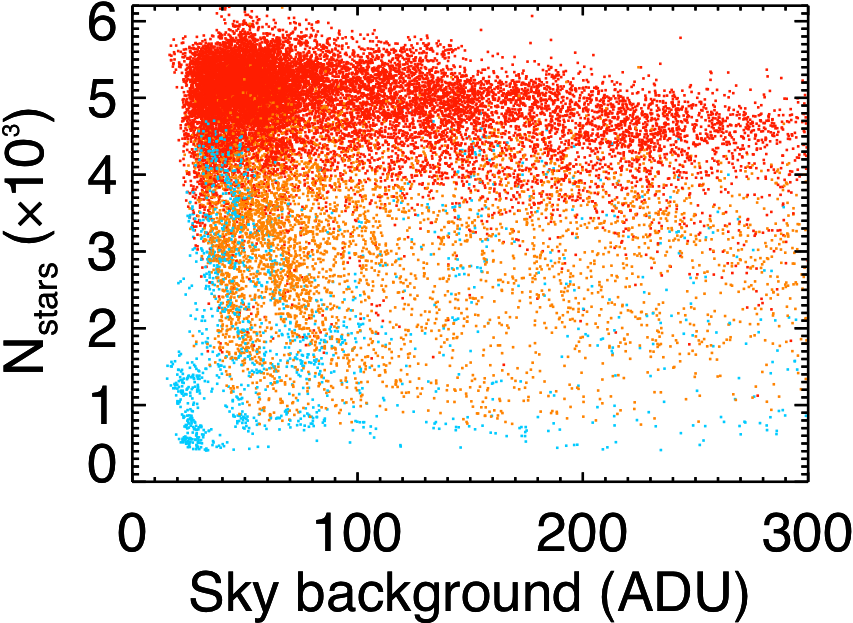}
   \includegraphics[width=4.5cm]{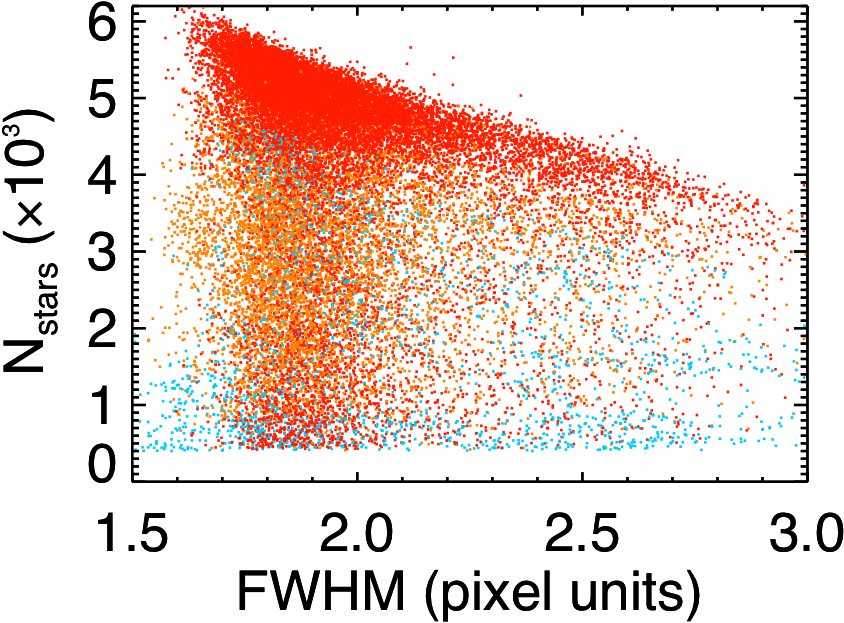}\hfill
   \includegraphics[width=4.4cm]{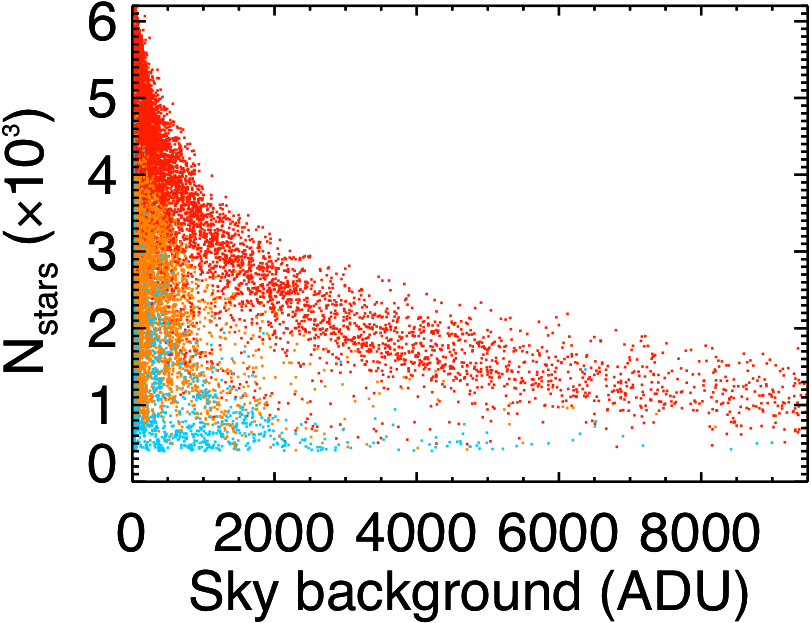}
   \caption{Correlations of the number of detected stars with $\sigma_{e}$ (top left), the FWHM (top right), and the sky background (bottom left and right, in two different scales) for the 2011 winter. Photometric and veiled epochs as inferred from our model are indicated in red and orange, respectively. Bad epochs are indicated in blue; most of them do not have a calculated $\sigma_{e}$.}
   \label{fig: number of stars}
\end{figure}

We validated the results of our method by verifying the epoch quality with respect to several parameters. To infer the quality of individual epochs, we calculated the ratio $\sigma_{e,\;m}/\sigma_{e,\;t}$ and we considered that an epoch is photometric or veiled if this ratio was below or above a certain limit, respectively. This limit was calculated for each winter to match the weather statistics. Figure~\ref{fig: fwhm sky validation} shows $\sigma_{e}$ as a function of the FHWM and of the sky background for 2010 and 2011 (same as Fig.~\ref{fig: fwhm sky} now showing the quality of each epoch). As expected, the bulk of data corresponds to photometric weather, and the variations of $\sigma_{e}$ with the FHWM and sky background are well taken into account in differentiating the photometric and veiled weather. We also show the number of detected stars as a function of these parameters for the 2011 winter (Fig.~\ref{fig: number of stars}). This number is a good proxy for the weather quality and was used by \citet{Crouzet2010} to derive the weather statistics at Dome~C for the 2008 winter, before extracting the lightcurves. As expected, the number of detected stars is correlated with $\sigma_{e}$. Interestingly, the correlation has two tails, one that extends the bulk of photometric weather and one that corresponds to veiled weather. This is also the case for 2010 but is less obvious for 2008 and 2009. This reflect the sky background distribution: the photometric (lower) tail corresponds to the correlation between large sky backgrounds and $\sigma_{e}$ whereas the veiled (upper) tail corresponds to low sky backgrounds for which the number of stars was affected by poor weather conditions. In these plots, the photometric, veiled, and bad epochs are distributed as expected. In particular, the bulk of photometric data inferred from our model matches very well the bulk of data with a nominal number of detected stars, which confirms that they correspond to excellent weather. A few epochs have almost nominal parameters (high number of stars, low sky background, nominal FWHM) but were classified as bad. These epochs were identified as outliers before calculating $\sigma_{e}$ (see Sect.~\ref{sec: Data selection}) and were often adjacent to or surrounded by bad weather periods; they were not suitable for photometry. Overall, these results validate our model and show that the approach used here and that of \citet{Crouzet2010} are consistent.

\section{Duty cycle}

\begin{figure*}[htbp]
   \centering
   \includegraphics[width=5.223cm]{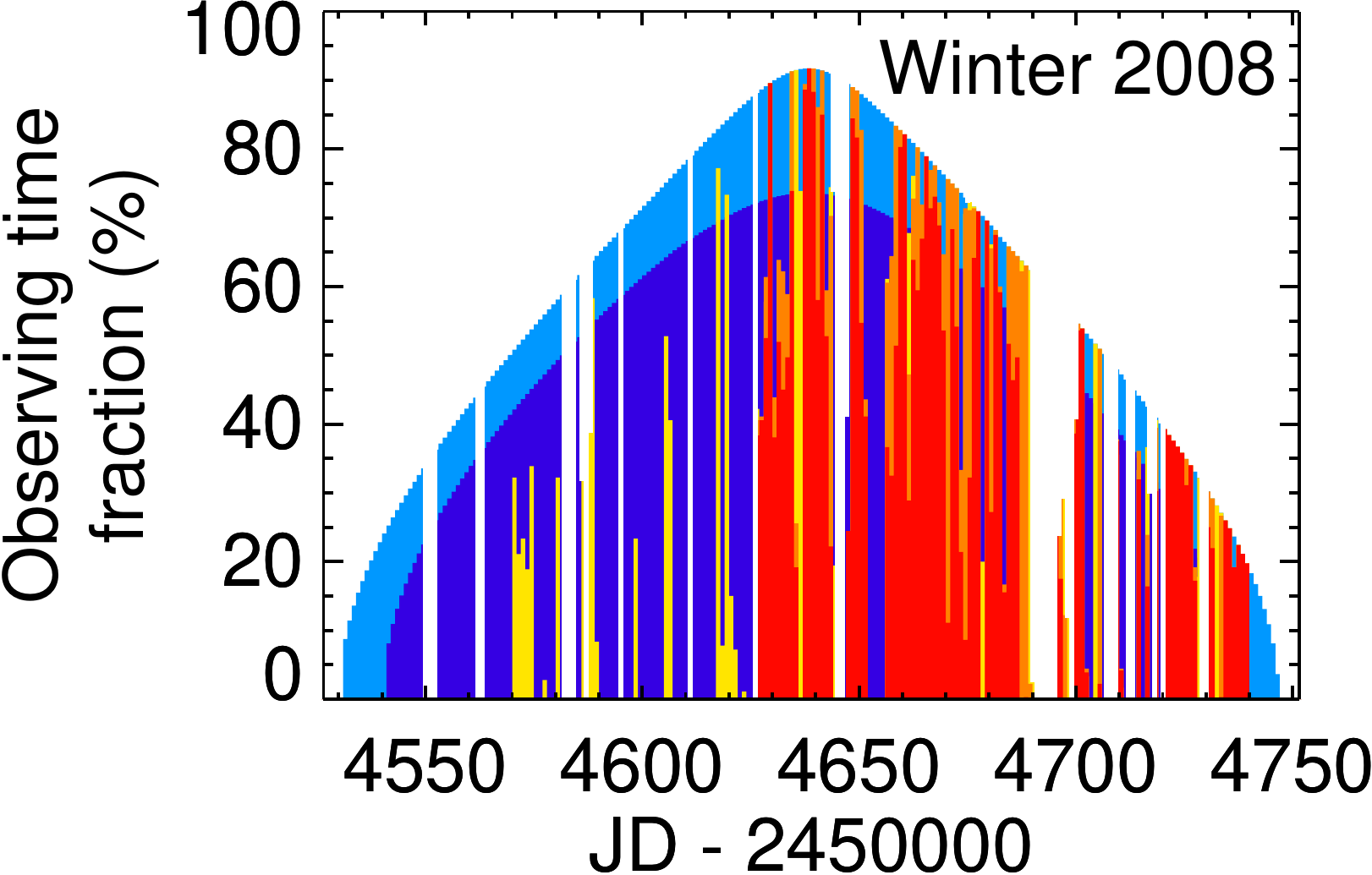}
   \hspace{-0.2cm}
   \includegraphics[width=4.25cm]{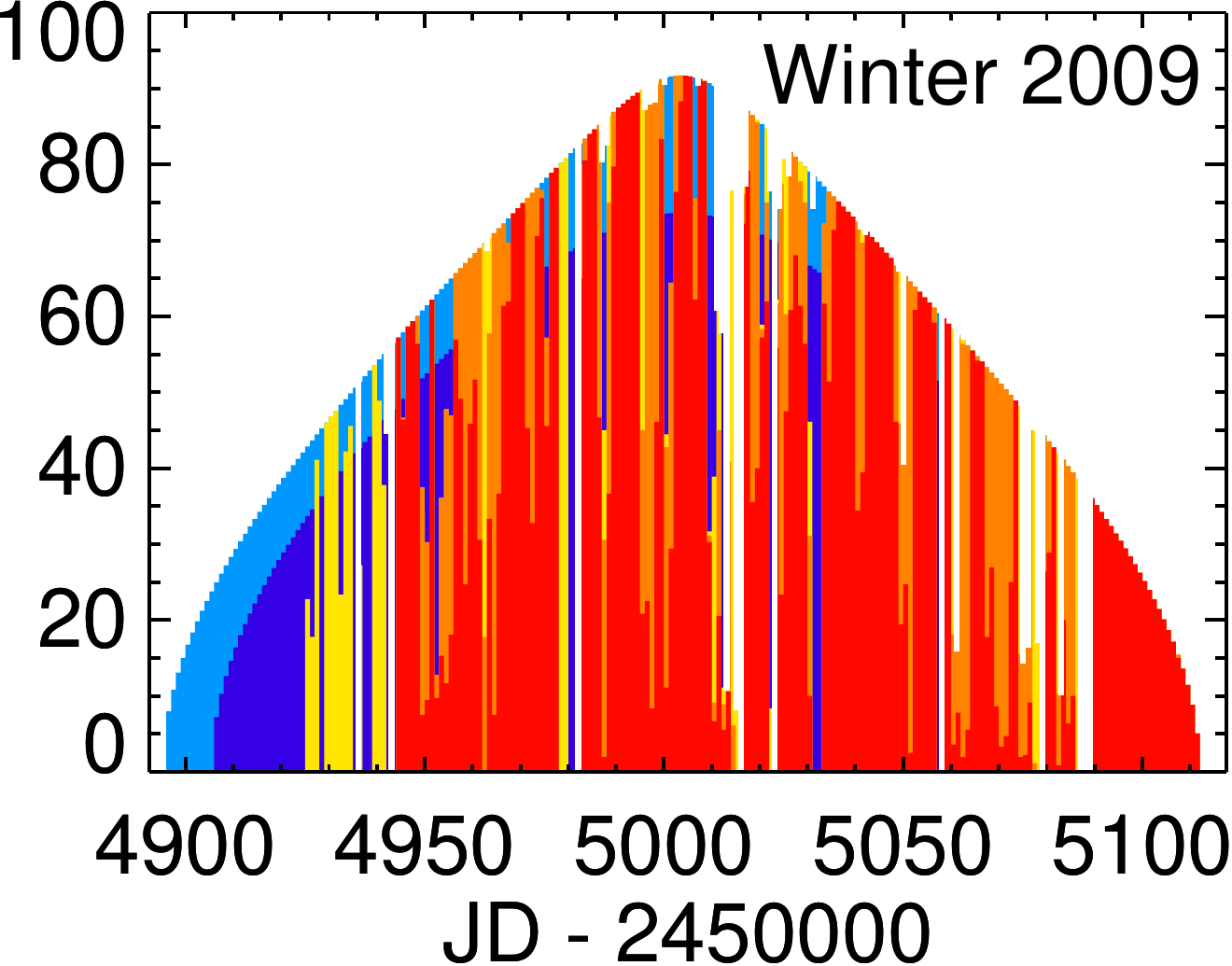}
  \hspace{0.05cm}
   \includegraphics[width=4.25cm]{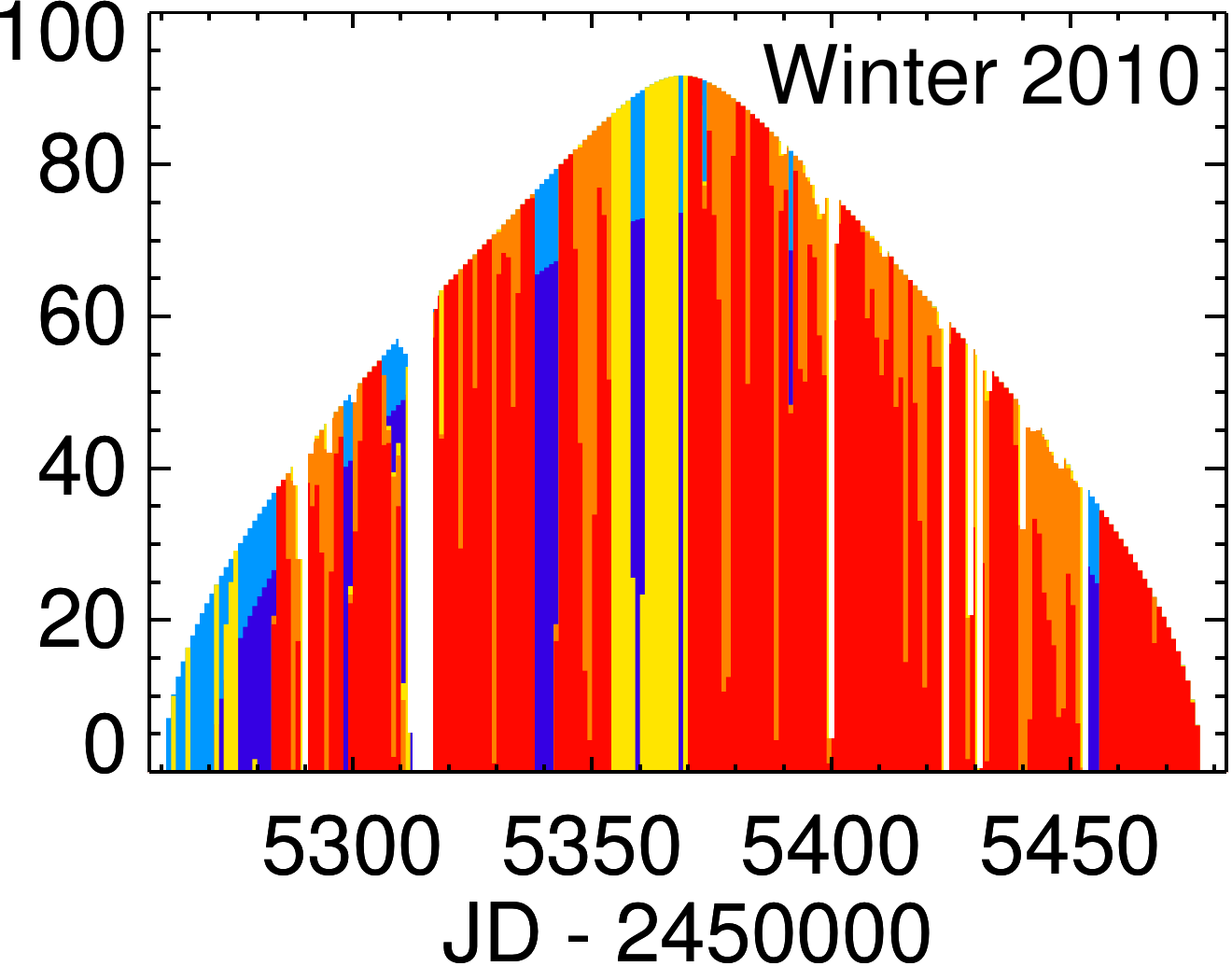}
   \hspace{0.05cm}
   \includegraphics[width=4.25cm]{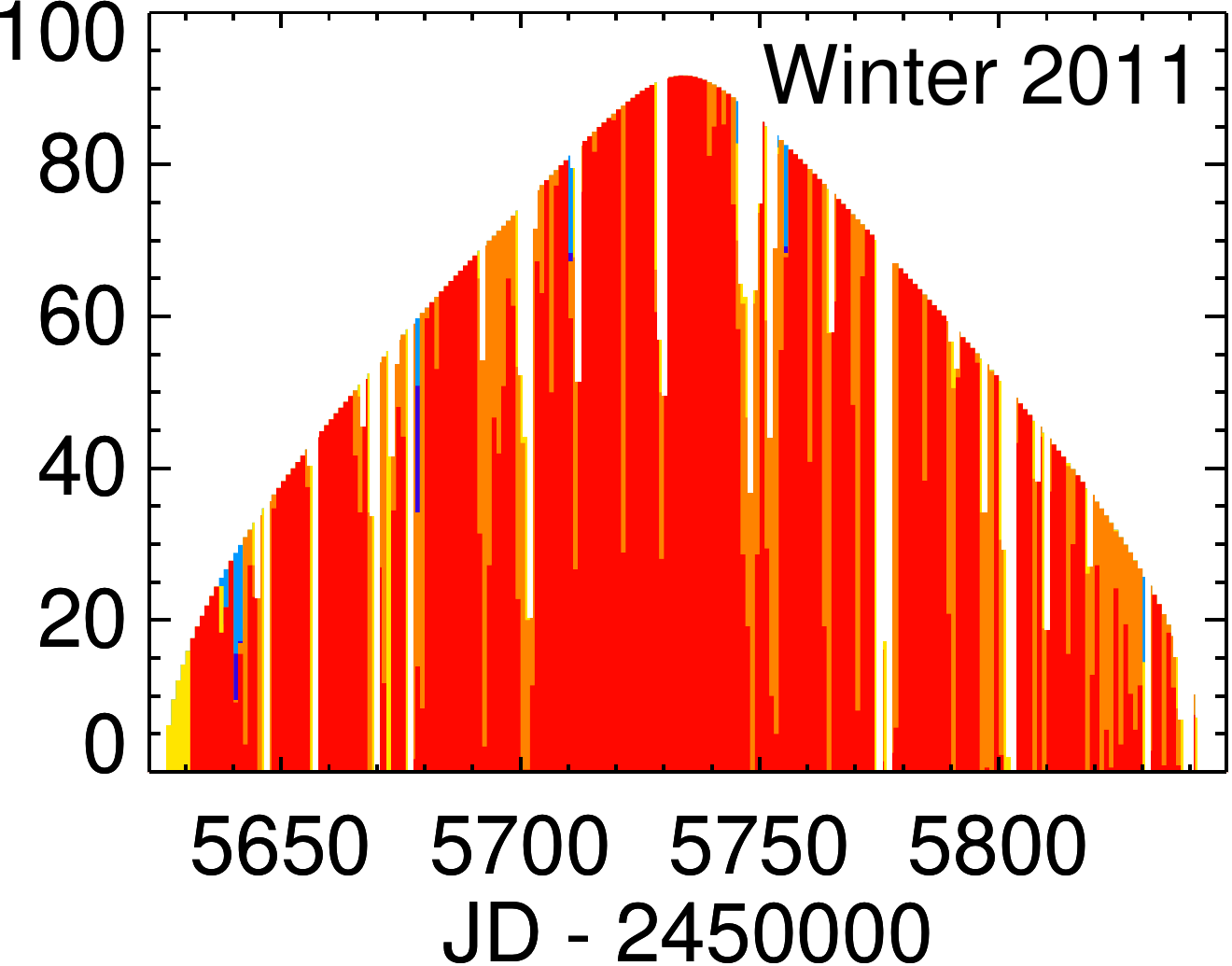}
      \caption{Daily observing time fraction for ASTEP South for the winters 2008, 2009, 2010, and 2011. The light blue and dark blue regions indicate the fraction of time with a Sun elevation lower than $-9^\circ$ and $-13^\circ$ (9\dgr and 13\dgr below the horizon), respectively. Periods of photometric and veiled weather as observed by ASTEP South are indicated in red and orange, respectively. Periods of bad weather including white-outs are indicated in white. Periods in yellow indicate that data were recorded but not used in the photometric analysis, either because they were taken during the instrument setup at the beginning of the winter or were lost during the data transfer.}
   \label{fig: obshist}
\end{figure*}

The duty cycle of ASTEP South for the winters 2008 to 2011 is represented in Fig.~\ref{fig: obshist}: the limit due to the Sun elevation, the fraction of time where ASTEP South was operating, and the photometric, veiled, and bad weather fractions including white-outs are shown for each day of the winter. In total, we acquired 11415.1 hours of data with ASTEP South including 6699.8 hours under photometric conditions and nominal functioning of the instrument. Considering only times with a Sun elevation lower than $-9$\dgr, these numbers are 9451.2 hours and 6308.5 hours respectively. The detail for each year is reported in Table~\ref{tab: observing time}. The increase of the operating time from the 2008 to the 2011 winter is evident and reflects the improvements of the system (hardware and software) made over the years. The first half of the 2008 winter was hampered by numerous technical issues while setting up the instrument. A long white out period occurred at the end of August and lasted for about ten days. In 2009, the beginning of the winter was dedicated to setting up the instrument and improving the thermalization; data acquired during this time were not used in the lightcurves. In 2010, we lost two weeks of data taken in June, either in the data transfer between computers and external disks or while shipping the disks from Concordia to Nice. In 2011, the installation was fast and the system acquired high quality data nearly continuously for the whole winter, with almost no human interventions. This demonstrates that a simple instrument can operate almost continuously for several winters at Dome~C. In addition, as noted in \citet{Crouzet2010}, periods of bad weather are generally concentrated in white-out episodes lasting from one to a few days; between these episodes, high quality photometric observations can be achieved for extended periods of time. Overall, the weather statistics and duty cycle at Dome~C demonstrate the high quality of this site for continuous photometric observations during the winter.

\begin{table*}
\begin{center}
\caption{Amount of data acquired by ASTEP South and subsample of data taken under photometric conditions and nominal functioning of the instrument, with or without the constraint of a Sun elevation lower than $-9$\dgr. The amount of time with a Sun elevation lower than $-9$\dgr for the whole year at Dome~C is indicated for comparison. All numbers are in hours.}
\label{tab: observing time}
\begin{tabular}{c|cc|cc|c}

\hline
\hline
 Winter    &  All data  &  Photometric data  &  All data   &  Photometric data  &  $h_{sun} < -9\dgr$  \\
          &            &				      & 	$h_{sun} < -9\dgr$  &  $h_{sun} < -9\dgr$ &  \\	
\hline
2008  &  1640.5  &  887.0    &    1376.3  &   864.8   & 3030.6   \\
2009  &  2977.1  &  1692.8  &    2453.4  &  1575.7  & 3030.9 \\
2010  &  3155.5  &  1829.9  &    2627.7  &  1749.2  & 3031.0 \\
2011  &  3642.0  &  2290.1  &    2993.8  &  2118.8  & 3031.0 \\
\hline
Total  & 11415.1  &  6699.8  &  9451.2  &    6308.5  &  12123.5\\

\hline
\hline

\end{tabular}
\end{center}
\end{table*}

\section{Study of variable stars and eclipsing binaries}
\label{sec: study of variable stars and eclipsing binaries}

\subsection{Search for periodic signals}

We searched for periodic signals in the lightcurves using the box least square (BLS) algorithm \citep{Kovacs2002}. This program searches for square shapes representing transit-like events, and also detects stellar variations. 
We performed the search over a wide period range ($0.4 < P < 100$ days) which we divided into 15 intervals with sizes following approximately a logarithmic scale. We built a frequency comb for each interval using a constant spacing in $\Delta f / f$, a number of frequencies set by the expected duty cycle, and we ran BLS in each interval separately.
During the search, we applied locally a refined comb with nine times more frequencies, first around the forty best frequencies, then around the best frequency. This adaptive comb yielded a more precise determination of the best frequency.
We ran BLS twice. After the first run, the histogram of periods showed peaks that were centered around integer or half-integer values, indicating harmonics of the ``day-night'' cycle (the sky background increases around noon even during the winter, see Sect.~\ref{sec: Sky background}). Another peak around 29 days was due to the Moon cycle. We excluded periods around these peaks for the second BLS run.
Finally, we combined the 15 intervals into three period ranges: 0.4 -- 1 day, 1 -- 10 days, and 10 -- 100 days. For each lightcurve, we compared the results obtained for the intervals within each range and kept only the period that yielded the largest power in the BLS spectrum. Every lightcurve ended up with a best period, but most of them did not contain a valid signal. We ranked them by decreasing signal to noise ratios, where the signal was calculated from the amplitude and duration of the variations and the noise was obtained from the standard deviation of the lightcurve residuals. Then, we inspected the lightcurves visually to identify variable stars, eclipsing binaries, and transiting exoplanet candidates. Examples of lightcurves of variable objects are shown in Fig.~\ref{fig: lightcurve examples}.

\subsection{Analysis of variable objects}
\label{sec: Analysis of variable objects}

We detected periodic signals in the lightcurves of 211 stars. A similar search was performed by the Chinese Small Telescope ARray \citep[CSTAR,][]{Yuan2008, Zhou2010} from Dome~A, Antarctica, \modif{using data acquired during the 2008, 2009, and 2010 winters} \citep{Wang2011, Wang2013, Oelkers2015, Yang2015, Wang2015}. \modif{We cross-matched our objects with the variable stars reported by CSTAR and by the GCVS \citep{Samus2017}, VSX \citep{Watson2006}, and ASAS \citep{Pojmanski2002, Richards2012} catalogs. The results are given in Table~\ref{tab: inventory}. We identified 96 objects that were not reported} and we analyzed their lightcurves with the Period04 software program \citep{Lenz2005}. First, we analyzed the 2011 data alone because they were more abundant and of better quality, then we analyzed the four winters together. In the Fourier spectra, we found systematic peaks around 1.003 and 0.034~$\rm d^{-1}$ caused by the day-night and Moon cycles, respectively (these frequencies were not completely avoided in our BLS search probably due to the frequency refinement procedure). We eliminated stars showing variations only at these frequencies. We also discarded stars with frequencies that were common to many stars, indicating non-astrophysical origin, and stars with signal to noise ratios that were too low to extract accurate periods and amplitudes. \modif{This left 42 objects with confirmed variability}. We analyzed them year by year to distinguish variable stars and eclipsing binaries: the amplitude of variations due to stellar activity varies from year to year but remains the same for eclipsing binaries. \modif{We also analyzed the lightcurves of five known objects: we investigated in detail \object{$\sigma$~Oct} and \object{HD~184465}, we improved the parameters of \object{EN~Oct}, and we re-classified two CSTAR variables as eclipsing binaries.}

In this study, we computed the frequency and amplitude uncertainties using the formulae proposed by \citet{Montgomery1999} providing 3$\sigma$ estimates. These uncertainties correspond to ideal cases (constant noise, regular spacing, no gaps) and underestimate the true uncertainties. Based on our experience, we used the 3$\sigma$ values for our uncertainties.

\modif{Finally, we counted the number of previously reported variables in the ASTEP South field of view that were missed by our search (that were not in our list of 211 potential variables). We missed 103 objects including 90 from CSTAR and 13 from GCVS, VSX, or ASAS. Some were not in our initial target list, some were too faint, and some were simply not identified as variable. CSTAR is composed of four Schmidt-Cassegrain telescopes, each with a pupil entrance aperture of 145 mm and a field of view of 4.5\dgr in diameter (20\dgrq in total) equipped with different filters (g, r, i, and no filter). The CSTAR target list contains around $18\,000$ objects. Although technical issues generally prevented the acquisition of data by the four telescopes simultaneously, each CSTAR telescope already collects about twice more flux than ASTEP South. This set up can explain that many variables were detected by CSTAR and missed by ASTEP South. The photometric data reduction pipelines are also different, and \citet{Oelkers2015} use differential images. In addition, our search for periodic signals was primarily oriented toward the detection of transiting exoplanet candidates and we used only the BLS algorithm to search for variability. Other algorithms or metrics that were used for CSTAR such as the Welch-Stetson variability index \citep{Stetson1996}, the analysis of variance method \citep{Schwarzenberg-Czerny1996}, or the Lomb-Scargle method \citep{Lomb1976, Scargle1982} may reveal more variables in the ASTEP South lightcurves.} Besides, we restrained the search range to periods larger than 0.4 days, and most $\delta$ Scuti were not identified by their main frequency: some were found from their harmonics and some may have been missed. \modif{In contrast, the longer coverage by ASTEP South compared to CSTAR may explain that some variables were detected only by ASTEP South.}

Our goal is not to detect all the variables in the ASTEP South field of view, which has already been studied extensively by CSTAR. Instead, we complement the CSTAR search and focus in the following on a few interesting objects that highlight the long photometric coverage that we achieved with ASTEP South. 


\begin{table}
\begin{center}
\caption{\modif{Summary of our search for variable objects in the ASTEP South lightcurves.}}
\label{tab: inventory}
\begin{tabular}{lc}
\hline
\hline
Number of target stars     &      5954    \\
\hline
Objects with detected variability     &      211    \\
\hspace{5mm} New objects     &      96    \\
\hspace{5mm}\hspace{5mm} Confirmed variability     &      42    \\
\hspace{5mm}\hspace{5mm}\hspace{5mm} New variable stars     &      34    \\
\hspace{5mm}\hspace{5mm}\hspace{5mm} New eclipsing binaries     &      8    \\
\hspace{5mm}\hspace{5mm} Unconfirmed variability     &      54     \\
\hspace{5mm} Known objects     &      115    \\
\hspace{5mm}\hspace{5mm} From CSTAR     &      95    \\
\hspace{5mm}\hspace{5mm} From GCVS, VSX, and ASAS     &      20    \\
\hline
Objects missed     &      103    \\
\hspace{5mm} From CSTAR     &      90    \\
\hspace{5mm} From GCVS, VSX, and ASAS     &      13    \\
\hline
\hline
\end{tabular}
\end{center}
\end{table}

\subsubsection{\object{$\sigma$~Oct}}

\begin{figure}[htbp]
   \centering
   \includegraphics[width=8.5cm]{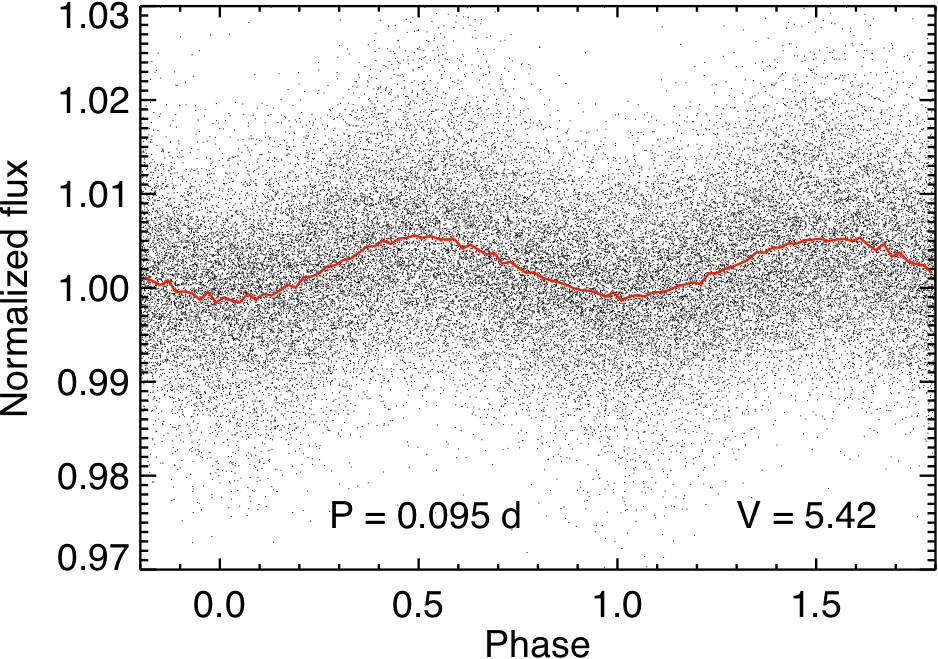}
   \caption{Lightcurve of \object{$\sigma$~Oct} as observed with ASTEP South over four winters, phase-folded at a period of 0.0953 day. The black points correspond to a time sampling of ten minutes. The red line is a binning with 100 points over the phase range. The period and V magnitude are indicated on the plot.}
   \label{fig: sigma Oct lightcurve}
\end{figure}

\begin{figure}[htbp]
   \centering
   \includegraphics[width=8cm]{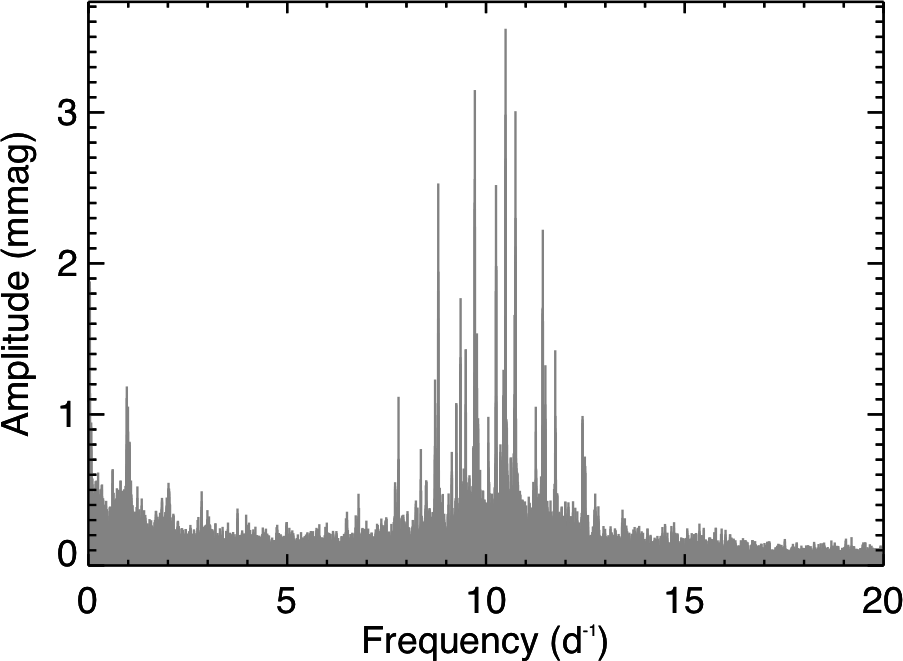}
   \caption{Fourier spectrum of \object{$\sigma$~Oct} obtained from its four winter lightcurve.}
   \label{fig: sigma Oct Fourier}
\end{figure}

\begin{figure}[htbp]
   \centering
   \includegraphics[width=8cm]{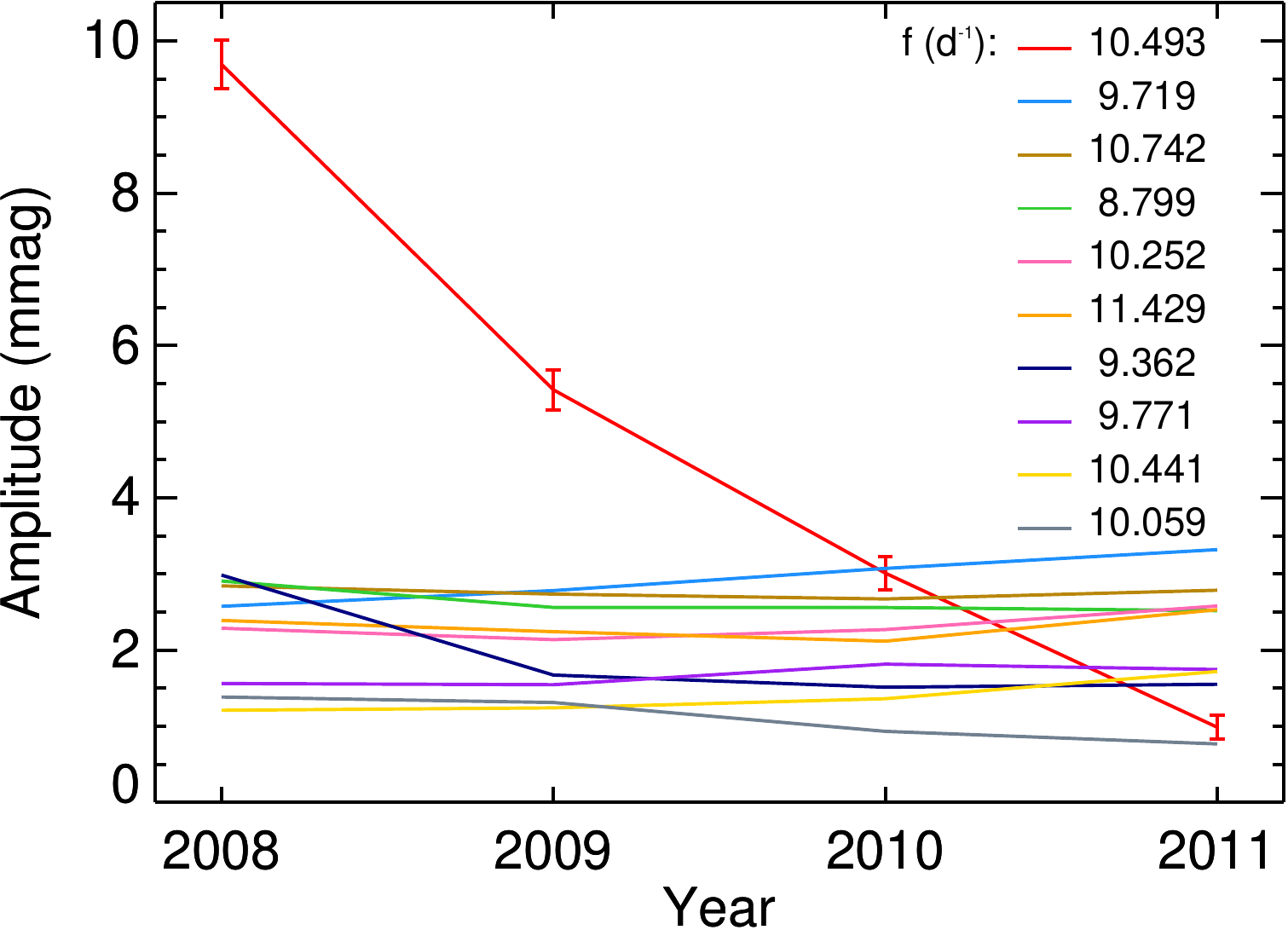}
    \caption{Amplitude variations of the ten main $\delta$ Scuti frequencies of \object{$\sigma$~Oct} from 2008 to 2011, with one point per winter. The 3$\sigma$ uncertainties are shown for the main frequency and are the same for all frequencies.}
   \label{fig: sigma Oct amplitudes}
\end{figure}

\object{$\sigma$~Oct}, also known as \object{HR 7228} and \object{HD 177482}, is the brightest star in the field (V = 5.42). It is a $\delta$ Scuti reported first by \citet{McInally1978} and re-observed only once by \citet{Coates1981}. Its properties are poorly known as it had been observed only four nights in total before the ASTEP South and CSTAR experiments. Its known pulsation period is on the order of 0.1 day with an amplitude of 15~mmag. The ASTEP South lightcurve of \object{$\sigma$~Oct} is shown in Fig.~\ref{fig: sigma Oct lightcurve}; we note that this star was saturating the CCD so the measured amplitudes should be taken with caution. We analyzed the data with the Period04 software program and detected 21 frequencies with signal to noise ratios larger than four (Fig.~\ref{fig: sigma Oct Fourier}, Table~\ref{tab: sigma Oct 21 freq}). Seventeen of them are between 8.023 and 11.756~$\rm d^{-1}$. The dominant peak is at 10.493~$\rm d^{-1}$ and corresponds to $\delta$ Scuti-type oscillations with an average amplitude of 3.73~mmag. Interestingly, four other frequencies are between 0.612 and 2.848~$\rm d^{-1}$ with amplitudes below 0.7~mmag, and could correspond to $\gamma$ Dor-type pulsations. Thus, \object{$\sigma$~Oct} is apparently an hybrid $\delta$ Scuti -- $\gamma$ Dor with very low amplitudes for the $\gamma$ Dor modes. We analyzed the ten main $\delta$ Scuti frequencies year by year (Fig.~\ref{fig: sigma Oct amplitudes}, Table~\ref{tab: sigma Oct}). Four of them show amplitude variations. In particular, the amplitude of the dominant frequency decreases from 9.74 to 2.76~mmag (a factor of 3.5) between 2008 and 2011, and this decrease is not compensated by an equivalent increase of the amplitudes of other frequencies. Thus, the cause is not an energy transfer between oscillation modes but is an actual diminution of the total amplitude of the star's oscillations. A similar behavior was found in 603 $\delta$ Scuti out of 983 using four year lightcurves from the \mbox{\textit{Kepler}} mission and no explanation is known to date \citep{Bowman2016}.

\subsubsection{\object{HD~184465}}

\begin{figure}[htbp]
   \centering
   \includegraphics[width=8.5cm]{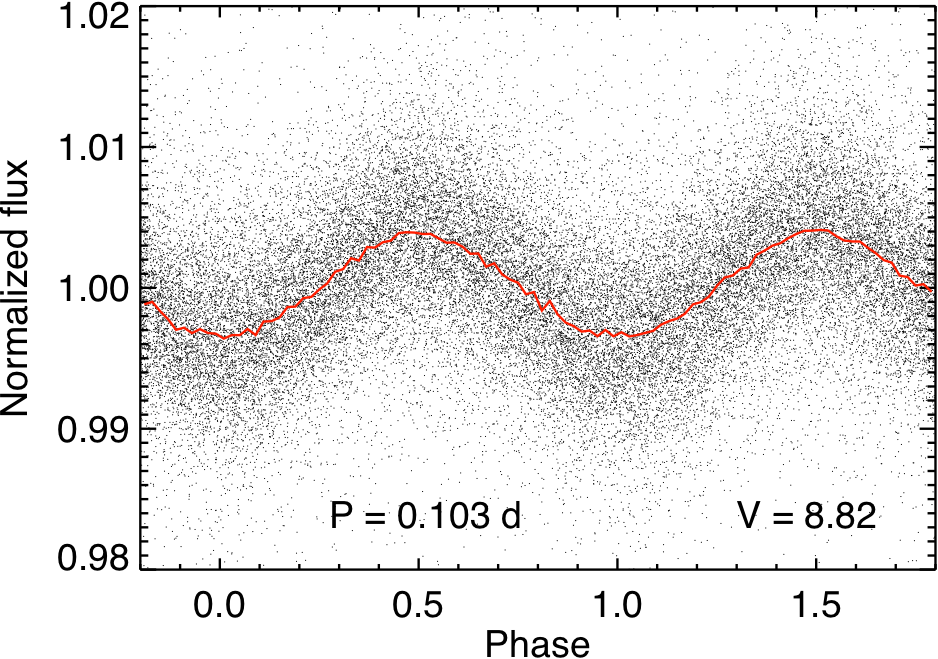}
   \caption{Lightcurve of \object{HD~184465} as observed with ASTEP South over four winters, phase-folded at a period of 0.1034 day. The black points correspond to a time sampling of ten minutes. The red line is a binning with 100 points over the phase range. The period and V magnitude are indicated on the plot.}
   \label{fig: HD 184465 lightcurve}
\end{figure}

\begin{figure}[htbp]
   \centering
   \includegraphics[width=8cm]{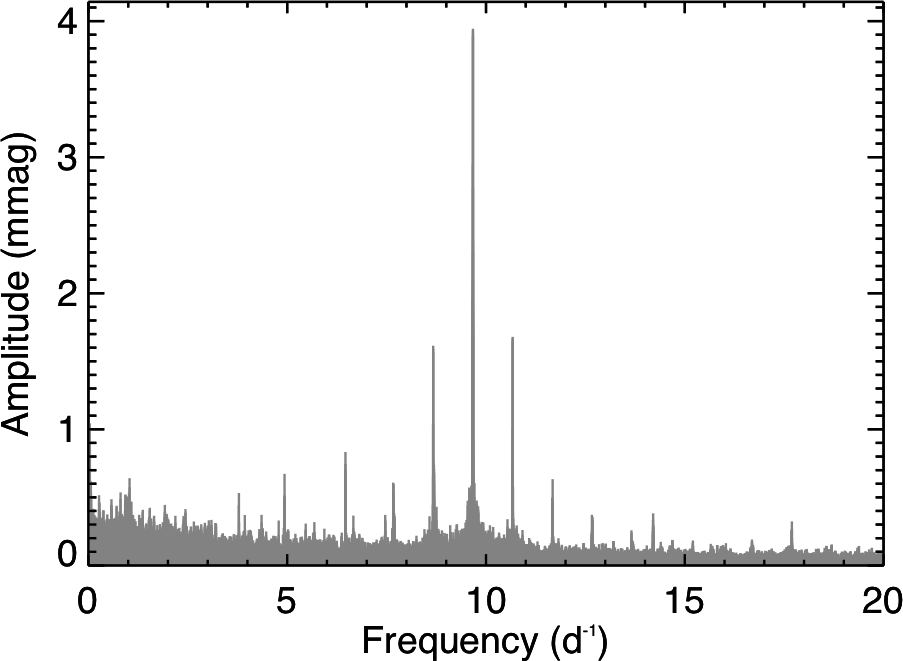}
   \caption{Fourier spectrum of \object{HD~184465} obtained from its four winter lightcurve.}
   \label{fig: HD 184465 Fourier}
\end{figure}

\begin{figure}[htbp]
   \centering
   \includegraphics[width=8cm]{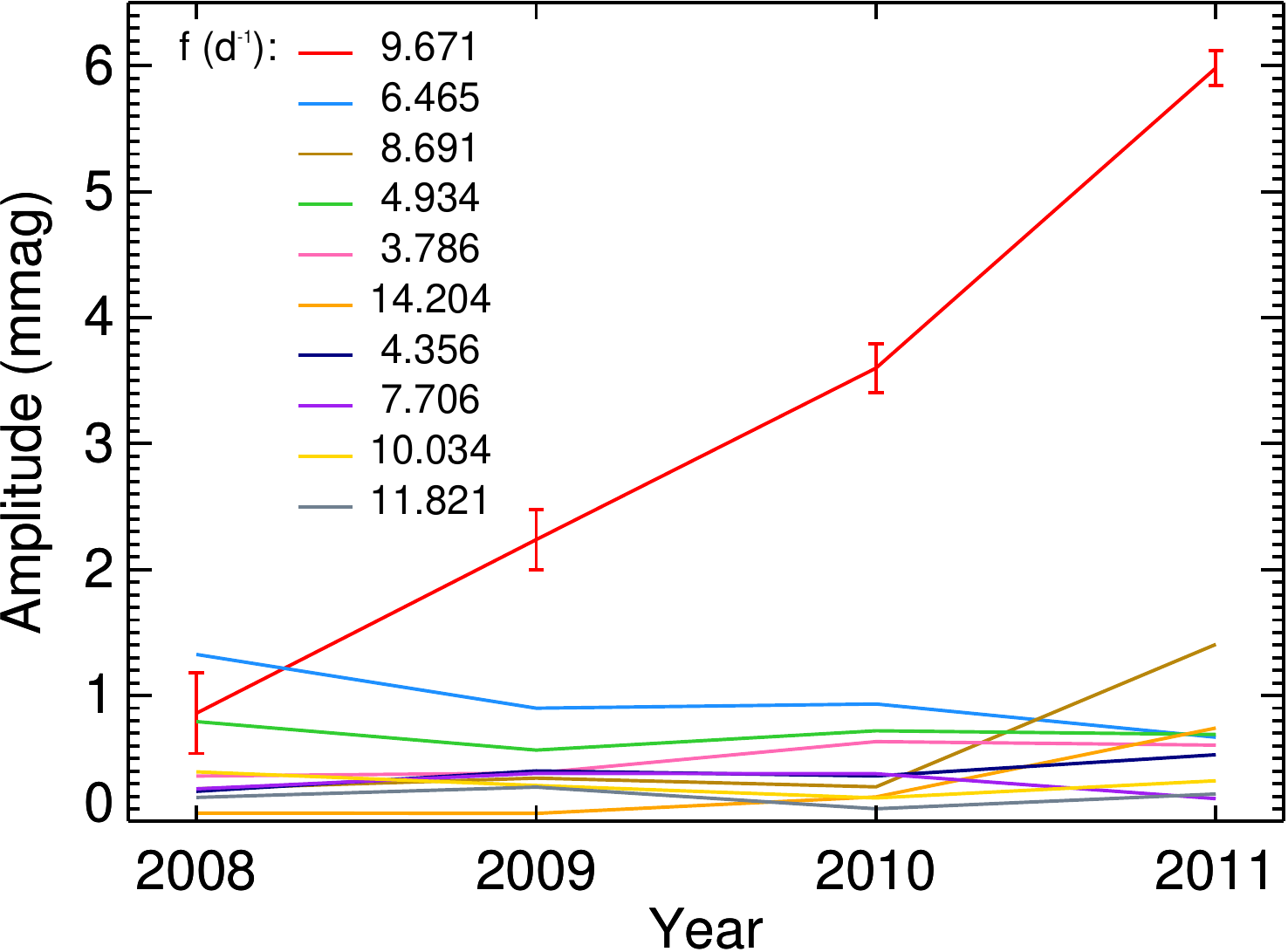}\vspace{3mm}
   \includegraphics[width=8cm]{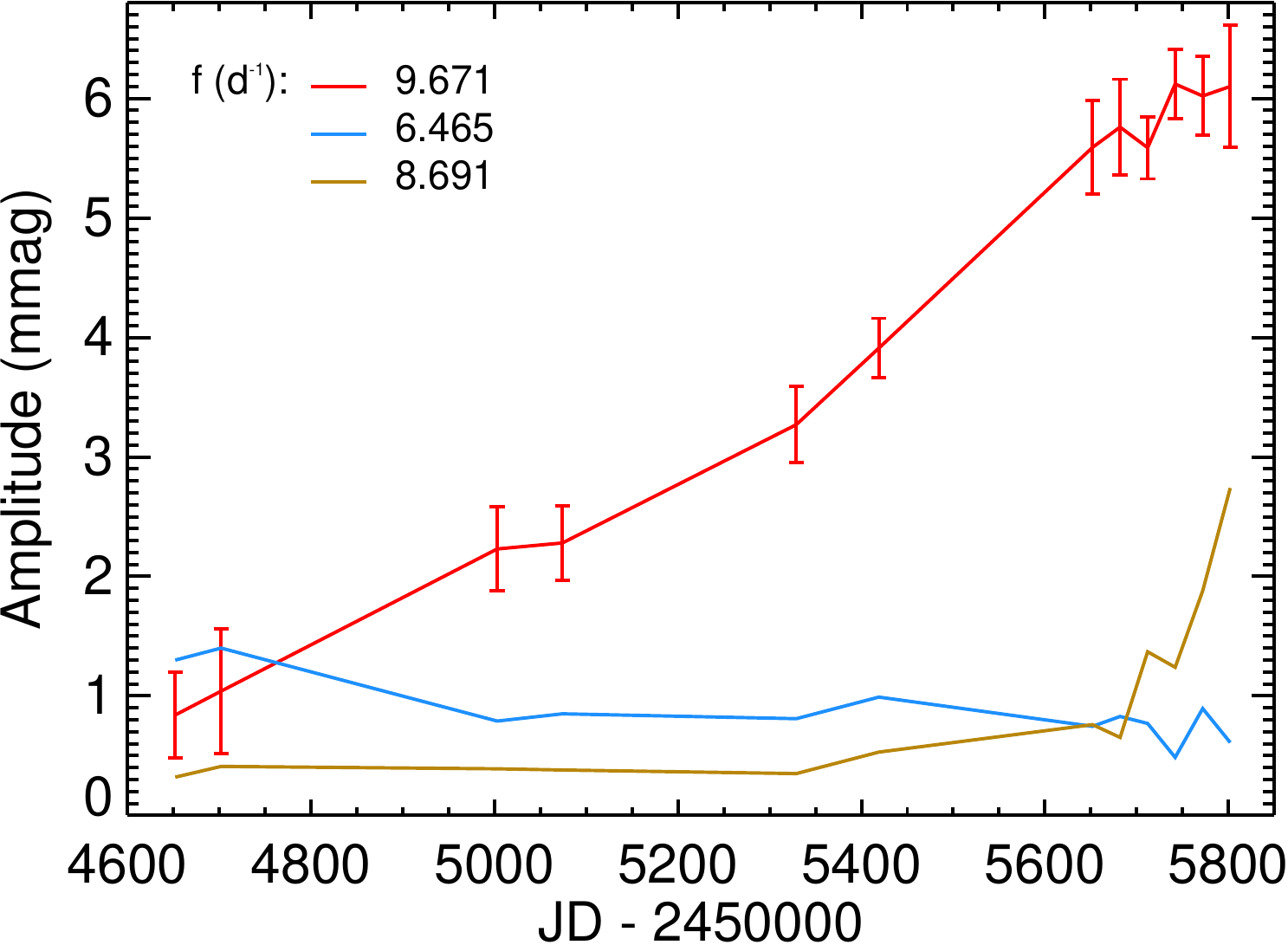}
   \caption{Top: Amplitude variations of the ten $\delta$ Scuti frequencies of \object{HD~184465} with signal to noise ratios greater than four from 2008 to 2011, with one point per winter. Bottom: Amplitude variations of three frequencies with two time intervals for each of the 2008, 2009, and 2010 winters and six 30-day intervals for the 2011 winter. The 3$\sigma$ uncertainties are shown for the main frequency and are the same for all frequencies.}
   \label{fig: HD 184465 amplitudes}
\end{figure}

\object{HD~184465} is a star of $V$ magnitude 8.82 and spectral type A1 V, located in the $\delta$ Scuti instability strip. Its ASTEP South lightcurve and Fourier spectrum are shown in Figs.~\ref{fig: HD 184465 lightcurve} and \ref{fig: HD 184465 Fourier}. We detected ten $\delta$ Scuti-type frequencies with signal to noise ratios greater than four and we analyzed them year by year (Fig.~\ref{fig: HD 184465 amplitudes}, Table~\ref{tab: HD 184465 - 1}). The 9.671~$\rm d^{-1}$ frequency largely dominates the Fourier spectrum, and its amplitude increases from 0.9 to 6.02~mmag (a factor of 6.7) between 2008 and 2011. This increase is not compensated by an equivalent decrease of the amplitudes of other frequencies. The amplitudes of other frequencies are generally too small to detect significant variations between years. We analyzed the 2011 data by 30-day segments (Table~\ref{tab: HD 184465 - 2}, Fig.~\ref{fig: HD 184465 amplitudes}). The amplitudes of the 9.671~$\rm d^{-1}$ and 6.465~$\rm d^{-1}$ frequencies vary at similar rates in 2011 compared to the four winters together. Interestingly, the amplitude of the 8.691~$\rm d^{-1}$ frequency increases in 2011 from 0.7 to 2.7~mmag (a factor of 3.8). Overall, this star does not show energy transfer between oscillations modes, but shows actual variations of the amplitude of its oscillations on several timescales.

\subsubsection{New variable stars and eclipsing binaries}

We identified 34 new variable stars (Table~\ref{tab: variables}). Their amplitudes are typically between 1 and 5~mmag and several have periods longer than ten days. More than half are active stars with variations induced by their rotation. In many cases, the signal to noise ratio is too low to identify the precise type of variable. We also identified eight new eclipsing binaries and reclassified two CSTAR variables as eclipsing binaries (Table~\ref{tab: EBs}). Six of them have orbital periods between 23 and 81 days, which are much longer than those usually detected from the ground. In addition, we refined the period and eclipse depth of the known eclipsing binary \object{EN~Oct} compared to previous studies \citep{Knigge1967, Otero2004}. We also detected variable stars and eclipsing binaries already reported by CSTAR and other catalogs, so we do not report them here. Figure~\ref{fig: lightcurve examples} shows examples of lightcurves of variable stars and eclipsing binaries obtained with ASTEP South. Despite challenges in the photometric reduction due to the unusual observing mode of ASTEP South, these lightcurves are of good quality, and the large amount of data enables the detection of variations over a wide range of magnitudes and periods (in Fig.~\ref{fig: lightcurve examples}, the V magnitudes range from 6.52 to 15.04 and the periods range from 0.12 to 40.53 days). Overall, these detections demonstrate the benefit of the long photometric coverage that can be achieved during the winter at Dome~C.

\subsubsection{\modif{Inventory of variable objects}}

\modif{We gathered the new variable stars and eclipsing binaries detected by ASTEP South with those reported by CSTAR, GCVS, VSX, and ASAS to make an inventory of variable objects in the ASTEP South field of view. Their classification is given in Table~\ref{tab: classification}. Further details on these objects can be found in \citet{Wang2011, Wang2013, Oelkers2015, Yang2015, Wang2015, Samus2017, Watson2006, Pojmanski2002, Richards2012} and in Tables~\ref{tab: variables} and \ref{tab: EBs} of this paper.}

\begin{table}
\begin{center}
\caption{\modif{Inventory of variable objects in the ASTEP South field of view ($ \rm Dec < -88$\dgr) including those detected by ASTEP South and those reported by CSTAR, GCVS, VSX, and ASAS.}}
\label{tab: classification}
\begin{tabular}{lc}
\hline
\hline
Total number of variable objects    &         260     \\
\hline
Variable stars     &    209     \\
\hspace{5mm} VRot    &          96     \\
\hspace{5mm} Ceph    &           1     \\
\hspace{5mm} Delta Scuti    &          19     \\
\hspace{5mm} Gamma Dor    &          10     \\
\hspace{5mm} RR Lyrae    &           4     \\
\hspace{5mm} Mira    &           2     \\
\hspace{5mm} Multiperiodic    &           7     \\
\hspace{5mm} Irregular    &          38     \\
\hspace{5mm} Unclassified    &          32     \\
\hline
Eclipsing binaries     &    47     \\
\hspace{5mm} EA    &          19     \\
\hspace{5mm} EB    &          10     \\
\hspace{5mm} EW    &          18     \\
\hline
Transit-like    &           4     \\
\hline
\hline
\end{tabular}
\end{center}
Notes. ``VRot'' are rotational variables and include the ACV, BY, SP, ELL classes and the ``VRot'' and ``VRot?'' of Table~\ref{tab: variables}. ``Unclassified'' include the ``Var'' of Table~\ref{tab: variables}. ``Transit-like'' objects are those reported by CSTAR; the transit candidates identified by ASTEP South are not reported here and are under investigation. This table does not include the 54 objects initially suspected as variable with ASTEP South that were not confirmed by a finer analysis (see Sect.~\ref{sec: Analysis of variable objects}). 
\end{table}

\subsubsection{A 75.6 day period G--M eclipsing binary}
\label{sec: asud-1184}

Eclipsing binaries are the most common by-products of transiting exoplanet searches. In our search for transiting objects, we detected an eclipsing binary with a 75.6 day orbital period and a 13.6 hour primary eclipse (star ID: \object{UCAC3 2-3056}, fmag: 12.76; see Table~\ref{tab: EBs}). These period and duration are much longer than those usually detected from the ground, which shows the benefit of the exceptional phase coverage that we achieved from Dome~C. This object was not detected by Period04 because of its too low signal to noise ratio but it was among the \modif{transit candidates} that we identified from the BLS search and visual inspection. We conducted follow-up observations in radial velocity and photometry, which confirmed the orbital period and the presence of the eclipses. This system is a G9--M2.5 eclipsing binary with an eccentric orbit ($e = 0.26$) \modif{and a relatively large semi-major axis ($a = 0.42\rm \,au$). Thus, tidal coupling between both stars is weak and the M dwarf should behave as a single star. This object will provide an ideal benchmark to constrain the evolution models of low mass stars, and will be presented in a forthcoming paper \citep{Crouzet-inprep}}.

\section{Conclusion}

We presented the full analysis of four winters of observations collected with ASTEP South, a 10~cm refractor installed at Dome~C, Antarctica. The instrument ran continuously for most of the winters. The functioning of the instrument and the lightcurve quality improved over the years. A main limitation was shutter malfunctions that started after two winters and required specific temperature adjustments and image calibration methods. Another factor that affected the data quality was imperfect thermalisation, in particular during the first half of the 2009 winter, caused by deficient air circulation leading to variable temperature and gradients inside the box. This resulted in unstable PSFs and poor quality lightcurves for these periods. By acquiring knowledge of the instrument's behavior over the years and improving the experiment (hardware and software), we were able to observe longer and longer, from half of the winter in 2008 to the full winter with almost no intervention in 2011. As previously shown for the 2008 winter, we confirmed that the instrument was very stable over the ice and that PSF stability was achieved when the thermalisation was nominal. We also measured the sky background magnitude and its variations over the winters.

We developed a specific data reduction pipeline to take into account the motion and elongated shape of the PSFs on the CCD. We built the lightcurves of nearly 6000 stars over four winters, which are available publicly. The experimental design was particularly sensitive to flat-field errors, which we mitigated by removing the one-sidereal day period variations in the lightcurves. Most of the data are of good quality. The final lightcurve RMS is standard for a 10~cm instrument, and the large quantity of data yields a drastic improvement of the precision when folding the lightcurves at given periods and binning the data points. We measured the photometric quality of Dome~C from the lightcurves themselves using a method that we developed specifically for this work, which may be applied to other experiments.

We detected periodic signal in the lightcurves of around two hundred objects. Half of them were already reported by the CSTAR experiment. We discovered new variable stars and eclipsing binaries, and conducted a detailed study of a few objects that was enabled by the long and continuous coverage of our observations. These results demonstrate the high quality of Dome~C for photometry in the visible and the technical feasibility of running astronomical instruments for several years at this site.

\begin{acknowledgements}

The ASTEP project was funded by the Agence Nationale de la Recherche (ANR), the Institut National des Sciences de l'Univers (INSU), the Programme National de Plan\'etologie (PNP), and the Plan Pluri-Formation OPERA between the Observatoire de la C\^ote d'Azur and the Universit\'e de Nice-Sophia Antipolis. The logistics at Concordia is handled by the French Institut Paul-Emile Victor (IPEV) and the Italian Programma Nazionale di Ricerche in Antartide (PNRA). This work made use of the SIMBAD and VizieR databases operated at CDS, Strasbourg, France, the NASA's Astrophysics Data System bibliographic services, and the NASA's IDL astronomy library. Software: Period04 \citep{Lenz2005}.

\end{acknowledgements}

\appendix

\onecolumn

\noindent\begin{minipage}{\textwidth}

\section{Lightcurve examples}

\vspace{-0.15mm}

  \centering

   \includegraphics[width=3.825cm]{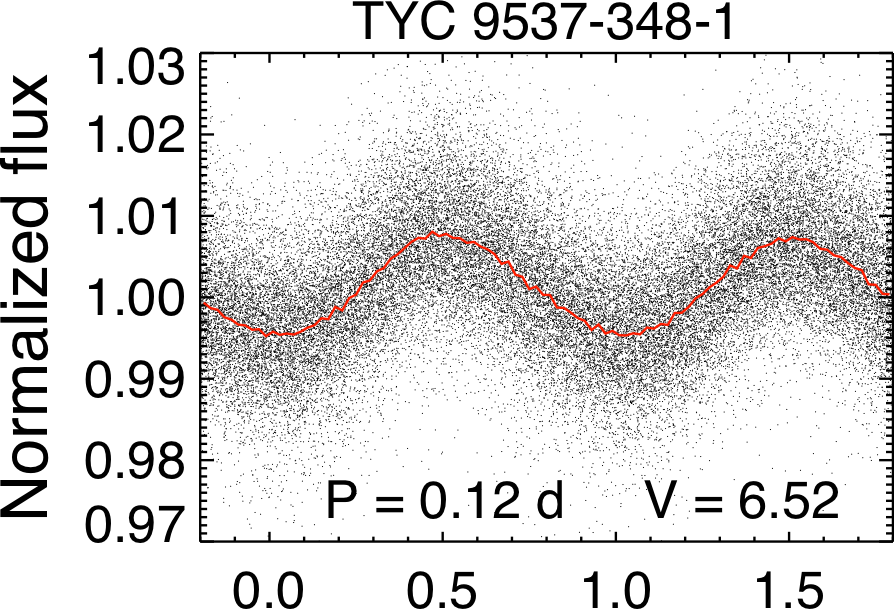}\hspace{2mm}
   \includegraphics[width=3.45cm]{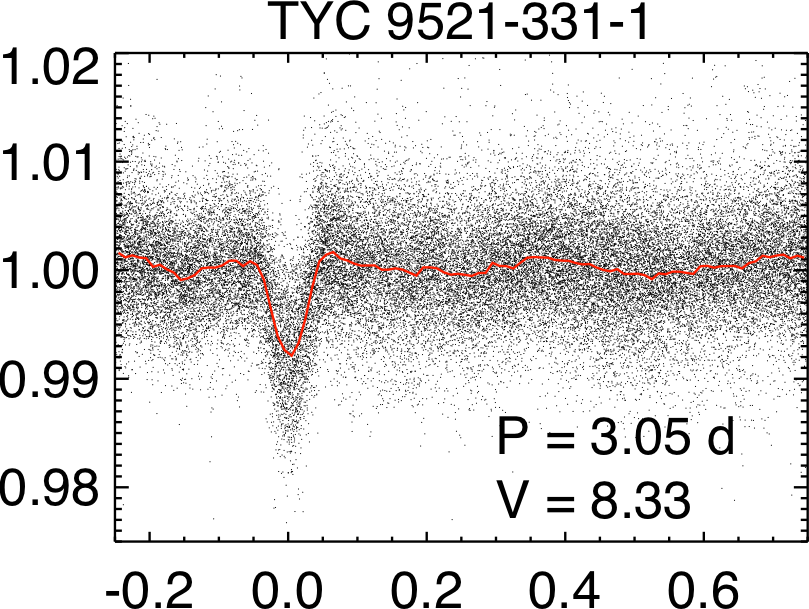}\hspace{2mm}
   \includegraphics[width=3.45cm]{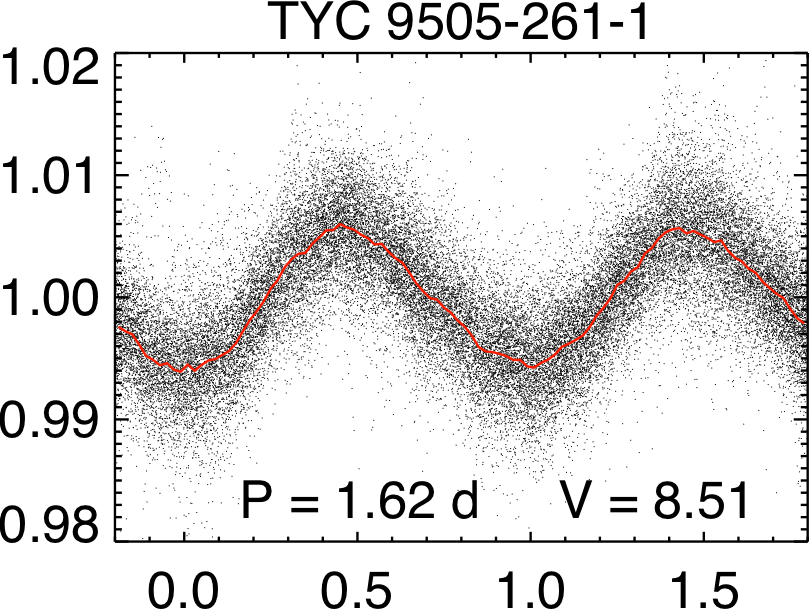}\hspace{2mm}
   \includegraphics[width=3.45cm]{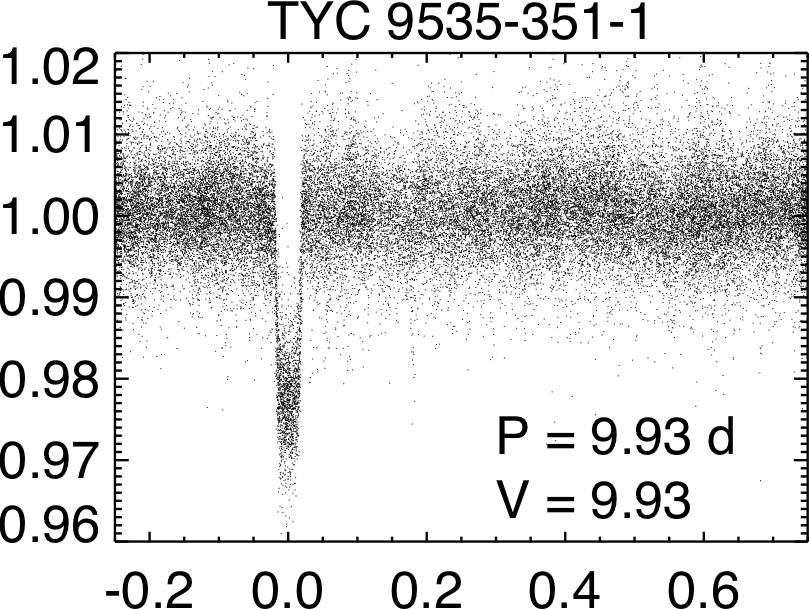}\vspace{0.6mm}
   \includegraphics[width=3.825cm]{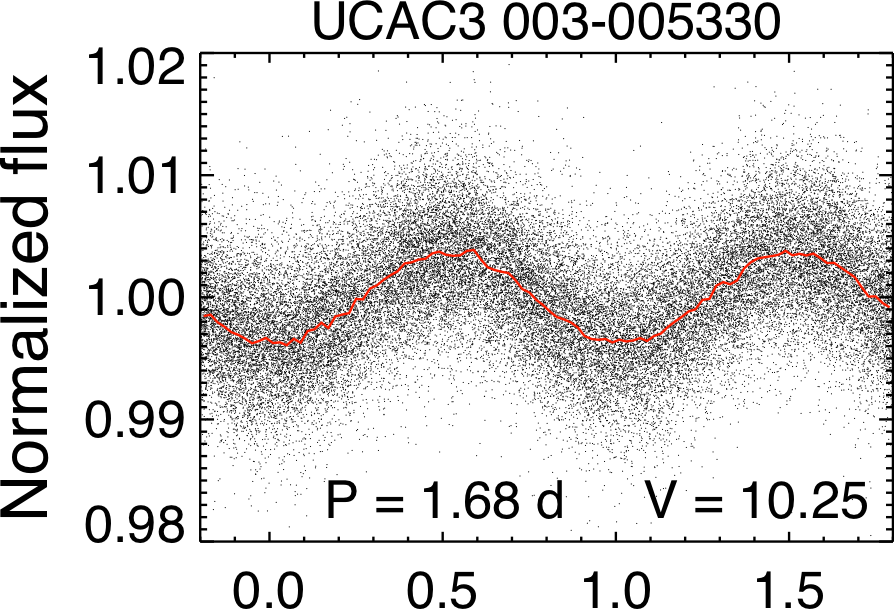}\hspace{2mm}
   \includegraphics[width=3.45cm]{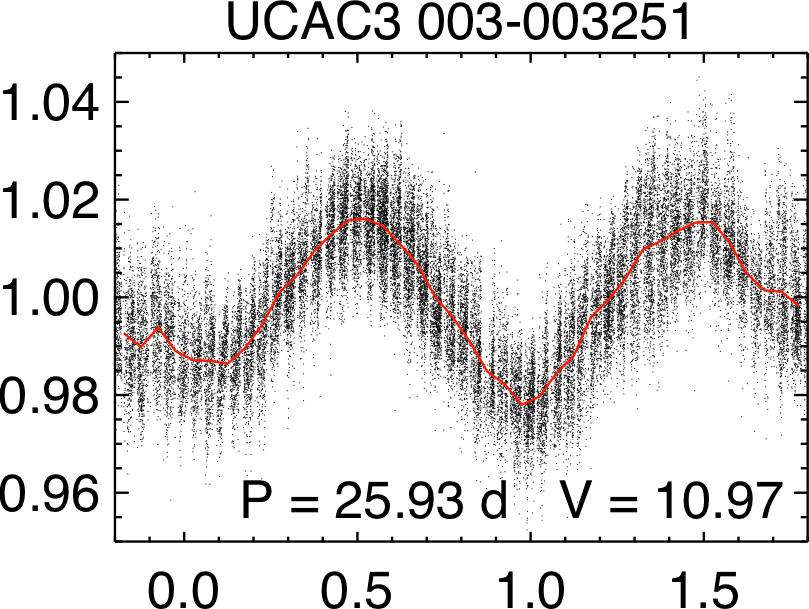}\hspace{2mm}
   \includegraphics[width=3.45cm]{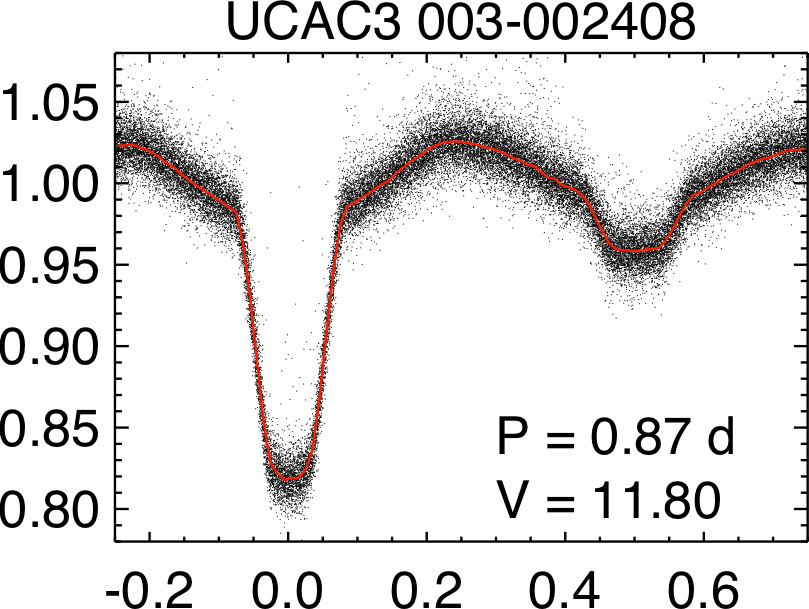}\hspace{2mm}
   \includegraphics[width=3.45cm]{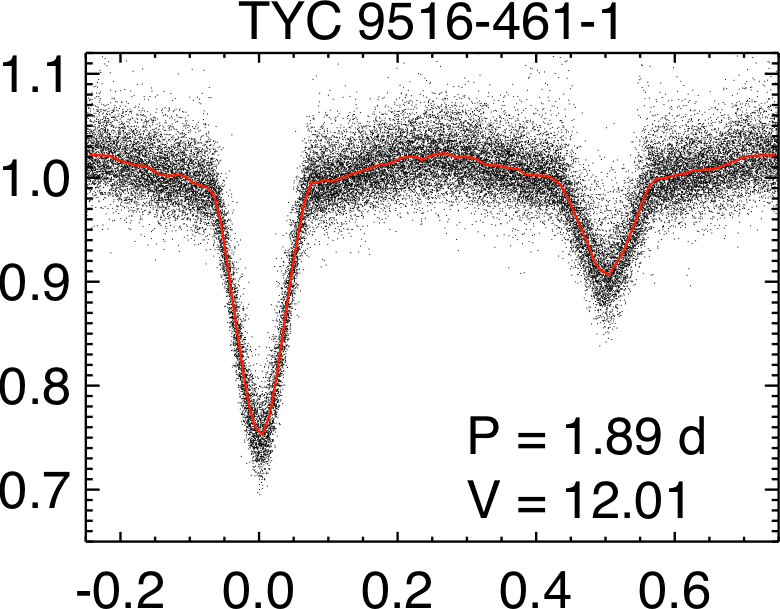}\vspace{0.6mm}
   \includegraphics[width=3.825cm]{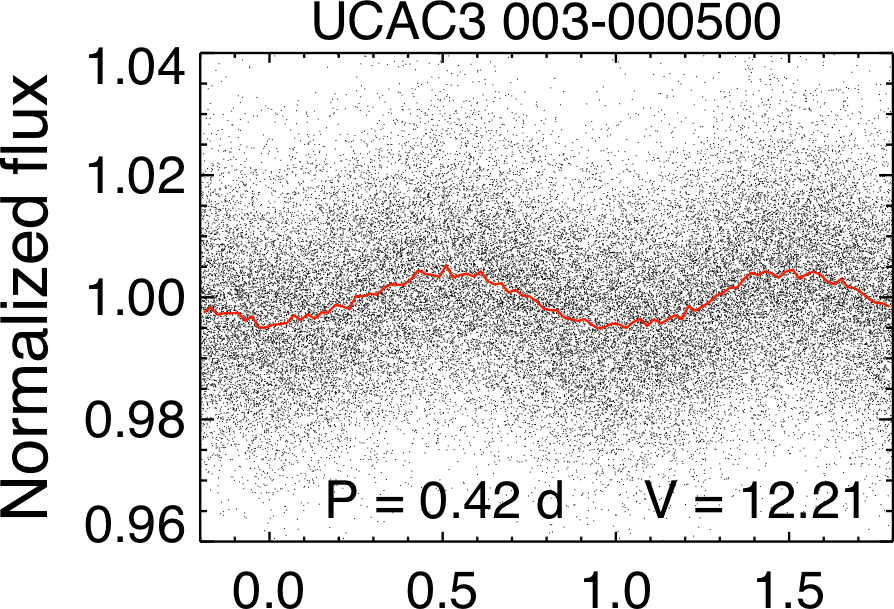}\hspace{2mm}
   \includegraphics[width=3.45cm]{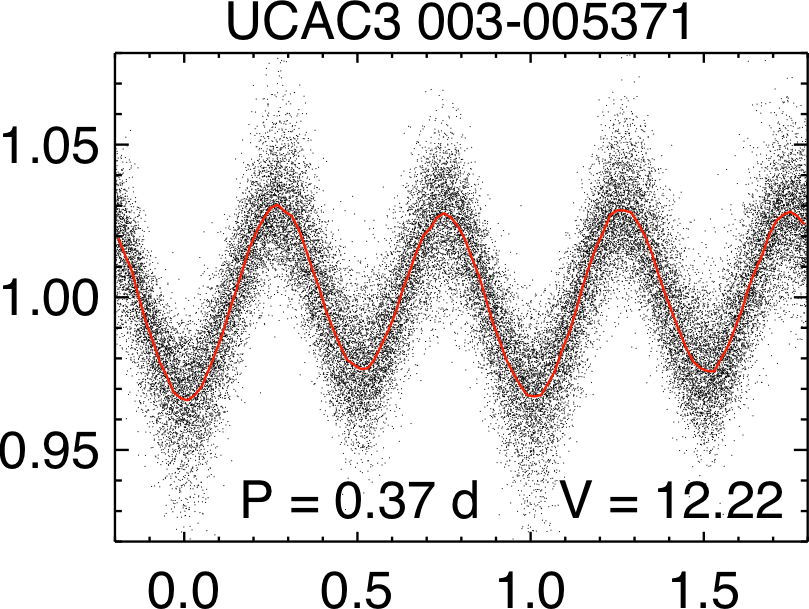}\hspace{2mm}
   \includegraphics[width=3.45cm]{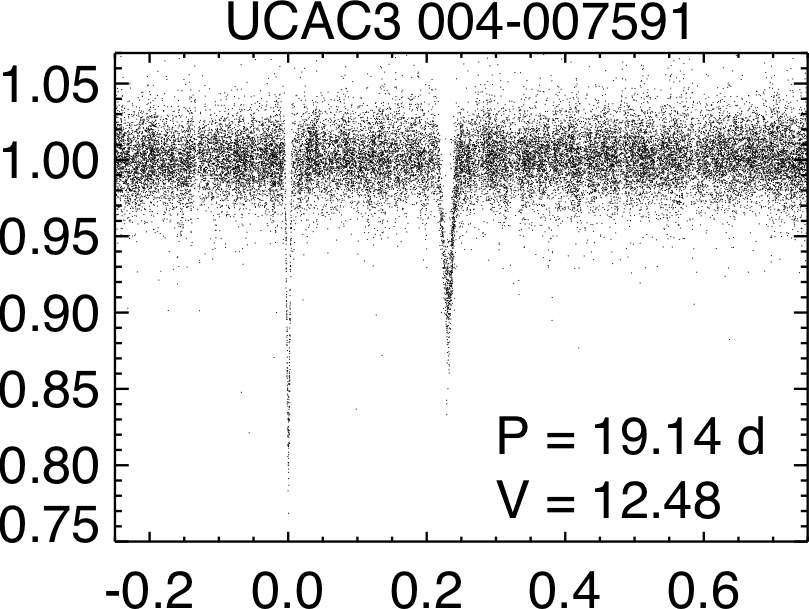}\hspace{2mm}
   \includegraphics[width=3.45cm]{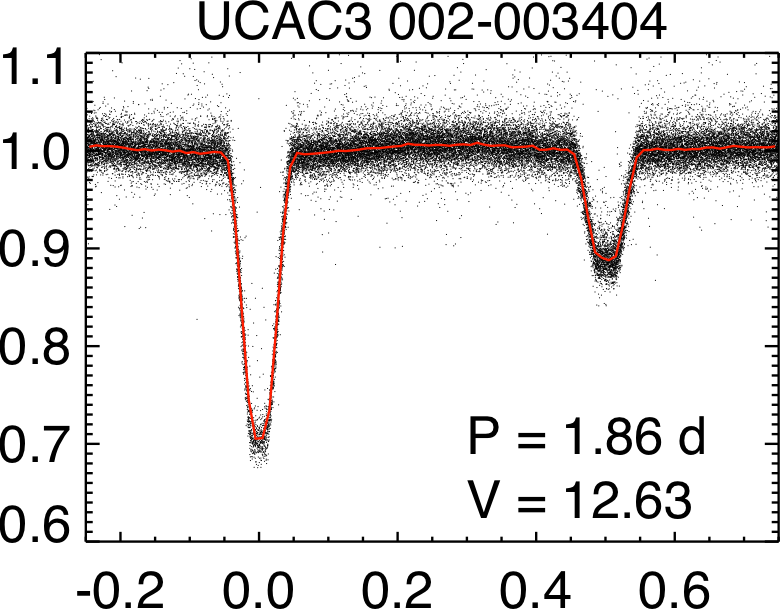}\vspace{0.6mm}
   \includegraphics[width=3.825cm]{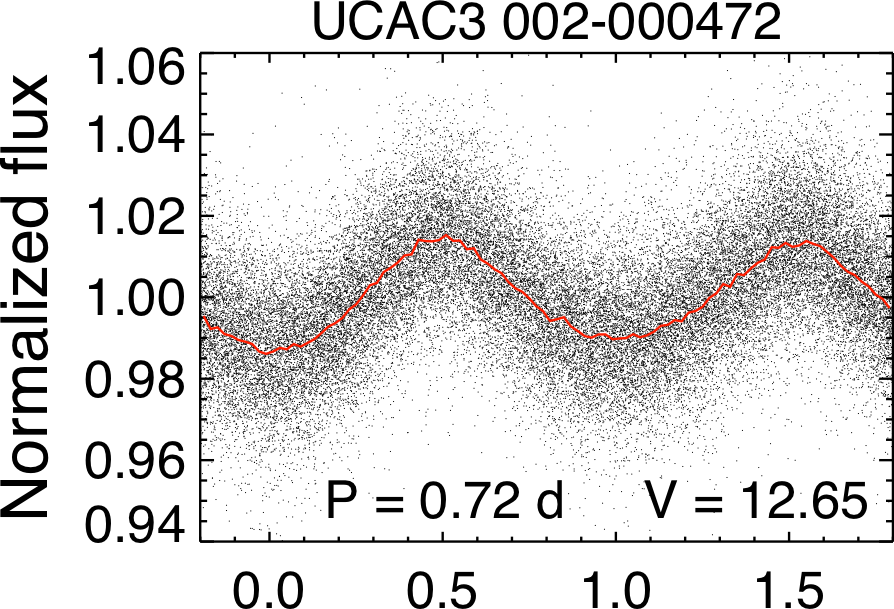}\hspace{2mm}
   \includegraphics[width=3.45cm]{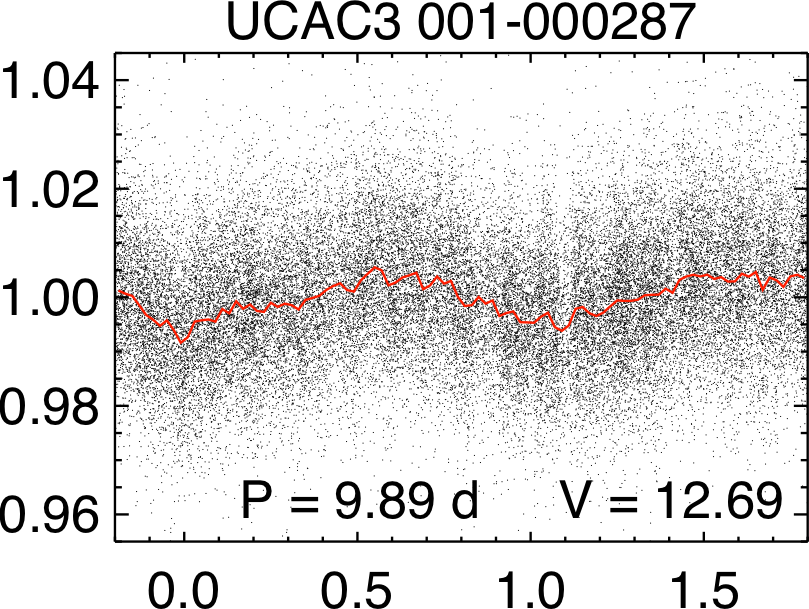}\hspace{2mm}
   \includegraphics[width=3.45cm]{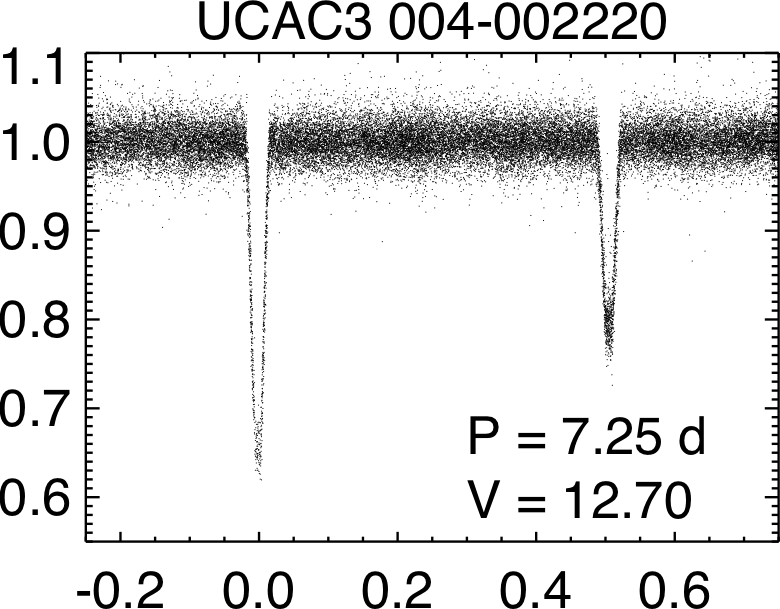}\hspace{2mm}
   \includegraphics[width=3.45cm]{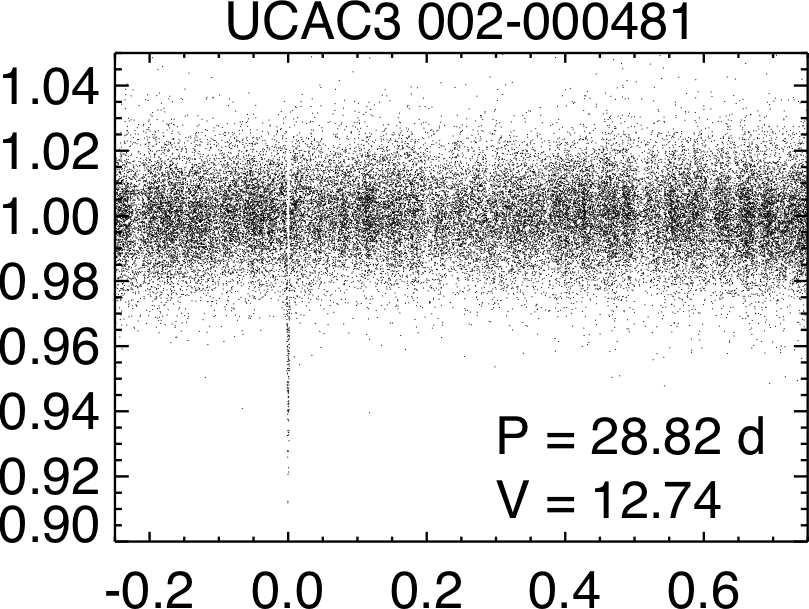}\vspace{0.6mm}
   \includegraphics[width=3.825cm]{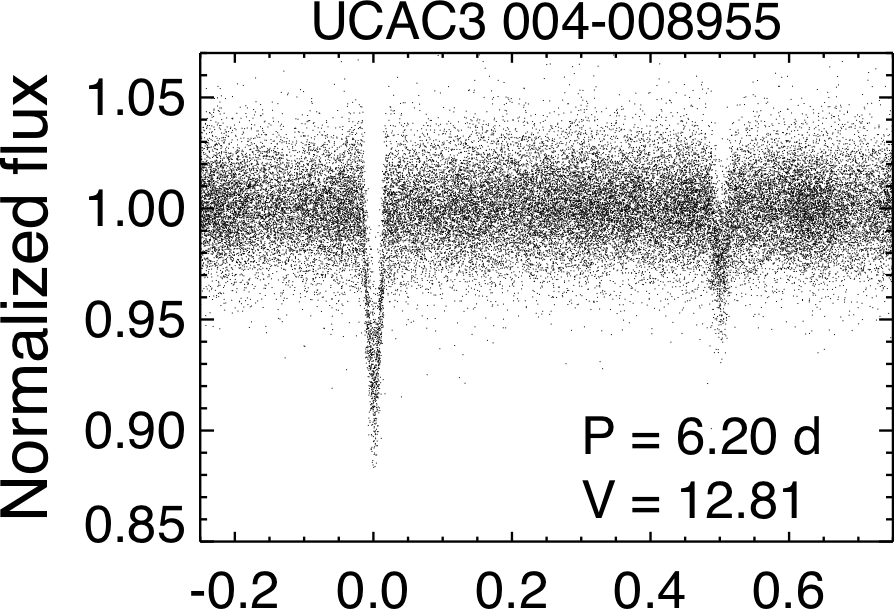}\hspace{2mm}
   \includegraphics[width=3.45cm]{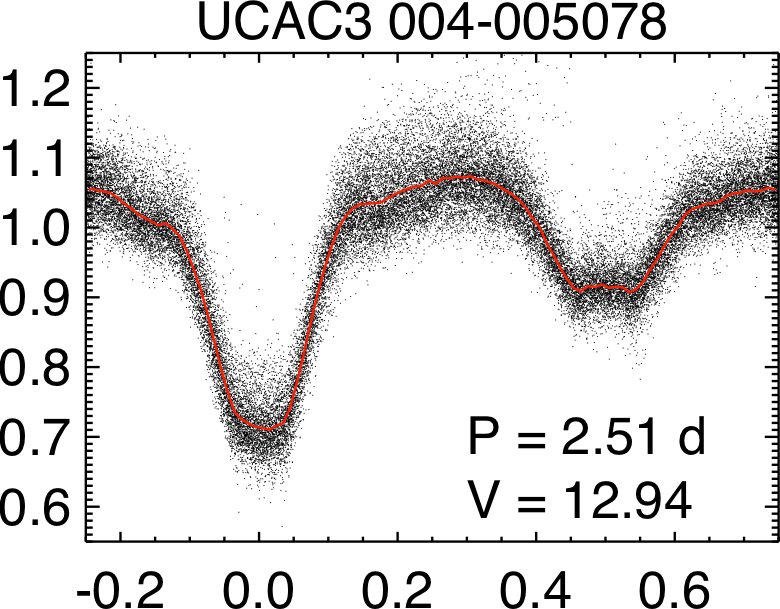}\hspace{2mm}
   \includegraphics[width=3.45cm]{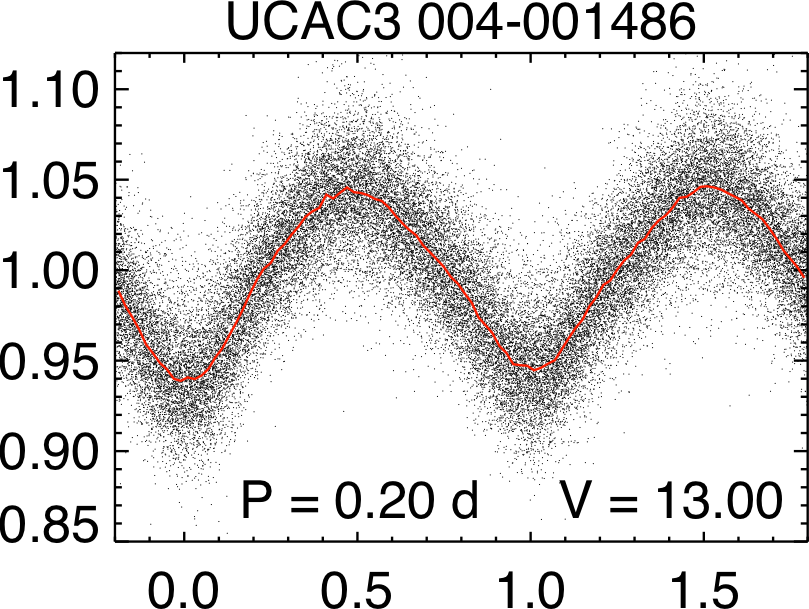}\hspace{2mm}
   \includegraphics[width=3.45cm]{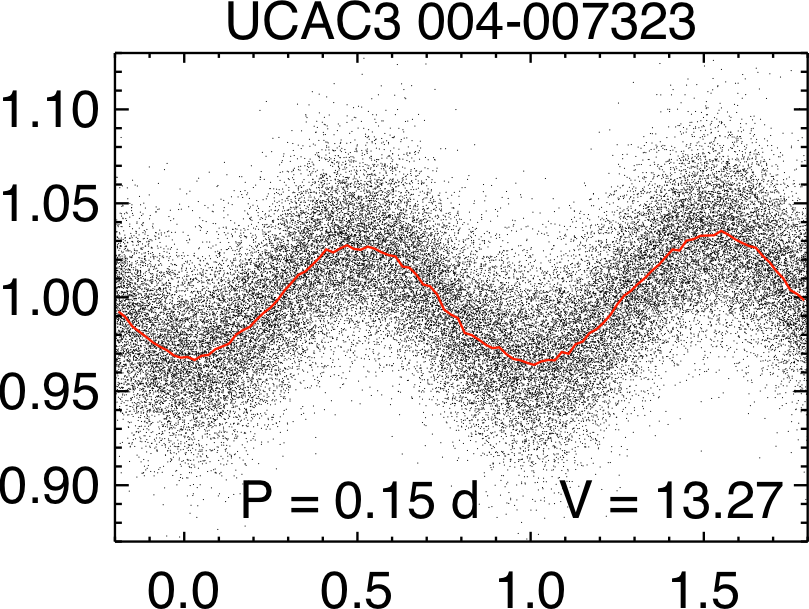}\vspace{0.6mm}
   \includegraphics[width=3.825cm]{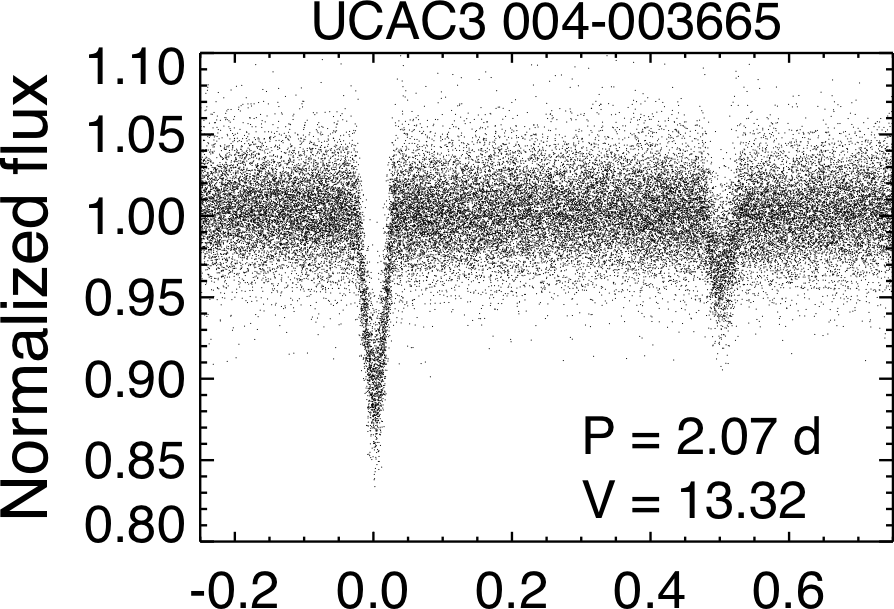}\hspace{2mm}
   \includegraphics[width=3.45cm]{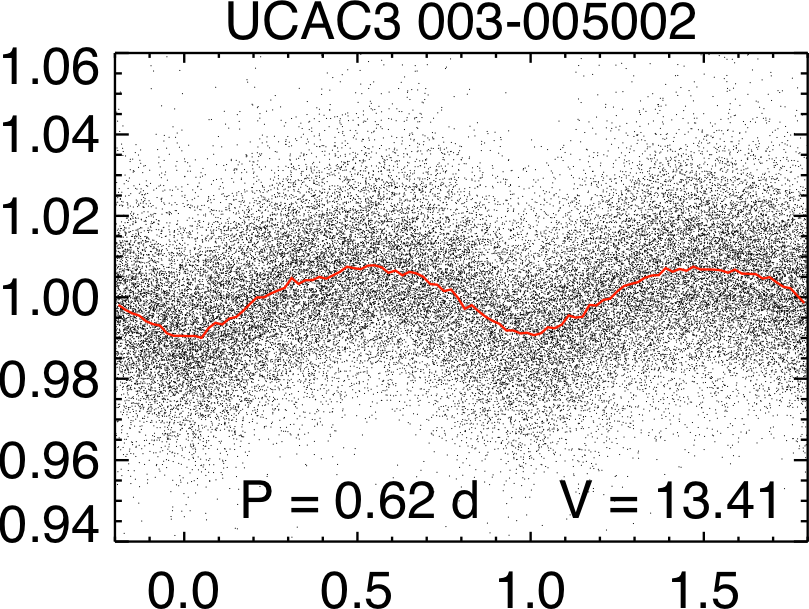}\hspace{2mm}
   \includegraphics[width=3.45cm]{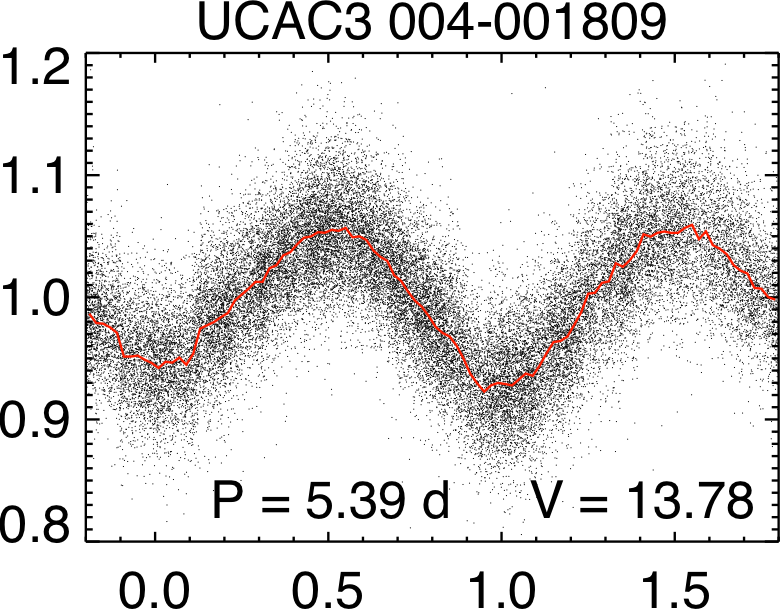}\hspace{2mm}
   \includegraphics[width=3.45cm]{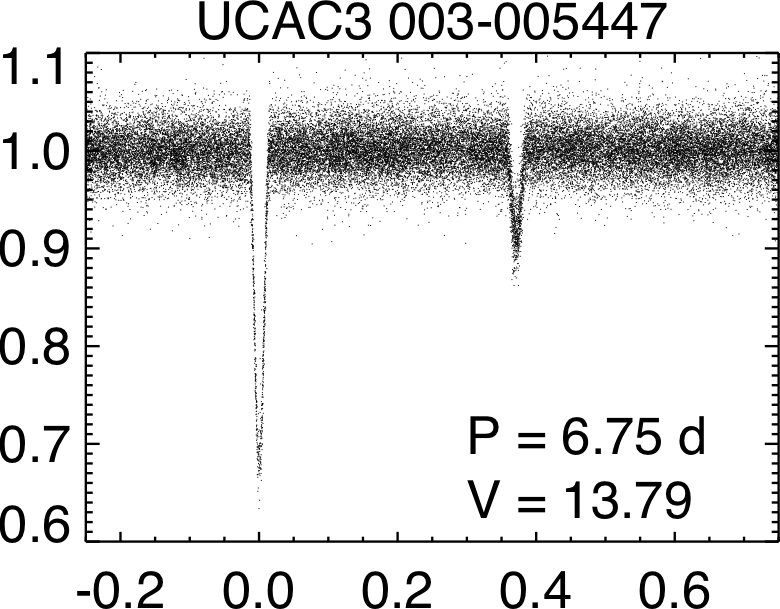}\vspace{0.6mm}
   \includegraphics[width=3.825cm]{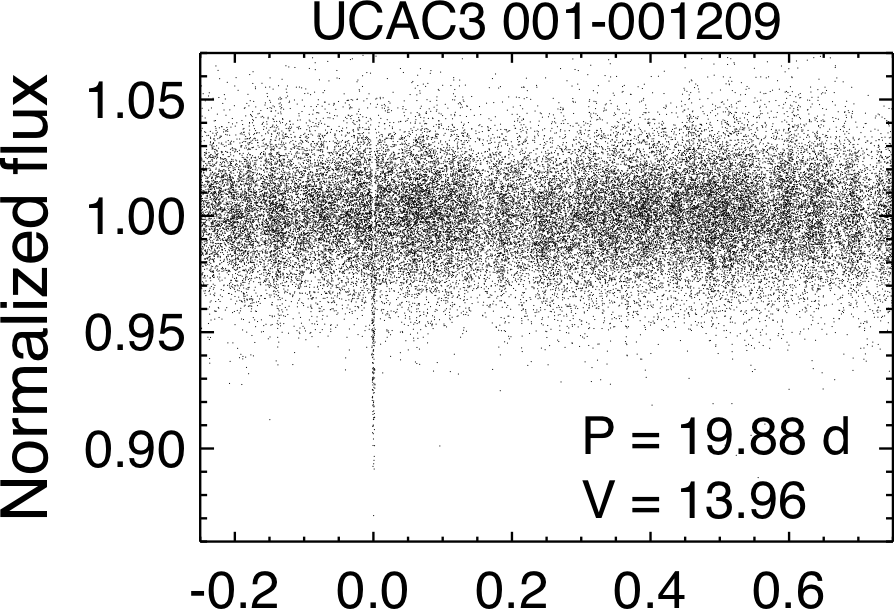}\hspace{2mm}
   \includegraphics[width=3.45cm]{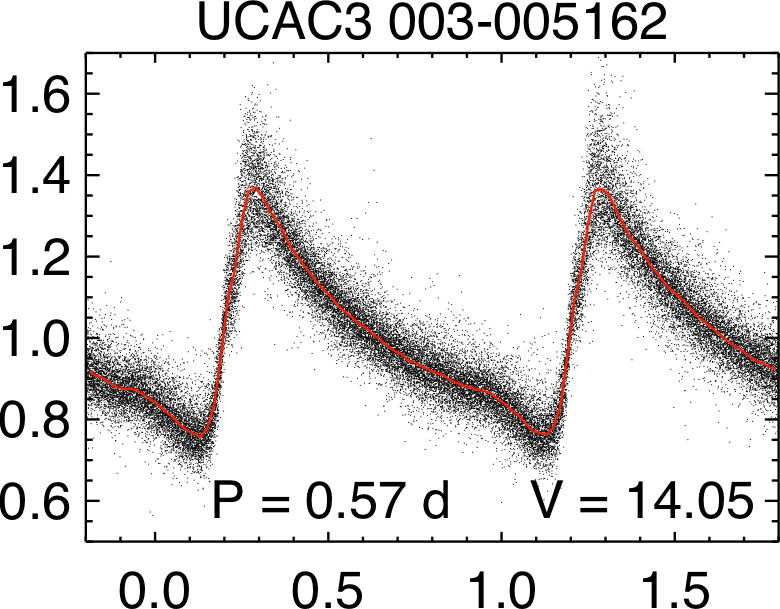}\hspace{2mm}
   \includegraphics[width=3.45cm]{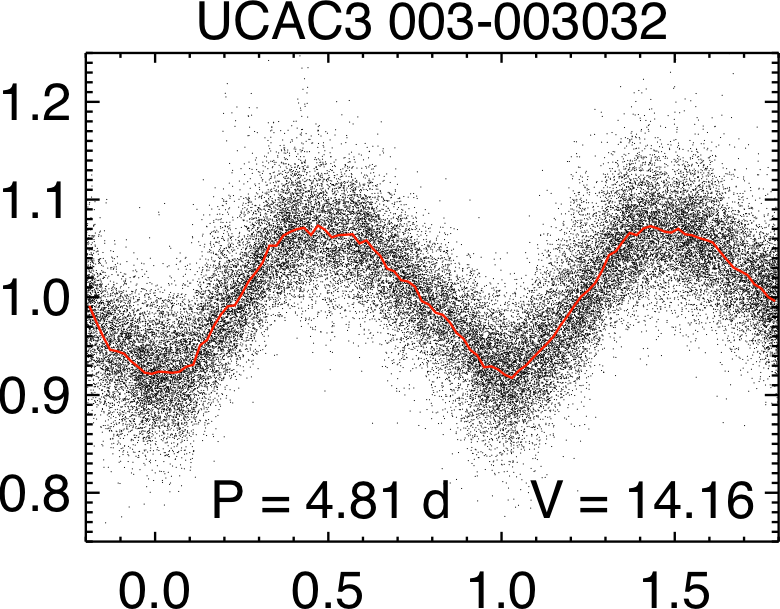}\hspace{2mm}
   \includegraphics[width=3.45cm]{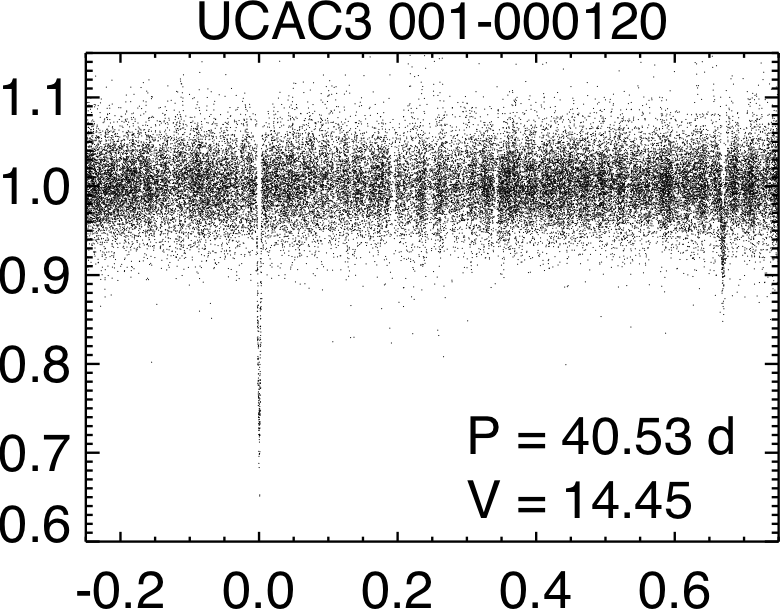}\vspace{0.6mm}
   \includegraphics[width=3.825cm]{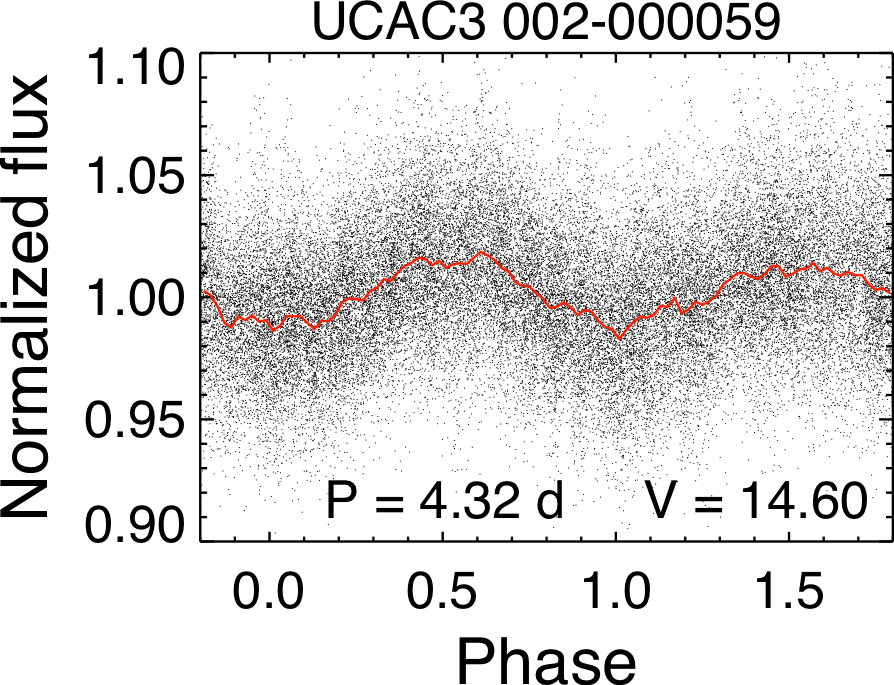}\hspace{2mm}
   \includegraphics[width=3.45cm]{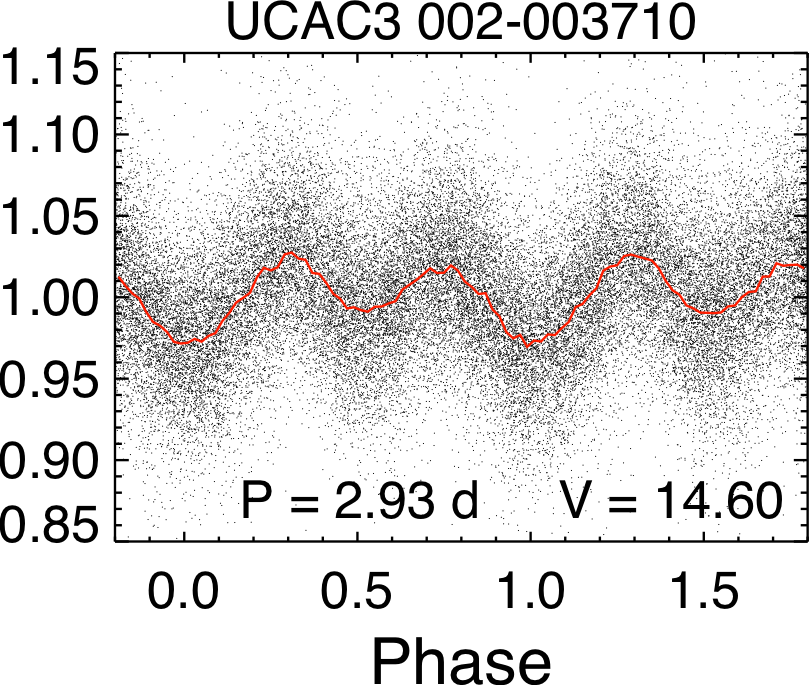}\hspace{2mm}
   \includegraphics[width=3.45cm]{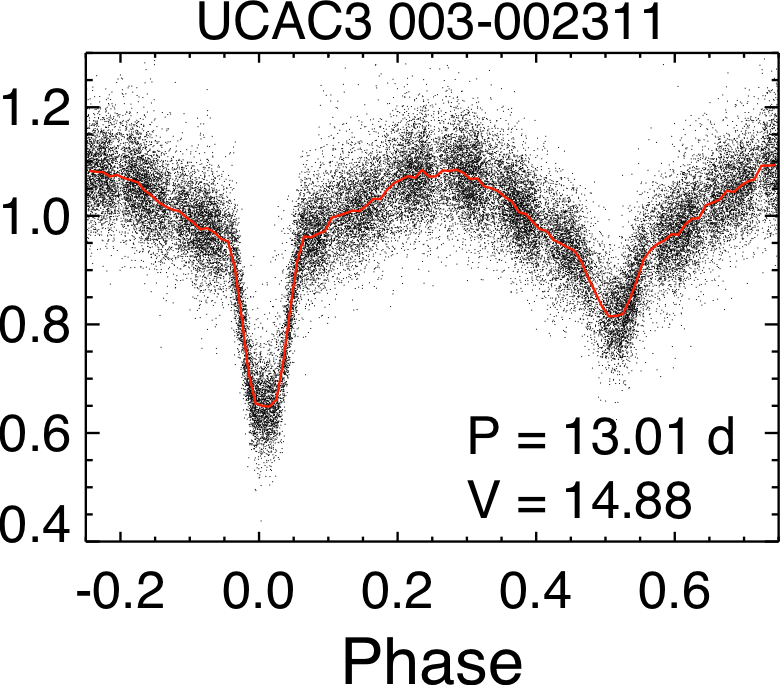}\hspace{2mm}
   \includegraphics[width=3.45cm]{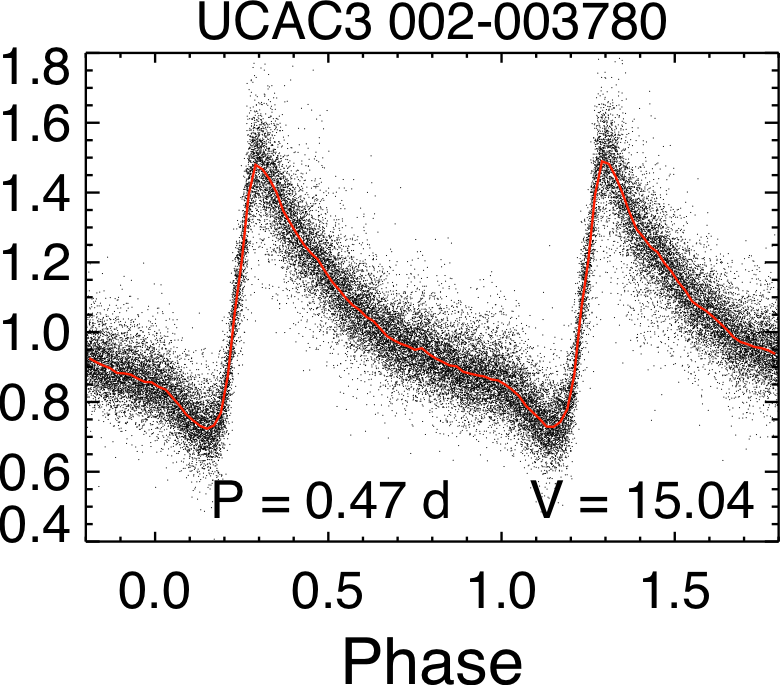}
 
   \captionof{figure}{Examples of phase-folded lightcurves of variable stars and eclipsing binaries obtained with ASTEP South, sorted by magnitude. The black points correspond to a time sampling of ten minutes. A binning with 100 points over the phase range is shown as a red line, except for eclipsing binaries with narrow eclipses. The star ID, the period in days, and the V magnitude (UCAC3 fmag) are indicated.}
   \label{fig: lightcurve examples}
   
\end{minipage}

\section{Analysis of $\delta$ Scuti pulsations of \object{$\sigma$~Oct} and \object{HD~184465}}

\begin{table}[ht]
\begin{center}
\caption{Twenty-one main frequencies detected in the Fourier spectrum of \object{$\sigma$~Oct} with their amplitudes and phases. The reference for the phase is BJD 2454600. Numbers in parentheses indicate the 3$\sigma$ uncertainties on the last two digits.}
\label{tab: sigma Oct 21 freq}
\begin{tabular}{rcc}
\hline
\hline
\multicolumn{1}{c}{Frequency}  &  Amplitude  &  Phase  \\
\multicolumn{1}{c}{[$\rm d^{-1}$]}  &  [mmag]  &  [rad]  \\
\hline
10.492693 (13)  &  3.73 (11)  &  0.449 (30)  \\
 9.719016 (16)  &  3.11 (11)  &  2.972 (36)  \\
10.741796 (18)  &  2.77 (11)  &  2.953 (40)  \\
 8.799398 (19)  &  2.61 (11)  &  4.475 (42)  \\
10.251893 (20)  &  2.47 (11)  &  4.517 (45)  \\
11.429052 (22)  &  2.27 (11)  &  2.818 (49)  \\
 9.361590 (28)  &  1.80 (11)  &  2.375 (61)  \\
 9.771399 (30)  &  1.65 (11)  &  1.183 (67)  \\
10.440784 (37)  &  1.37 (11)  &  6.276 (81)  \\
10.058734 (52)  &  0.96 (11)  &  1.24 (12)  \\
11.756385 (69)  &  0.73 (11)  &  4.06 (15)  \\
 9.139954 (75)  &  0.67 (11)  &  1.88 (17)  \\
 0.960644 (81)  &  0.62 (11)  &  3.39 (18)  \\
 0.611525 (82)  &  0.61 (11)  &  3.31 (18)  \\
 0.898446 (85)  &  0.59 (11)  &  4.89 (19)  \\
11.189760 (93)  &  0.54 (11)  &  5.06 (21)  \\
 2.848108 (99)  &  0.50 (11)  &  2.09 (22)  \\
11.26188 (10)  &  0.49 (11)  &  4.08 (23)  \\
 8.23509 (11)  &  0.47 (11)  &  5.52 (24)  \\
 8.02296 (14)  &  0.35 (11)  &  2.15 (32)  \\
11.02859 (16)  &  0.32 (11)  &  5.41 (35)  \\
\hline
\hline
\end{tabular}
\end{center}
\end{table}

\begin{table}[ht]
\begin{center}
\caption{Amplitude of the ten main $\delta$ Scuti frequencies found in the Fourier spectrum of \object{$\sigma$~Oct} for each winter and for the four winters together. Numbers in parentheses indicate the 3$\sigma$ uncertainties on the last two digits. The amplitudes are in mmag.}
\label{tab: sigma Oct}
\begin{tabular}{r|ccccc}
\hline
\hline
Frequency $\rm [d^{-1}]$ & 2008 & 2009 & 2010 & 2011 & All \\
\hline
10.492693 (13)  &  9.69 (32)  &  5.42 (26)  &  3.01 (22)  &  0.99 (16)  &  3.73 (11)  \\
 9.719016 (16)  &  2.58 (32)  &  2.78 (26)  &  3.07 (22)  &  3.32 (16)  &  3.11 (11)  \\
10.741796 (18)  &  2.84 (32)  &  2.74 (26)  &  2.67 (22)  &  2.79 (16)  &  2.77 (11)  \\
 8.799398 (19)  &  2.91 (32)  &  2.56 (26)  &  2.56 (22)  &  2.52 (16)  &  2.61 (11)  \\
10.251893 (20)  &  2.29 (32)  &  2.14 (26)  &  2.27 (22)  &  2.58 (16)  &  2.47 (11)  \\
11.429052 (22)  &  2.39 (32)  &  2.24 (26)  &  2.12 (22)  &  2.53 (16)  &  2.27 (11)  \\
 9.361590 (28)  &  2.99 (32)  &  1.67 (26)  &  1.51 (22)  &  1.55 (16)  &  1.80 (11)  \\
 9.771399 (30)  &  1.56 (32)  &  1.55 (26)  &  1.82 (22)  &  1.75 (16)  &  1.65 (11)  \\
10.440784 (37)  &  1.21 (32)  &  1.24 (26)  &  1.36 (22)  &  1.72 (16)  &  1.37 (11)  \\
10.058734 (52)  &  1.38 (32)  &  1.31 (26)  &  0.93 (22)  &  0.77 (16)  &  0.96 (11)  \\
\hline
\hline
\end{tabular}
\end{center}
\end{table}

\begin{table}[ht]
\begin{center}
\caption{Amplitude of the ten $\delta$ Scuti frequencies with signal to noise ratios greater than four found in the Fourier spectrum of \object{HD~184465} for each winter and for the four winters together. Numbers in parentheses indicate the 3$\sigma$ uncertainties on the last two digits. The amplitudes are in mmag.}
\label{tab: HD 184465 - 1}
\begin{tabular}{r|ccccc}
\hline
\hline
Frequency [$\rm d^{-1}$] & 2008 & 2009 & 2010 & 2011 & All \\
\hline
 9.671458 (12)  &  0.86 (32)  &  2.24 (24)  &  3.60 (19)  &  5.98 (14)  &  3.94 (10)  \\
 6.465094 (56)  &  1.33 (32)  &  0.90 (24)  &  0.93 (19)  &  0.67 (14)  &  0.85 (10)  \\
 8.691272 (64)  &  0.26 (32)  &  0.34 (24)  &  0.28 (19)  &  1.41 (14)  &  0.74 (10)  \\
 4.934047 (69)  &  0.79 (32)  &  0.57 (24)  &  0.72 (19)  &  0.69 (14)  &  0.68 (10)  \\
 3.785687 (87)  &  0.36 (32)  &  0.38 (24)  &  0.63 (19)  &  0.61 (14)  &  0.55 (10)  \\
14.20371 (12)  &  0.07 (32)  &  0.06 (24)  &  0.19 (19)  &  0.74 (14)  &  0.40 (10)  \\
 4.35554 (13)  &  0.24 (32)  &  0.40 (24)  &  0.36 (19)  &  0.53 (14)  &  0.37 (10)  \\
 7.70589 (18)  &  0.26 (32)  &  0.38 (24)  &  0.38 (19)  &  0.18 (14)  &  0.26 (10)  \\
10.03363 (25)  &  0.39 (32)  &  0.29 (24)  &  0.19 (19)  &  0.32 (14)  &  0.19 (10)  \\
11.82126 (26)  &  0.19 (32)  &  0.27 (24)  &  0.10 (19)  &  0.22 (14)  &  0.19 (10)  \\
\hline
\hline
\end{tabular}
\end{center}
\end{table}

\begin{table}[ht]
\begin{center}
\caption{Amplitude variations of three $\delta$ Scuti frequencies of \object{HD~184465} during the four winters, with two time intervals for each of the 2008, 2009, and 2010 winters and six 30-day intervals for the 2011 winter. Dates are the central time of each interval. Numbers in parentheses indicate the 3$\sigma$ uncertainties on the last two digits for the amplitudes (they are the same for all amplitudes at a given time interval). The amplitudes are in mmag.}
\label{tab: HD 184465 - 2}
\begin{tabular}{c|cccc}
\hline
\hline
JD - 2450000  & \multicolumn{3}{c}{Frequency [$\rm d^{-1}$]} &  \\
      &  9.671    &  8.692    &  6.465  &   \\
\hline
4652  &  0.84  &  0.32  &  1.30  &  (36) \\
4702  &  1.04  &  0.41  &  1.40  &  (52) \\
5003  &  2.23  &  0.39  &  0.79  &  (35) \\
5074  &  2.28  &  0.38  &  0.85  &  (31) \\
5329  &  3.27  &  0.35  &  0.81  &  (32) \\
5419  &  3.91  &  0.53  &  0.99  &  (25) \\
5652  &  5.59  &  0.76  &  0.75  &  (39) \\
5682  &  5.76  &  0.65  &  0.83  &  (40) \\
5712  &  5.59  &  1.37  &  0.77  &  (26) \\
5742  &  6.12  &  1.24  &  0.49  &  (29) \\
5772  &  6.02  &  1.88  &  0.89  &  (33) \\
5802  &  6.10  &  2.74  &  0.61  &  (51) \\
\hline
\hline
\end{tabular}
\end{center}
\end{table}

\clearpage
   
\section{New variables stars and eclipsing binaries identified with ASTEP South}

\begin{table}[ht]
\begin{center}
\caption{New variable stars identified with ASTEP South. The ID, coordinates, and magnitudes fmag are extracted from the UCAC3 catalog. For each object, we report the frequency, the corresponding period, the semi-amplitude of the variations, and the type of variable. Numbers in parentheses indicate the 3$\sigma$ uncertainties on the last digit for the frequencies, and the 3$\sigma$ uncertainties for the amplitudes. ``Var'' indicates that the type of variable cannot be determined and ``?'' indicates a tentative classification.}
\label{tab: variables}
\begin{tabular}{ccccrrrc}
\hline
\hline
ID & RA & Dec & fmag & Frequency & Period & Amplitude & Type  \\
(UCAC3) & (J2000) & (J2000) & [mag] & [$\rm d^{-1}$] & [d] & [mmag] &   \\
\hline
 \object{004-005139} & 13:31:52.604 & -88:22:17.75 & 10.23 & 0.14252 (3) &  7.02 &  2.3 (0.1) & Var \\
 \object{003-002049} & 07:54:08.982 & -88:31:04.01 & 10.91 & 5.83351 (6) &  0.17 &  1.1 (0.1) & $\delta$ Scuti \\
 \object{003-004378} & 15:31:11.257 & -88:31:17.40 & 10.93 & 0.10227 (3) &  9.78 &  3.7 (0.2) & VRot? \\
 \object{004-006257} & 16:13:29.743 & -88:12:25.00 & 11.82 & 0.08001 (2) & 12.50 &  4.5 (0.2) & Var \\
 \object{002-000843} & 05:12:39.072 & -89:10:02.73 & 12.18 & 0.13586 (3) &  7.36 &  3.0 (0.5) & VRot \\
 \object{002-000214} & 01:15:41.760 & -89:00:19.72 & 12.21 & 0.02587 (2) & 38.65 &  6.1 (0.2) & VRot \\
 \object{002-000853} & 05:17:21.064 & -89:21:19.78 & 12.25 & 1.05540 (6) &  0.95 &  1.6 (0.2) & VRot? \\
 \object{004-000689} & 01:57:52.063 & -88:13:15.56 & 12.55 & 0.12161 (4) &  8.22 &  3.8 (0.3) & VRot? \\
 \object{003-002304} & 08:39:17.699 & -88:44:20.12 & 12.59 & 0.08060 (2) & 12.41 &  5.3 (0.3) & Var \\
 \object{003-004226} & 15:05:46.232 & -88:39:23.33 & 12.67 & 0.02679 (3) & 37.33 &  3.9 (0.3) & VRot \\
 \object{003-005314} & 19:03:58.030 & -88:59:04.65 & 12.69 & 0.02492 (3) & 40.13 &  3.4 (0.2) & VRot? \\
 \object{001-000287} & 05:43:45.773 & -89:33:57.57 & 12.69 & 0.10116 (3) &  9.89 &  4.5 (0.2) & Var \\
 \object{001-000176} & 03:35:19.894 & -89:41:08.84 & 12.80 & 0.04475 (3) & 22.35 &  4.0 (0.2) & Var \\
 \object{003-001672} & 06:30:27.586 & -88:59:11.08 & 12.83 & 0.04459 (2) & 22.43 &  5.6 (0.2) & Var \\
 \object{003-003813} & 13:42:59.863 & -88:53:36.27 & 12.96 & 0.07838 (2) & 12.76 &  6.7 (0.3) & VRot \\
 \object{003-000054} & 00:16:06.515 & -88:34:08.37 & 12.98 & 0.01286 (1) & 77.76 &  2.1 (0.5) & VRot \\
 \object{002-000558} & 03:32:17.485 & -89:15:06.15 & 13.51 & 0.09741 (4) & 10.27 &  4.6 (0.4) & VRot? \\
 \object{003-001584} & 06:12:58.011 & -88:35:13.08 & 13.52 & 0.12314 (4) &  8.12 &  6.5 (0.5) & VRot? \\
 \object{003-000486} & 01:59:49.979 & -88:54:05.91 & 13.64 & 0.02878 (2) & 34.75 & 13.3 (0.6) & VRot \\
 \object{002-002489} & 14:58:37.642 & -89:07:37.09 & 13.81 & 0.04184 (2) & 23.90 & 18.5 (0.6) & VRot \\
 \object{003-004679} & 16:35:40.811 & -88:43:46.12 & 13.83 & 0.02809 (5) & 35.60 & 88.9 (1.0) & VRot \\
 \object{003-004666} & 16:32:18.171 & -88:36:33.72 & 14.21 & 0.02950 (3) & 33.90 & 18.4 (1.1) & VRot \\
 \object{004-004408} & 11:55:09.767 & -88:23:58.50 & 14.32 & 0.23194 (4) &  4.31 &  7.7 (0.7) & VRot? \\
 \object{004-004603} & 12:17:46.280 & -88:05:49.99 & 14.44 & 1.03020 (5) &  0.97 &  9.6 (1.1) & Var \\
 \object{004-007265} & 19:04:41.581 & -88:29:17.88 & 14.46 & 2.11961 (4) &  0.47 &  5.4 (0.5) & Var \\
 \object{003-003146} & 11:28:48.149 & -88:50:41.89 & 14.48 & 0.42148 (4) &  2.37 & 11.5 (1.0) & Var \\
 \object{002-001119} & 06:54:02.979 & -89:05:10.41 & 14.54 & 5.89217 (4) &  0.17 & 10.8 (1.0) & Var \\
 \object{003-006325} & 23:21:41.025 & -88:50:24.29 & 14.91 & 0.64068 (2) &  1.56 & 35.6 (1.4) & Var \\
 \object{001-000349} & 06:48:03.486 & -89:32:05.52 & 14.99 & 5.64772 (6) &  0.18 & 97.0 (1.3) & $\delta$ Scuti \\
 \object{001-000856} & 15:36:30.014 & -89:47:17.82 & 15.05 & 0.80258 (3) &  1.25 & 18.9 (1.2) & VRot \\
 \object{002-001114} & 06:52:18.556 & -89:01:33.13 & 15.09 & 0.12765 (2) &  7.83 & 14.9 (0.6) & Var \\
 \object{002-003252} & 19:53:39.345 & -89:01:49.63 & 15.20 & 1.82823 (2) &  0.55 & 43.0 (1.4) & VRot \\
 \object{002-001915} & 11:32:03.119 & -89:04:57.12 & 15.21 & 0.95488 (2) &  1.05 & 28.6 (1.4) & Var \\
 \object{002-001216} & 07:28:33.245 & -89:03:48.27 & 15.69 & 0.77189 (3) &  1.30 & 14.1 (0.9) & Var \\
\hline
\hline
\end{tabular}
\end{center}
\end{table}

\begin{table}[ht]
\begin{center}
\caption{New eclipsing binaries identified with ASTEP South and known objects with improved classification and parameters. The ID, coordinates, and magnitudes fmag are extracted from the UCAC3 catalog. For each object, we report the period, the mid-eclipse date of the first detected eclipse, the depth of the primary eclipse (or of the secondary eclipse in the potential case of eccentric binaries with no primary eclipses), and the type of eclipsing binary. Numbers in parentheses indicate the 3$\sigma$ uncertainties on the last two digits for the periods and ephemerides, and the 3$\sigma$ uncertainties for the depths.}
\label{tab: EBs}
\begin{tabular}{ccccrrrc}
\hline
\hline
ID & RA & Dec & fmag & \multicolumn{1}{c}{Period} & \multicolumn{1}{c}{$\rm T_0$} & \multicolumn{1}{c}{Depth} & Type  \\
(UCAC3) & (J2000) & (J2000) & [mag] & \multicolumn{1}{c}{[d]} & \multicolumn{1}{c}{[BJD]} & \multicolumn{1}{c}{[mmag]} &   \\
\hline
\hspace{0.5mm} \object{003-001765}$^*$ & 06:53:22.343 & -88:34:04.28 & 10.04 & 0.4574110 (48) & 4627.4151 (34) &  2.6 (0.4) & EB \\
 \object{002-002671} & 16:04:20.122 & -89:00:40.74 & 10.99 & 13.4892 (25) & 4627.70 (13) &  5.7 (0.4) & EB \\
 \object{002-003763} & 23:23:51.240 & -89:25:17.62 & 11.25 & 23.3853 (53) & 4636.35 (18) &  8.6 (0.4) & EB \\
 \object{004-006956} & 18:03:31.003 & -88:16:36.79 & 11.86 & 36.263 (24) & 4634.18 (23) &  5.9 (0.4) & EB \\
\hspace{0.3mm} \object{004-000272}$^{**}$ & 00:45:26.547 & -88:25:21.03 & 11.98 & 34.559 (19) & 4627.69 (19) & 15.0 (0.7) & EB \\
\hspace{0.5mm} \object{001-000823}$^\dagger$ & 14:50:14.729 & -89:46:58.96 & 12.18 & 0.28891080 (20) & 4627.2346 (02) & 479 (8) & EW \\
 \object{001-000701} & 12:41:48.571 & -89:42:30.30 & 12.69 & 54.133 (15) & 4668.37 (20) & 38 (2) & EB \\
 \hspace{0.3mm} \object{002-003056}$^{\dagger\dagger}$ & 18:36:30.712 & -89:11:02.84 & 12.76 & 75.57626 (20) & 4629.9895 (60) & 30 (2) & EA \\
 \object{001-001090} & 20:37:01.418 & -89:44:34.88 & 12.82 & 23.7052 (70) & 4632.38 (10) & 10 (1) & EW \\
 \object{001-000120} & 02:26:38.190 & -89:31:31.72 & 14.45 & 81.051070 (20) & 5058.9223 (24) & 335 (5) & EA \\
 \object{001-000682} & 12:27:06.862 & -89:48:58.91 & 14.59 & 1.048938 (30) & 4627.2738 (97) & 28.0 (0.8) & EB \\
\hline
\hline
\end{tabular}
\end{center}
Notes. \modif{$^*$This object was reported as a rotating ellipsoidal variable by \citet{Wang2015}. $^{**}$This object was reported as a variable of unknown type by \citet{Wang2011} and was not identified by \citet{Wang2013}.} $^\dagger$This object is \object{EN~Oct}: we improved its period and eclipse depth compared to previous studies by \citet{Knigge1967, Otero2004}. $^{\dagger\dagger}$This object is discussed in Sect.~\ref{sec: asud-1184}.
\end{table}

\end{document}